\documentclass{aa}
\usepackage{graphicx}
\usepackage{natbib}
\bibpunct{(}{)}{;}{a}{}{,} 
\usepackage{rotating}
\usepackage{amsmath}
\usepackage{amssymb}
\usepackage{array}
\usepackage{txfonts}
\newcounter{one}   
\newcounter{two}   
\newcounter{three} 
\newcounter{four}  
\newcounter{five}  
\newcounter{six}   

\newcommand{\rxj}{\mbox{RX J1347$-$1145 }}

\begin{document}
	 
\title{The mass distribution of \rxj from strong lensing\thanks{Based on observations made with the NASA/ESA Hubble
    Space Telescope, obtained from the data archives at the Space
    Telescope European Coordinating Facility and the Space Telescope
    Science Institute, which is operated by the Association of
    Universities for Research in Astronomy, Inc., under NASA contract
    NAS 5-26555.}}
\author{A. Halkola\inst{1}\thanks{E-mail:halkola@astro.uni-bonn.de}
  \and H. Hildebrandt\inst{1}
  \and T. Schrabback\inst{1}
  \and M. Lombardi\inst{2}
  \and M. Brada{\v c}\inst{3,4}
  \and \\T. Erben\inst{1}
  \and P. Schneider\inst{1}
  \and D. Wuttke\inst{1}
}
\institute{ Argelander-Institut f\"ur Astronomie, Auf dem H\"ugel 71,
  53121 Bonn, Germany
\and European Southern Observatory, Karl-Schwarzschild-Strasse 2
85748 Garching bei M\"unchen, Germany
\and Kavli Institute for Particle Astrophysics and Cosmology,2575 Sand
Hill Rd. MS29, Menlo Park, CA 94025, USA
\and Department of Physics, University of California, Santa
Barbara, CA 93106, USA
}

  \date{Received October 2007 / Accepted January 2008}

  \abstract{}
  {We determine the central mass distribution of galaxy cluster \rxj
    using strong gravitational lensing.}
  {High resolution HST/ACS images of the galaxy cluster \rxj have
    enabled us to identify several new multiple image candidates in
    the cluster, including a 5 image system with a central image. The
    multiple images allow us to construct an accurate 2-dimensional
    mass map of the central part of the cluster. The modelling of the
    cluster mass includes the most prominent cluster galaxies modelled
    as truncated isothermal spheres and a smooth halo component that
    is described with 2 parametric profiles. The mass reconstruction is
    done using a Markov chain Monte Carlo method that provides us with
    a total projected mass density as well as estimates for the
    parameters of interest and their respective errors.}
  {Inside the Einstein radius of the cluster ($\sim$35$''$, or
    $\sim$200 kpc, for a source at redshift 1.8) we obtain a total
    mass of \mbox{(2.6 $\pm$ 0.1) $\times$ 10$^{14}$ M$_{\odot}$}.
    The mass profile of the cluster is well fitted by both a Navarro,
    Frenk and White profile with a moderate concentration of $c$ =
    5.3$^{+0.4}_{-0.6}$ and $r_{200}$ = 3.3$^{+0.2}_{-0.1}$ Mpc, or a
    non-singular isothermal sphere with velocity dispersion $\sigma$ =
    1949 $\pm$ 40 km/s and a core radius of $r_{c}$ = 20 $\pm$ 2 $''$.
    The mass profile is in reasonable agreement with previous mass
    estimates based on the X-ray emission from the hot intra-cluster
    gas, however the X-ray mass estimates are systematically lower
    than what we obtain with gravitational lensing.} {}
     

  \maketitle

%

\section{Introduction}
\label{sec:intro}

A precise knowledge of the masses and mass profiles of galaxy clusters
is a key to better understand what the mass and energy in the universe
is made of. The standard model of cosmology that has emerged over the
past 100 years is remarkably successful in reproducing a wealth of
observational phenomena. These include the formation and evolution of
structure from the tiny temperature variations seen in the cosmic
microwave background to galaxies and clusters of galaxies, the element
abundances produced in the Big Bang nucleosynthesis and the apparent
acceleration of the expansion of the universe as seen from the
brightnesses of distant supernovae. The greatest drawback of the
standard model is the need to include two dark components in the
energy density of the Universe. The first is dark matter which is
needed to explain the formation of structure and the dynamics of both
galaxies and clusters of galaxies. The second is dark energy which in
turn is needed to make the Universe flat and to explain the supernova
data. Both of these dark components are still unexplained although
candidate particles for dark matter exist.

Clusters of galaxies are of prime interest when trying to constrain
the properties of both dark matter and dark energy. The mass budget
and therefore the dynamics of clusters are dominated by dark matter,
whereas the mass function and power spectrum of clusters is influenced
by both dark matter and dark energy. In order to obtain strong
constraints on the various cosmological parameters such as
$\Omega_{\mathrm{m}}$, $\sigma_{8}$ and the equation of state $w$ of
dark energy, accurate mass estimates for many clusters are necessary.
Gravitational lensing is a powerful tool for determining cluster
masses since the deflection of light is independent of the nature and
dynamical state of the matter in clusters and can therefore provide
unbiased estimates of the total masses of galaxy clusters.

An important prediction of numerical simulations with cold dark matter
is the so called universal mass profile. For large radii (greater than
the scale radius) the density $\rho$ depends on $r$ like $\rho(r)
\propto r^{-3}$, the exact value of the slope at small radii (inside
the scale radius) is unknown but is believed to be somewhere between
$-1.5$ and $-1$.  Although there is strong support for the universal
mass profile, other mass profiles such as the family of isothermal
sphere profiles have not been ruled out.  By studying the mass
profiles of galaxy clusters we can potentially constrain the nature of
dark matter through comparison with simulations.

As the most luminous X-ray cluster \citep{schindler:95} known to date
\rxj is obviously of great interest. It has been studied
spectroscopically by \citet{cohen:02} and extensively with X-rays and
Sunyaev-Zel'dovich effect
\citep{schindler:97,pointecouteau:99,pointecouteau:01,allen:02,gitti:04,gitti:07}.
Gravitational lensing based only on the arcs has also been used to
estimate the mass \citep{sahu:98} as well as weak lensing
\citep{fischer:97,kling:05}. Recently a combination of both the weak
and strong lensing has also been used \citep{bradac:05}. The lack of
space-based observations with the Hubble Space Telescope (HST) has so
far limited the number of multiple images available for the strong
lensing modelling. In this paper we use deep HST images taken with the
Advanced Camera for Surveys (ACS) to identify new multiple image
candidates that enable us to derive well constrained mass maps for the
central regions of \rxj.  We have obtained spectra of some of the
multiple images with the FORS2 spectrograph on the Very Large
Telescope in Chile. The spectroscopic redshifts obtained are crucial
in fixing the mass scale of the cluster.

The cosmology used throughout this paper is
$\Omega_{\mathrm{m}}$=0.30, $\Omega_{\Lambda}$=0.70 and
H$_{0}$=70~km/s/Mpc, unless otherwise stated. With this cosmology an
arcsecond at the redshift of the cluster ($z_{\mathrm{cl}}$=0.451)
corresponds to 5.77 kpc.


\section{Data}
\label{sec:data}

\subsection{HST Imaging}
\label{sec:data:imaging}

Strong gravitational lensing requires deep, high quality and high
resolution images to facilitate the identification of multiple images
in galaxy clusters. The best instrument for this is the ACS/WFC on
board the HST.  Following the strong and weak lensing analysis of
\citet{bradac:05}, we have acquired imaging data in filter bands
F475W, F814W and F850LP, 5280s in each band (PI: T.  Erben, proposal
no. 10492). This gives us not only high resolution images but also
limited colour information for the galaxies.

The data reduction is based on the bias and flat-field corrected
images obtained with the ACS calibration pipeline CALACS. Before the
standard pipeline reduction proceeds we compute a noise model for each
exposure. These are later used to appropriately weight the individual
exposures when coadding the exposures. At this point also the badpixel
masks are updated. For more details on the noise modelling and bad
pixel masking see Marshall et al. (in preparation).

The two CCDs on the ACS have 2 readout ports each, and this can
sometimes lead to residual bias level between the four image
quadrants. We have therefore subtracted the sky background separately
in each of the four image quadrants.

We finally use MultiDrizzle \citep{koekemoer:02} to coadd all the
individual exposures of each filter. This step also corrects for
geometric distortion caused by the telescope optics and detectors, and
rejects the cosmic rays present in individual frames. In order to
accurately align all the images taken in the 3 bands, we select high
$S/N$ objects in the individual exposures and compare their positions
to find out any residual shifts and rotations between the images. The
necessary corrections given to MultiDrizzle are calculated using the
geomap task in IRAF. In the final coaddition we use the noise models
calculated earlier to weight the individual exposures using inverse
variance weighting. We apply the "square" drizzle kernel in
MultiDrizzle in combination with a reduced pixel scale of 0\farcs03
and slightly shrunk pixels (pixfrac=0.9). More details on the data
reduction can be found in Marshall et al. (in prepararion).

\subsection{Photometric redshifts}
\label{sec:data:photoz}

The photometric redshifts used in this work are calculated from the 5
band catalogue based on images taken with the VLT/FORS in the $U$,
$B$, $V$, $R$, and $I$ bands. For more details on the data and data
reduction see \citet{bradac:05}.The photometric redshifts are
calculated using the Hyperz package \citep{bolzonella:00}. We do not
use any prior on the magnitude which suppresses the probability
density at high redshifts.  The multiple images we are interested in
are often highly magnified which increases the flux coming from a
background galaxy. In many cases the multiple images are still very
faint however and it is very difficult to obtain good photometric
redshifts for the images. The three ACS bands on the other hand are
not enough to give a photometric redshift estimate on their own, and
combining them with the ground based date can introduce
systematics/biases. To check this, we have also includes the ISAAC
$Ks$-band data in the photometric redshift estimation. The best-fit
photo-$z$'s of the multiple images are barely affected and
additionally some of the probability distributions even get broader
and less well defined when using the $Ks$-band. This might be due to
slight photometric miscalibrations and difficulties in the
equalisation of the point spread function of the different images
leading to systematic colour errors. The photometric redshifts
probability distributions are compared to the lensing ones in Sects.
\ref{sec:results:lens_z} and \ref{sec:results:all_images:lens_z}.

\subsection{VLT Spectroscopy}
\label{sec:data:spectroscopy}

Spectra were taken for potential cluster members, a magnitude limited
sample of galaxies (to be used as a calibration sample for photometric
redshifts) as well as multiple image candidates. The spectra were
taken with the FORS2 spectrograph at the Very Large Telescope of the
European Southern Observatory.  Reduction of the spectra and detailed
results will be presented in a forthcoming publication (Lombardi et
al. 2008, in preparation). In this work we only use the spectra and
spectroscopic redshifts obtained for multiple image systems as
detailed in later sections.

\section{Components of the Strong Lensing Models}
\label{sec:SLcomponents}

In this section we describe the strong lensing modelling of the
cluster in detail.

\subsection{Mass components}
\label{sec:SLcomponents:mass}

The masses of clusters of galaxies have three major (distinct)
components: the galaxies and their DM haloes, the hot intra cluster
gas and the dark matter (DM) halo of the cluster. Although the
galaxies contain only a small fraction of the total mass they have a
significant local effect on the multiple images and need to be
included in the modelling.  The galaxies in a cluster affect strongly
the formation of strong gravitational lensing features such as arcs
and multiple images \citep[e.g.][]{meneghetti:07}. It has also been
investigated how the increased cross section of galaxy-sized haloes in
denser group or cluster environments can potentially be used to
constrain the mass profiles of clusters and the dark matter content of
the galaxies themselves \citep{king:07,tu:07}. In this work, however,
we do not attempt to do this.  We assume that the hot \mbox{X-ray} gas
and the DM together can be described by two dark matter haloes.

\subsubsection{Cluster Galaxies}
\label{sec:SLcomponents:mass:galaxies}

\begin{figure}
  \centering
  \includegraphics[height=1.0\columnwidth]{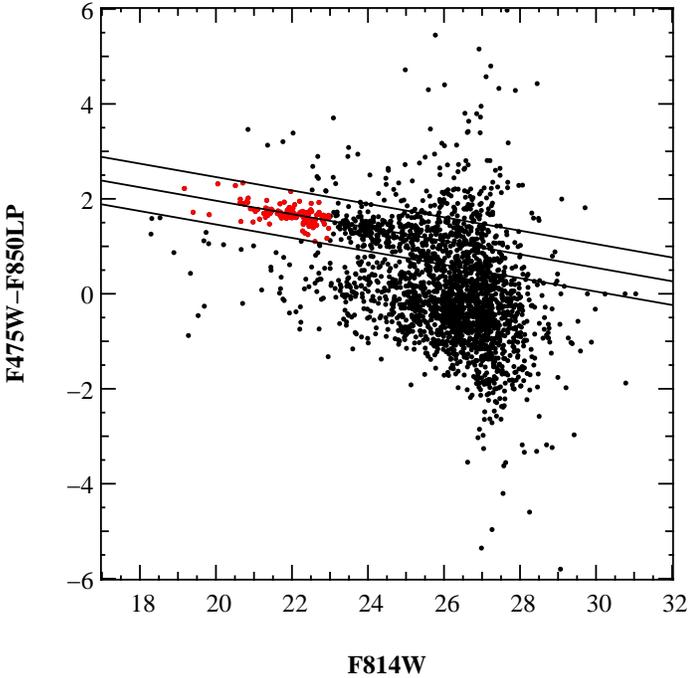}
  \caption{The red sequence of galaxy clusters in the F475W$-$F850LP,
    F814W colour-magnitude diagram. The galaxies that are included in
    the mass model are marked with red circles and have F475W$-$F850LP
    colour within 0.5 magnitudes from the mean of the red sequence.}
  \label{fig:redsequence}
\end{figure}

We have identified the cluster galaxies from the cluster red sequence.
The red sequence in the ACS F475W$-$F850LP, F814W colour-magnitude
diagram is shown in Fig. \ref{fig:redsequence}. We have selected
lensing galaxies based on their magnitude and F475W$-$F850LP colour.
All galaxies brighter than 23 (AB) in the F814W passband with
F475W$-$F850LP colour within 0.5 magnitudes from the mean of the red
sequence are included. We have additionally included all
spectroscopically confirmed cluster galaxies from \citet{cohen:02}.
This results in a total of 119 galaxy lenses in our lensing modelling.

The profiles of the cluster galaxies are based on a truncated
isothermal sphere profile \citep{brainerd:96}. The profile is
characterised by a velocity dispersion $\sigma$ and a truncation
radius $s$. The 3 dimensional mass density of a truncated isothermal
sphere is given by:

\begin{equation}
  \rho_{\mathrm{bbs}}(r) = \frac{\sigma^2}{2 \pi G r^2}\frac{s^2}{r^2+s^2}.
\end{equation}
The galaxy haloes can be elliptical with the ellipticity implemented
following the method presented in \citep{golse:02b}. All the galaxies
included are treated in exactly the same way. This includes also the
two massive central galaxies and the galaxies near multiple images.

The velocity dispersions of the galaxies are estimated from their
F814W magnitude using the Faber-Jackson relation. Since we have been
unable to find a good reference point to a fundamental plane or a
Faber-Jackson relation at the redshift of the cluster we have used the
fundamental plane known for the galaxy clusters A2218 and A2390
\citep{ziegler:01,fritz:05} at redshifts 0.18 and 0.23 respectively.
These same clusters were used as reference points when estimating the
velocity dispersions of the galaxies in A1689 in \citet{halkola:06}.
To model the galaxies in \mbox{RX J1347} we have taken a model
spectrum of an early type elliptical galaxy with a formation redshift
of 5 from \citet{bruzual:03}, and passively evolved it from the
redshift of RX J1347 ($z=0.451$) to 0.2.  This model is used to
calculate both the K-correction from the observed F814W to the F775W
of the Faber-Jackson relation shown in \citet{halkola:06} for clusters
at redshift 0.2, and the evolution of the galaxy colours. The measured
F814W magnitudes are corrected for galactic extinction according to
\citet{schlegel:98} obtained using the NASA/IPAC Extragalactic
Database (NED).

We assume a linear relation between the truncation radius $s$ of a
galaxy and its velocity dispersion $\sigma$. This is expected on
theoretical grounds \citep{merritt:83}.  Recently this topic has also
been investigated using simulations that take into account the effect
of baryons by \citet{limousin:07b}. These N-body simulations were done
in order to study the tidal stripping of dark matter haloes of
galaxies in a cluster, including star formation so that the stellar
components of the galaxies can be followed. The simulations agree with
trends observed using galaxy-galaxy lensing in galaxy clusters,
e.g. \citet{natarajan:02,limousin:07a}.  It was additionally shown in
\citet{halkola:07} that for the strong lensing analysis of Abell 1689
(which has much stronger constraints on the mass from the multiple
image systems identified in the cluster) it was not possible to
constrain the slope of the power-law relation between $s$ and
$\sigma$. Since the multiple images in \mbox{RX J1347} do not
constrain the sizes of the galaxies significantly we assume the same
truncation law for the galaxies in \mbox{RX J1347} as was found for
galaxies in the galaxy cluster A1689 \citep{halkola:07} as the
clusters are both similarly massive.

\subsubsection{Smooth cluster mass}
\label{sec:SLcomponents:mass:smooth}

The X-ray measurements as well as the combined strong and weak lensing
analysis of the cluster indicate that there is possibly a mass
extension to the south-east of the cluster centre. The mass maps are
otherwise fairly smooth and we therefore model the smooth mass
component of the cluster with two parametric profiles. The second halo
is also necessary in order to accurately reproduce the multiple images
observed. Using parametric haloes to reconstruct the mass of galaxy
clusters, as opposed to pixelated mass reconstructions for example,
have several advantages but also disadvantages. The simple description
of the mass distribution gives generally very good convergence
properties when optimising the model parameters. On the other hand the
mass profile is also relatively restricted in the mass distributions
that can be recovered. Although the second halo could in principle be
connected to a physical structure in the cluster, it is also possible
that a secondary halo is necessary in the modelling in order to
accurately model the mass distribution of the cluster with only one
distinct halo. These two haloes are both described by either
Navarro-Frenk-White \citep[NFW][]{navarro:97} or by non-singular
isothermal ellipsoid (NSIE) haloes.

The NFW halo is a prediction of cosmological numerical simulations and
its 3-dimensional mass density is given by

\begin{equation}
  \rho_{\rm{nfw}}(r) = \frac{ \rho_{\rm{c}} }{(r/r_{\rm{s}})(1+r/r_{\rm{s}})^2 },
\end{equation}
where $\rho_{\rm{c}}$ is the characteristic density of the cluster and
$r_{\rm{s}}$ is the scale radius which marks the transition from $\rho
\propto r^{-1}$ to $\rho \propto r^{-3}$.

The NSIE profile is based on the popular isothermal sphere profile
with the inclusion of a constant density core. The 3 dimensional mass
density of the NSIE profile is given by

\begin{equation}
  \rho_{\rm{nsie}}(r) = \frac{\sigma^2}{2 \pi G} \frac{1}{(r^2+r_{\rm{c}}^2)},
\end{equation}
where $\sigma$ is the velocity dispersion of the isothermal sphere and
$r_{\rm{c}}$ is the core radius of the sphere.

Both mass profiles have 6 free parameters: their position on the sky
($\alpha$, $\delta$), the ratio of their semi-minor and semi-major
axes $r=b/a$, the position angle $\theta$, and the two parameters
described above that define the density profile.

\subsection{Multiple images}
\label{sec:SLcomponents:images}

\begin{figure*}
  \centering
  \includegraphics[height=2.0\columnwidth]{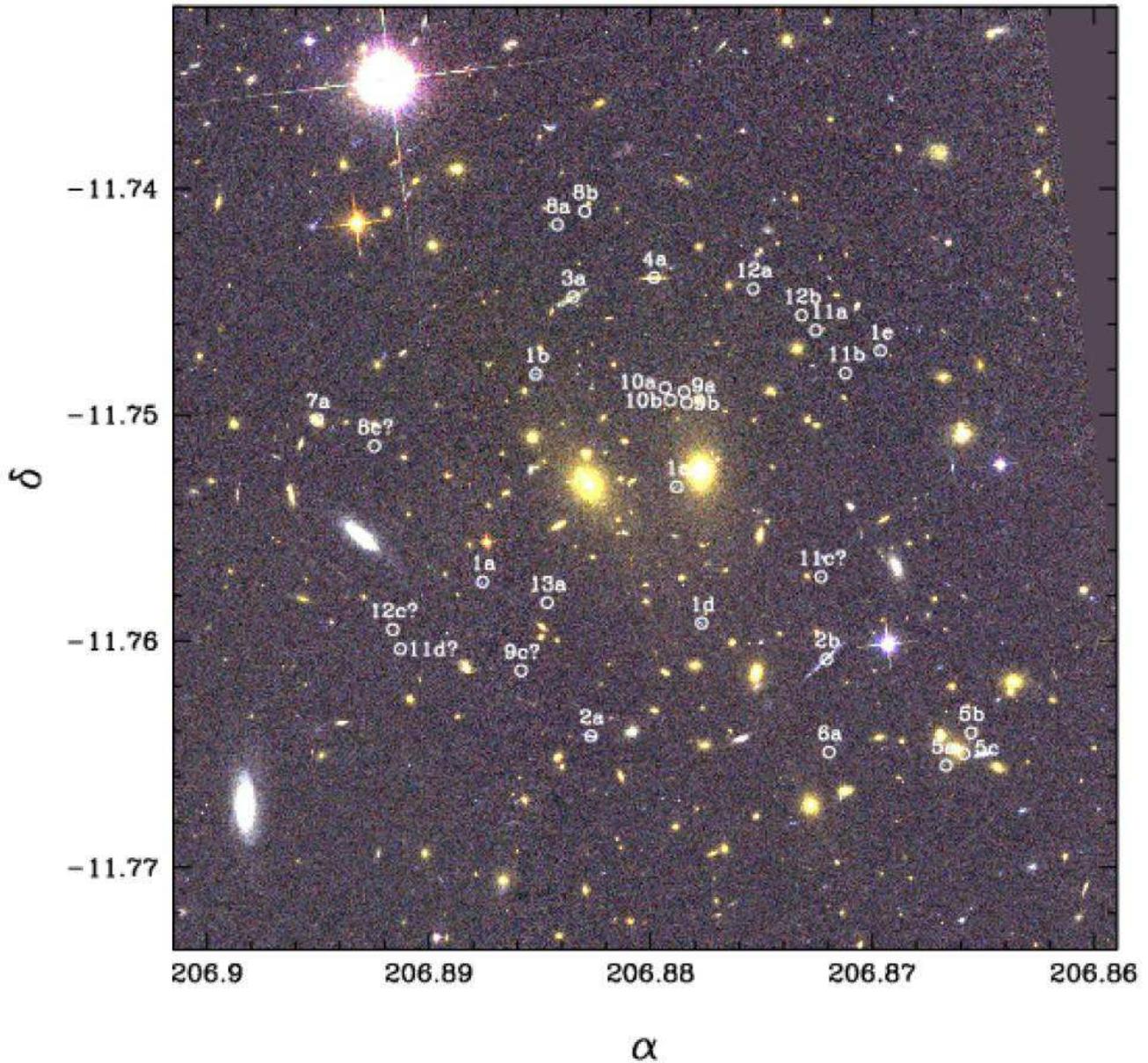}
  \caption{The multiple image systems identified in \rxj overlaid on a
    colour image of the cluster made from the images taken with the
    ACS in filter F475W, F814W, and F850LP.  The box size is $\sim$880
    kpc, north is up, east is left. Uncertain image candidates are
    marked with a question mark. Only image systems 1 and 2
    considerably constrain the mass profile of the cluster.}
  \label{fig:images}
\end{figure*}

In the literature there are several published multiple image
candidates in RX J1347 \citep[e.g.][]{sahu:98,bradac:05}. In addition
to these we have identified several new ones in the HST ACS images of
the cluster, bringing the total number of candidate lensing system to
13. The images are shown in Fig. \ref{fig:images} and some of their
properties are shown in Tab. \ref{tab:images} and thumbnail images in
Figs. \ref{fig:images:1} - \ref{fig:images:12-13} of the appendix. The
more tentative image candidates in a system are marked with a question
mark in Tab. \ref{tab:images} and Fig. \ref{fig:images}. The following
sections discuss the multiple image systems in some detail.

The main analysis in this paper is based on image systems 1 and 2
only. The locations of the 5 images in image system 1 provide strong
constraints on the distribution of mass in the cluster while the
spectroscopic redshift of image system 2 is important in fixing the
total mass scale. In section \ref{sec:results:all_images} we repeat
the analysis with all image systems to check the effect of including
all of them, and to obtain lensing redshift probability distributions
for the image systems.

\subsubsection{Image system 1}
\label{sec:SLcomponents:images:1}

This is a multiple image system with 5 images on 4 sides of the
cluster and a central image. The identification is based on the
similar colours and morphology of the images, all images have a
brighter knot at one end. Also the photometric redshift probability
distributions of the multiple images in this system are very similar.
They are flat and fairly broad and place the multiple image system in
a redshift range 0.5-2.0. The \mbox{M$_{\rm{F475W}}$ $-$
  M$_{\rm{F814W}}$} and \mbox{M$_{\rm{F814W}}$ $-$ M$_{\rm{F850LP}}$}
colours of the multiple images are also in good agreement. We have
obtained spectra for all the multiple images, but have not been able
to identify any emission lines in the individual spectra. The many
skylines in the spectra also hinder us from cross correlating the
individual spectra to confirm the compatibility of the spectra of the
multiple images. This mean that we cannot confirm the multiple image
nature of the source from the spectra. Although the redshift cannot be
fixed, the geometry of the images provide significant constraints on
the mass profile.

\subsubsection{Image system 2}
\label{sec:SLcomponents:images:2}

Our image system 2 is composed of two images. The brighter of the two
arcs was already identified by \citet{schindler:95} as the south-west
arc. The fainter arc was later found by \citet{sahu:98}. We have taken
spectra of the two arcs and obtained a redshift of z$_{\rm{spec}}$ =
1.75 for the two of them; for more details see Lombardi et al. (2008,
in prep.).  The redshifts of the images fixes the overall mass scale
of the cluster and provides strong support for them being multiple
images.  In \citet{bradac:05} these correspond to images A4 and
A5. Our models do not predict any additional detectable images for
this system.

\subsubsection{Image system 3}
\label{sec:SLcomponents:images:3}

Image system 3 consists of only one distorted image. The spectroscopic
redshift is z$_{\rm{spec}}$ = 0.806 \citep{ravindranath:02}, so this
galaxy is a background galaxy and must be lensed. In the absence of
counter images it cannot be effectively used to constrain the mass
profile. Also the lack of multiple images can constrain the mass
distribution, see \citet{jullo:07}. This is only effective if the
singly lensed images are relatively close to critical regions so that
multiple images are formed for model parameters that fulfill the other
constraints reasonably well. In our analysis we have only afterwards
checked that no additional images for this system are predicted.

\subsubsection{Image system 4}
\label{sec:SLcomponents:images:4}

Similarly to image system 3 this one consists of only one distorted
image and therefore cannot effectively be used to constrain the mass
model. The spectroscopic redshift of this galaxy is z$_{\rm{spec}}$ =
0.785 \citep{ravindranath:02}. No additional images for this system
are predicted.

\subsubsection{Image system 5}
\label{sec:SLcomponents:images:5}

This image system consists of 3 faint images that are split by a
cluster galaxy. The images are very sensitive to the mass in the
galaxy which can therefore influence the total mass distribution
significantly when this image system is included in the analysis.

\subsubsection{Image system 6}
\label{sec:SLcomponents:images:6}

A faint thin arc at a cluster centric radius of $\sim$12'' larger than
image system 2, and hence expected to be at a higher redshift. No
counter images found. This is image system C in \citet{bradac:05}.

\subsubsection{Image system 7}
\label{sec:SLcomponents:images:7}

An arc that nicely bends around the galaxy at $ \alpha$=13:47:34.8,
$\delta$=$-$11:45:01.5. The exact shape of the arc is sensitive to the
mass of the galaxy.

\subsubsection{Image system 8}
\label{sec:SLcomponents:images:8}

This is a 2-image system merging on a tangential critical line, also
identified in \citet{bradac:05} as a possible arc candidate. There is
a tentative third image on the east side of the cluster but a positive
identification is difficult due to the faintness of the source.

\subsubsection{Image system 9}
\label{sec:SLcomponents:images:9}

2 images merging on a radial critical line. We have identified a
possible third image in the south-east part of the cluster.

\subsubsection{Image system 10}
\label{sec:SLcomponents:images:10}

A similar merging of 2 images at a radial critical line, close to
image system 9. This one is much fainter though and less prominent and
consequently less certain. A third image is expected near the third
image of image system 9 but as this source is significantly fainter we
have not identified one.

\subsubsection{Image systems 11 and 12}
\label{sec:SLcomponents:images:11and12}

In the north-west of the cluster there are several images that define
a tangential critical line. The colours and morphology of the images
vary, suggesting that the images originate from several different
sources at a similar redshift. Possible counter images on the
south-east and south west sides of the cluster are identified for
system 11. The fainter image system 12 is more challenging in this
respect although several candidates have been identified. In
\citet{bradac:05} identified as D1-D4.

\subsubsection{Image system 13}
\label{sec:SLcomponents:images:13}

A bluish galaxy oriented tangentially with respect to the cluster
centre. It is potentially lensed although we have not been able to
identify any counter images to this one.

\section{Estimating the cluster mass distribution and its error}
\label{sec:SLoptimising}

In order to estimate the mass distribution of the cluster we use
Bayesian statistics with Markov chain Monte Carlo methods to estimate
the probability distribution function of the free parameters of the
cluster mass model.

Our mass models have 13 free parameters in total, the redshift of
multiple image system 1 and 2$\times$6 parameters for the 2 smooth
haloes for the global mass distribution.

\subsubsection{Model $\chi^2$}
\label{sec:SLoptimising:chi2}

Both the positional information as well as the relative magnifications
of the images in an image system are used in calculating the $\chi^2$
for the models. The $\chi^2$ is given by

\begin{equation}
  \chi^2 = \sum_{i=1}^{N} \sum_{j=1}^{n_i} \frac{|\ \vec{\beta}_{i,j}\ -\
    \langle \vec{\beta} \rangle _i |^2}{(\sigma_{i,j}/\sqrt{\mu_{i,j}})^2}
  + \sum_{i=1}^{N} \sum_{j=2}^{n_i} \frac{ (\
    f_{i,1}/\mu_{i,1}-f_{i,j}/\mu_{i,j}\
    )^2}{(e_{i,1}/\mu_{i,1})^2+(e_{i,j}/\mu_{i,j})^2}.
  \label{eqn:SLoptimising:chi2}
\end{equation}
In equation \ref{eqn:SLoptimising:chi2}, $N$ is the number of
multiple image systems, $n_i$ the number of images in system $i$ and
$\vec{\beta}_{i,j}$ denotes the source position of image $j$ in image system
$i$, $\mu_{i,j}$ is the model magnification at the image position.
$f_{i,j}$ and $e_{i,j}$ represent the isophotal flux and its estimated
error of image $j$ in image system $i$. The first double summation in
equation \ref{eqn:SLoptimising:chi2} therefore represents the scatter
of the source positions weighted by the model magnification. The
errors are assumed to scale with the square root of the magnification
since the magnification is given by the ratio of an area of the image
in the image plane to that of the unlensed image in the source plane.
It was shown in \citet{halkola:06} that this is a good estimate for
the $\chi^2$ calculated in the image plane.  The second double
summation includes the relative magnifications between the images of
an image system in the modelling. This is calculated by comparing the
unlensed source fluxes of the multiple images. We use the first
multiple image in each system as a flux reference.

\subsubsection{Bayesian statistics}
\label{sec:SLoptimising:bayes}

Bayes' theorem deals with conditional probability distributions. We
are specifically interested in constraining a model with parameters
$w$ given data $d$ (and possibly some other prior knowledge of the
system). The probability of a parameter set $w$ given the data $d$ can
be written as

\begin{equation}
  P(w|d) = \frac{P(d|w)P(w)}{P(d)},
  \label{eqn:SLoptimising:bayes:theorem}
\end{equation}
where $P(d|w)$ is the probability of the data $d$ given the parameters
$w$, $P(w)$ is the prior on the parameters $w$ and $P(d)$ is the
probability of obtaining the data in the first place, also known as
evidence. $P(d|w)$ is measured by the $\chi^2$ of our models and we
will assume that it is proportional to $\rm{exp}(-\chi^2/2)$. We
assume flat priors $P(w)$ for the parameters that are reasonable for
DM haloes of galaxy clusters. $P(d)$ is treated as a normalising
constant. With these simplifying assumptions it only remains to
estimate $P(w|d)$. We do this with a Markov chain Monte Carlo method
described below.

\subsubsection{Markov chain Monte Carlo}
\label{sec:SLoptimising:mcmc}

The Markov chain Monte Carlo method is a popular way to estimate the
errors in model parameters. It allows one to estimate the conditional
probability distribution of parameters by statistically sampling the
parameter space. In the Metropolis-Hasting algorithm the parameter
space is explored in a random walk. It is a rejection sampling
algorithm which means that the next set of random parameters is
accepted with a probability that depends only on the difference in
probability between successive sets of parameters. Given parameter
sets $w_i$ with $\chi^2(w_i)$ and a new perturbed parameter set
$w_{i+1}$ with $\chi^2(w_{i+1})$ the probability $p$ of accepting the
new parameter set $w_{i+1}$ is given by the ratio of the likelihoods
of the parameter sets:

\begin{equation}
  2\ \ln(\,p\,) = \left\{\begin{array}{ll}
      \chi^2(w_i) - \chi^2(w_{i+1})&,\ \ for\ \chi^2(w_{i+1}) > \chi^2(w_i) \\
      0&,\ \ for\ \chi^2(w_{i+1}) \le \chi^2(w_i) \\
      \end{array} \right..
    \label{eqn:SLoptimising:mcmc}
\end{equation}
The new perturbed parameter sets $w_{i+1}$ are generated by random
selection from Gaussian distributions centred on the individual
parameters in $w_{i}$, and the widths of the Gaussian distributions
are matched so that when each parameter is varied individually the
effect on the acceptance rates converge to the same value of
$\sim$60\%. In the end all the widths are scaled so that an acceptance
rate of $\sim$60\% is achieved also when all the parameters are varied
simultaneously. Another possibility would be to assume that the width
is given by the error in the parameters. This would however require
prior knowledge on the errors of the parameters.  By tuning the widths
according to the acceptance rate we ensure that all free parameters of
the model have the same weight in the Markov chain. The acceptance
rate of 60\% is chosen as a compromise between denser local sampling
of the parameter space at the cost of poorer global sampling (smaller
width, higher acceptance rate), and an attempt to sample quickly a
large volume of the parameter space that is hindered by low
acceptance.

Each Markov chain is started with a mass model that has mass only in
the galaxies. The two haloes describing the smooth mass component are
set to be circular and to have no mass (\mbox{$\sigma$ = 0 km/s} for
the NSIE profiles, and \mbox{$\rho_{c}$ = 0 M$_{\odot} /
  \rm{kpc}^{3}$} for the ENFW profiles). The two haloes are placed
between the two bright central galaxies.

The Markov chains have a total length of \mbox{250 000}. After the
first \mbox{$\sim$10 000} parameter sets $w_{i}$, the acceptance rate
has stabilised at 60\% as well as the scatter in the $\chi^2$ values
for the parameter sets has stabilised. Although a `burning in' of a
Markov chain is strictly not necessary, we discard the first \mbox{10
  000} parameter sets as these still show some memory on the initial
conditions where haloes had no mass.

Each Markov chain has a fixed description for the mass in the
galaxies. This means that individual Markov chains do not take into
account the uncertainties of the velocity dispersion estimates of the
galaxies. In order to account also for these uncertainties we
calculate a total of 200 Markov chains where each has a different
realisation of the galaxy component. In each Markov chain we have
reassigned galaxy velocity dispersions from a Gaussian distribution
centred on the measured value and with a width corresponding to the
estimated error. Half of the Markov chains have a smooth component
described by NSIE haloes and the other half by ENFW haloes.

All the remaining analysis is performed by combining the 200 Markov
chains and randomly selecting \mbox{1000} parameter sets (with
replacement) from the combined chain.

\section{Results}
\label{sec:results}

The multiple images are reproduced well both in position and in
morphology, although the mean separation between the observed image
position and that predicted by the models is still well above the
accuracy with which positions of galaxies can be measured. The mean
separation in the MCMC analysis is $\sim1''$ (for reference, the mean
separation between an observed image position and that predicted by
the models is $\sim3.5''$ when the smooth mass component is described
with only one halo). For comparison, the best models have separations
below $\sim0.5''$. Observed and predicted morphologies of the images
can be seen in Figs. \ref{fig:images:1} and \ref{fig:images:2} in the
appendix, where a discussion can also be found.

We explore the mass distribution of the cluster in two different ways.
First of all we can study the parameter space of the smooth mass
component.  Secondly, we look at the total 2-dimensional mass
distribution of the cluster as well as the averaged radial mass
profile.

\subsection{Parameters of the mass smooth component}
\label{sec:results:parameters}

\begin{figure*}
  \centering
  \includegraphics[height=1.6\columnwidth]{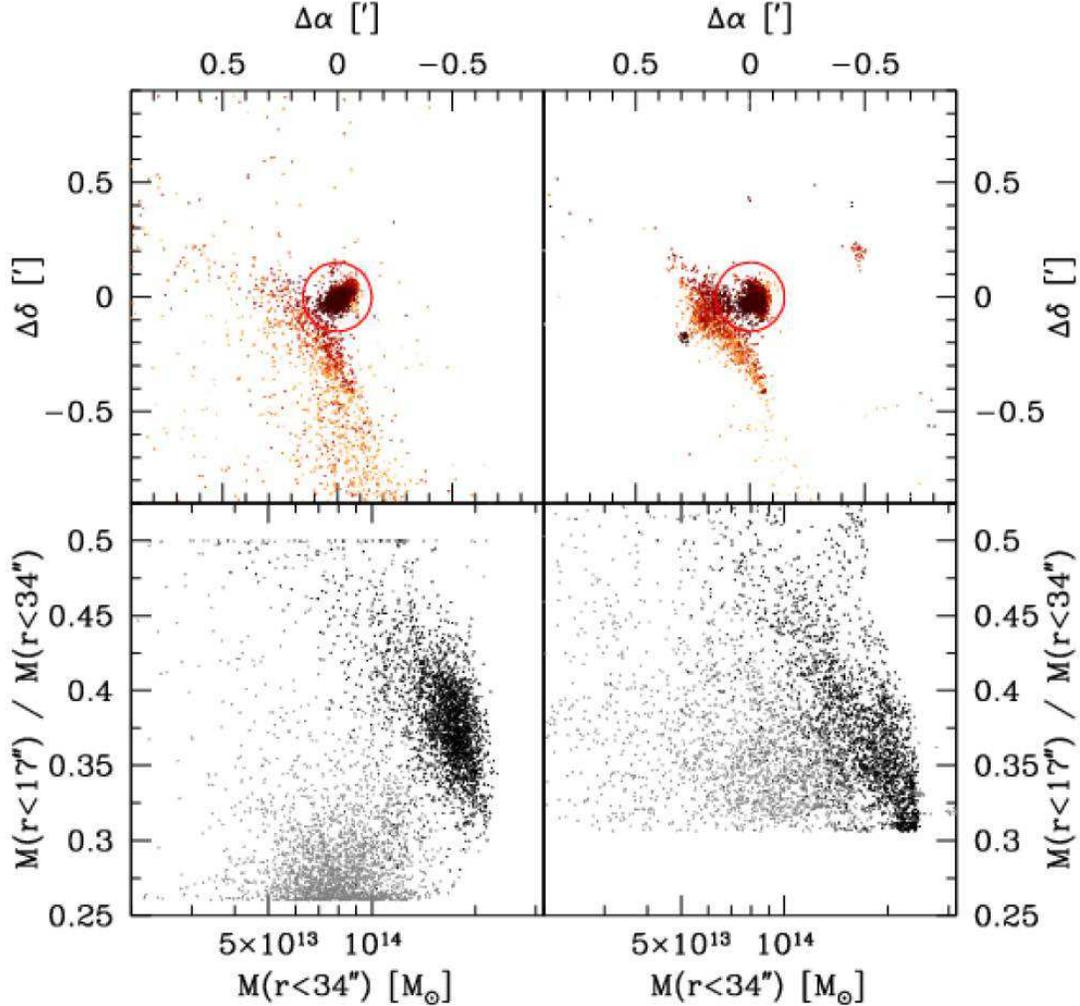}
  \caption{The distribution of the positions of the two smooth haloes
    in the MCMC sampling are shown on the two top panels, the NSIE
    profiles are on the left and the ENFW haloes on the right. One of
    the haloes resides preferentially in a relatively small region of
    the cluster centre, indicated by the red circle. This allows us to
    make a separation between the primary halo (within the circle) and
    the secondary halo (outside the circle).  The points are coloured
    according to the mass of the haloes. The smaller haloes with
    $\rm{M(r<34'') < 5 \times 10^{13} M_{\odot}}$ are light orange in
    colour and the massive haloes with $\rm{M(r<34'') > 2 \times
      10^{14} M_{\odot}}$ are coloured black.  In the bottom panels we
    plot the concentration as measured by the ratio of the halo mass
    at two different radii, $\rm{M(r<17'')} / \rm{M(r<34'')}$, against
    the mass within the Einstein radius of the cluster, $M(r<34'')$.
    The primary haloes (black) are generally both more massive and
    more concentrated then the corresponding secondary haloes (grey).
    The separation is clear for the NSIE haloes, where as the
    concentration of the ENFW profiles are the same for both the
    primary and secondary haloes. The positions on the upper panels
    are relative to $\alpha$ = 13:47:30.9, $\delta$ = $-$11:45:08.6,
    the centre of the circle. }
  \label{fig:mcmc_param3}
\end{figure*}

The distribution of the positions of the smooth DM haloes in the MCMC
analusis are shown in the two top panels in
Fig. \ref{fig:mcmc_param3}. The left panels are for the NSIE halos and
the right panels for the ENFW haloes. We have used the positions of
the haloes to separate the primary and secondary haloes. The primary
halo is defined as the one whose centroid lies within the red circle
shown in the top panels. In a small number of cases both the primary
and the secondary halo reside either inside or outside the red
circle. In these cases we have taken the halo with a larger
$M(r<34'')$ as the primary halo. The points are coloured according to
the mass of the haloes with smaller haloes, $M(r<34'') < 5 \times
10^{13} M_{\odot}$, being light orange and massive haloes, $M(r<34'')
> 2 \times 10^{14} M_{\odot}$, being black. It is clear from the plot
that the central haloes are also the more massive ones, and that the
less massive haloes populate the outer regions of the cluster centre.

The more massive haloes are spatially concentrated in a very small
region for both the NSIE and ENFW profiles. The less massive ones are
more spread out for both profiles, although the NSIE profiles are more
scattered than those described by ENFW profiles.

In order to be able to compare the masses and concentrations of the
NSIE and ENFW haloes we calculate the mass of the haloes at two
different radii ${r_{1}}$ and ${r_{2}}$, where ${r_{1}} < {r_{2}}$. We
can then define concentration ${c_{m}}$ as the ratio of the two masses
${c_{m}} = {M(r<r_{1})}\ /\ {M(r<r_{2})} = {M_{1}}\ /\ {M_{2}}$.  If
all the mass is already contained within ${r_{1}}$ then the mass ratio
is 1, where as for a mass sheet the ratio is given by
$({r_{1}}/{r_{2}})^2$.  For a singular isothermal sphere the maximum
value for ${c_{m}}$ is $({r_{1}}/{r_{2}})$, since ${M(r)} \propto r$.
The outer radius ${r_{2}}$ is taken to be 34 arcseconds which
corresponds roughly to Einstein radius of the cluster, while the inner
radius is arbitrarily defined as ${r_{1}} = {r_{2}} / 2$. The
separation in concentration and mass is not very sensitive to the
chosen radii, ${r_{1}}$ and ${r_{2}}$. On the bottom panels of Fig.
\ref{fig:mcmc_param3} we plot the concentration ${c_{m}}$ against
${M(r<34'')}$ for the NSIE (left) and the ENFW (right) haloes. The
primary haloes are coloured black, while the secondary haloes are
coloured grey. For the NSIE profiles the separation between the
primary and secondary haloes in the ${c_{m}}$ vs. ${M(r<34'')}$ space
is clear with primary haloes being both more massive and more
concentrated with a mean mass of ${M(r<34'')}$ = ( 16.1 $\pm$ 3.3 )
$\times$ $10^{13}$ ${M_{\odot}}$ and a mean concentration of ${c_{m}}$
= 0.38 $\pm$ 0.04 compared to the secondary haloes which have a mean
mass of ${M(r<34'')}$ = ( 8.9 $\pm$ 2.8 ) $\times$ $10^{13}$
${M_{\odot}}$ and a mean concentration of ${c_{m}}$ = 0.30 $\pm$ 0.06.
This separation is less pronounced for the ENFW profile where the
concentrations are virtually same for the primary and the secondary
haloes (${c_{m}}$ = 0.38 $\pm$ 0.06 vs.  ${c_{m}}$ = 0.38 $\pm$ 0.11).
The mean masses of the primary and the secondary haloes are
${M(r<34'')}$ = ( 16.3 $\pm$ 5.0 ) $\times$ $10^{13}$ ${M_{\odot}}$
and ${M(r<34'')}$ = ( 7.9 $\pm$ 5.4 ) $\times$ $10^{13}$ ${M_{\odot}}$
respectively. One should keep in mind that since the ENFW halo has
always an inner logarithmic slope of $-1$, the profile can never reach
as low a concentration as an NSIE profile can. This is also the reason
why the positions of the secondary haloes for the ENFW profile have
less scatter: the position of an ENFW halo is always well defined.

\subsection{Total mass distribution of \rxj}
\label{sec:results:mass}

\begin{figure*}
  \centering
  \includegraphics[width=2.0\columnwidth]{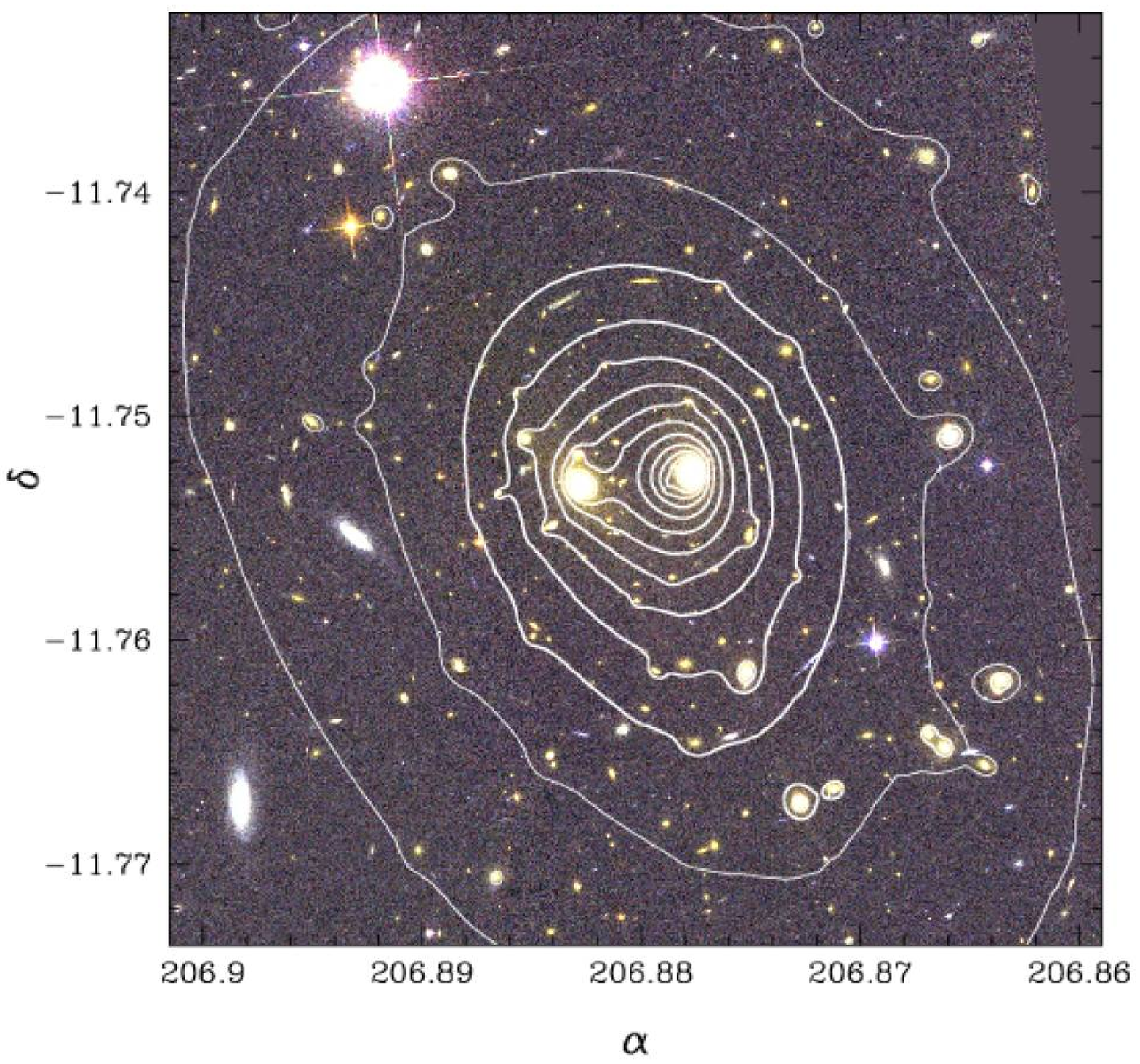}
  \caption{The average scaled surface mass density contours for a
    source at redshift 2. The thick contours show the $\kappa$ levels
    greater or equal to one, and the thin lines for $\kappa$ less than
    one (the contours are separated by $\Delta\kappa$=0.25, first
    thick contour level is at $\kappa$=1).}
  \label{fig:kappa_avg}
\end{figure*}

Due to the large scatter in the parameters of the smooth mass
distribution it is perhaps more interesting to look at the total
2-dimensional projected mass distribution of the cluster.  The
averaged 2-dimensional projected scaled mass distribution for a source
at redshift 2 is shown in Fig. \ref{fig:kappa_avg}.  Thick contours
lines are used for $\kappa \le 1$ levels, where as thin lines are used
for $\kappa<1$ contours (the contours are separated by
$\Delta\kappa$=0.25, the first thick contour level is at $\kappa$=1).

\begin{figure}
  \centering
  \includegraphics[width=1.0\columnwidth]{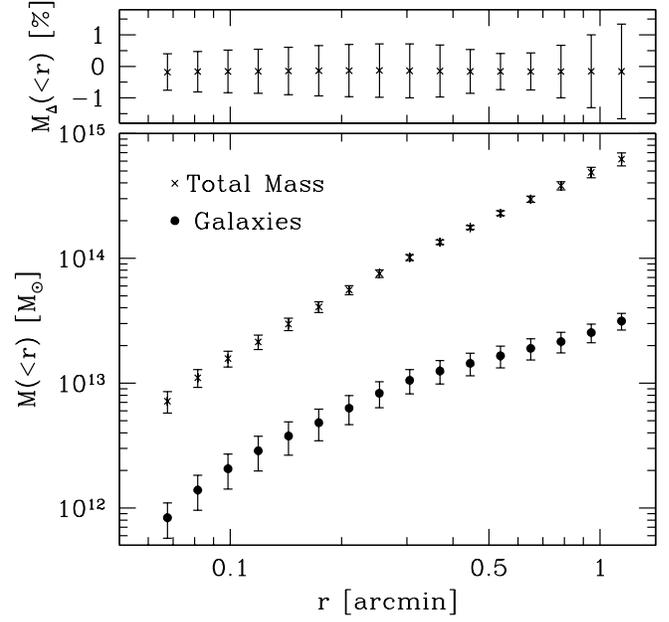}
  \caption{Mass profile of RX J1347. In the bottom panel we show the
    mass in the galaxies as circles and the total mass as crosses. The
    error bars show the 68\% confidence interval derived from the MCMC
    analysis. The top panel shows the residual mass needed to
    perfectly reproduce the image positions as a percentage of the
    total mass. It is consistent with 0\% at all radii.}
  \label{fig:mass}
\end{figure}

For each of the \mbox{1000} selected parameter sets from the MCMC
analysis we have also calculated separately the contributions to the
total mass from the galaxies and the smooth component. The radial mass
profiles of the galaxy mass component and that of the total cluster
mass are shown in Fig. \ref{fig:mass}. In the bottom panel the total
mass (galaxies and the smooth mass component) are shown as crosses,
and the mass in the galaxies as circles. The errors show the 1-sigma
errors derived from the MCMC analysis. The error in the total mass is
very small at ${r}\sim35''$ where the mass is fixed by the
spectroscopic redshift of image system 2.

\begin{figure}
  \centering
  \includegraphics[width=1.0\columnwidth,bb=38 174 592 718]{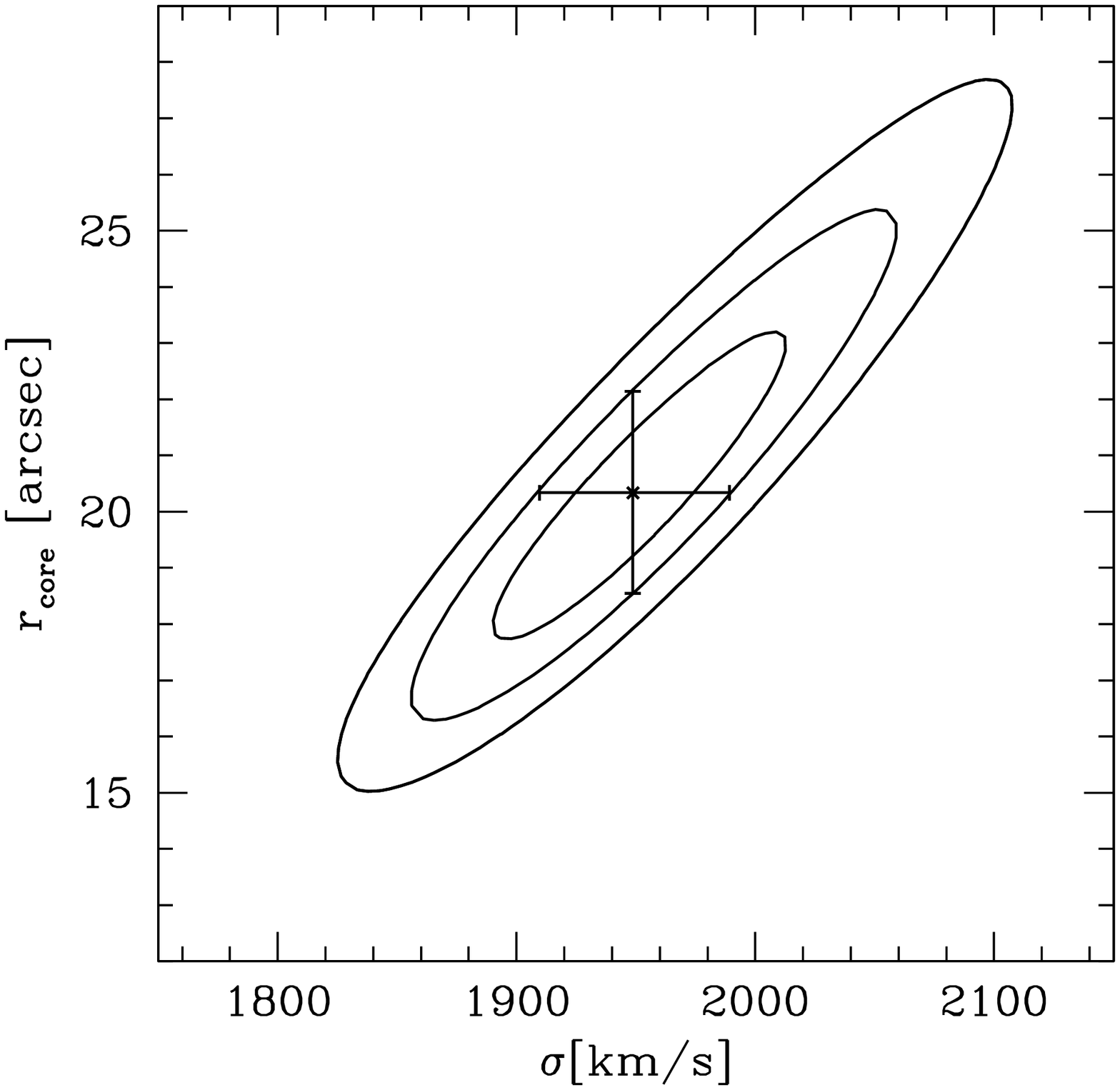}
  \caption{The 1-, 2- and 3-$\sigma$ confidence regions for a NSIS
    halo fitted to the total mass. The point with error bar shows the
    best fit value and the marginalised 1-sigma errors for the two
    parameters. The best fit parameters are $\sigma$=1949
    $_{-39}^{+40}$ km/s, and $r_{\rm{c}}$=20.3 $_{-1.8}^{+1.8}$\
    $''$.}
  \label{fig:nsie_conf}
\end{figure}

\begin{figure}
  \centering
  \includegraphics[width=1.0\columnwidth,bb=38 174 592 718]{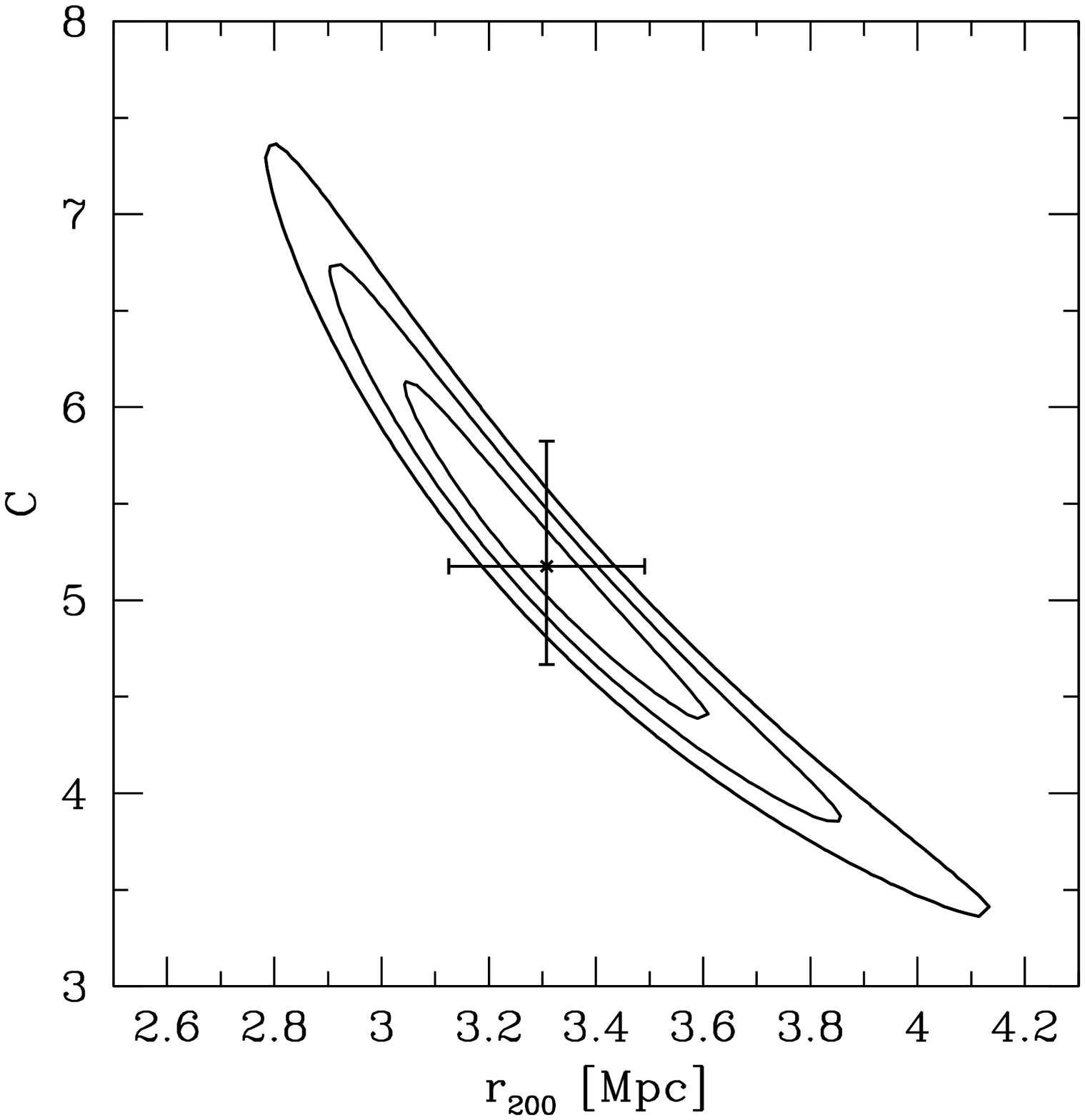}
  \caption{The 1-, 2- and 3-$\sigma$ confidence regions for an NFW
    halo fitted to the total mass. The point with error bar shows the
    best fit value and the marginalised 1-sigma errors for the two
    parameters. The best fit parameters are $r_{200}$ = 3.3 $\pm$ 0.2
    Mpc, and $c$ = 5.4 $_{-0.5}^{+0.7}$.}
  \label{fig:enfw_conf}
\end{figure}

We can estimate the NSIS and NFW parameters of the total mass
distribution by fitting a single halo to the total radial mass
profile. The 1-, 2- and 3-sigma contours for the NSIS and NFW
parameters are shown in Figs. \ref{fig:nsie_conf} and
\ref{fig:enfw_conf} respectively. We also plot in the figures the
error bars for the individual parameters when marginalised over the
other parameter. The best fit parameters of the NSIS profile are
$\sigma$=1949 $_{-39}^{+40}$ km/s, and $r_{\rm{c}}$=20.3
$_{-1.8}^{+1.8}$\ ''. The velocity dispersion is high due to the
presence of a large core. The NFW profile parameters are $r_{200}$ =
3.3 $\pm$ 0.2 Mpc, and \mbox{$c$ = 5.4 $_{-0.5}^{+0.7}$}. The
$\chi^{2}$ values for both of the profiles are very good (for 14
degrees of freedom $\chi^{2}$ values of 4.5 and 4.2 are obtained for
NSIS and NFW profiles respectively). The very good $\chi^{2}$ values
probably reflect the way the total mass profile was obtained using
haloes that had NSIE and ENFW profiles.

\subsubsection{\rxj as a merger}
\label{sec:results:mass:merger}

It is clear from the surface mass density contours that the mass
distribution of the cluster does not have any evident bimodality,
although two haloes were used in modelling the mass. If the cluster
was undergoing a merger along the line of sight, this could still
require two haloes in the modelling.  The fairly clean separation of
the two haloes in both mass and concentration as well as to a lesser
extent also in position, if taken seriously, would indicate that
{\mbox RX J1347} is a merger with a 2:1 mass ratio of the merger
progenitors. This supports well the results from both \mbox{X-rays}
\citep[e.g.][]{allen:02} and SZ-effect \citep[e.g.][]{komatsu:01} that
find X-ray shocks and complicated substructure in the cluster. This
was also seen in a combined strong and weak lensing analysis by
\citet{bradac:05}. Although the 2-dimensional mass map obtaned here
does not have a clear peak in the south-east part of the cluster where
the X-ray and SZ structures are seen, it is worth noting that it is
also in this part of the cluster where the second halo is
preferentially located.

We have added a third halo to the smooth mass component but have not
been able recover the south-east extension. The strong lensing
constraints in this part of the cluster are weak due to the lack of
multiple images in this region, and affects our ability to determine
the mass distribution accurately. This includes the position of the
secondary halo.

Interestingly the cluster has a very low velocity dispersion of only
910$\pm$130 km/s measured by \citet{cohen:02}. They discussed that
this could be explained if the merger took place perpendicular to the
line of sight and therefore would have increased the velocity
dispersion mostly in that direction.

It is important to keep in mind that in the strong lensing analysis
presented here we assume that the cluster has two haloes. More work is
necessary in order to establish with more confidence that the two
haloes in our modelling do correspond to physical haloes. It is also
possible that the need for two haloes in the modelling has to do less
with physically separated structures in the cluster than with a
complex morphology of one dominant halo. This could be, for example, a
varying ellipticity (both magnitude and position angle) as a function
of cluster centric radius, slightly asymmetric mass profile, or a mass
profile that is neither an ENFW nor an NSIE profile but can be
modelled as a superposition of two such haloes.

\subsubsection{Residual mass maps}
\label{sec:results:mass:residuals}

We have taken inspiration from LensPerfect by Dan Coe that, as its
name implies, allows one to find a mass distribution that exactly
reproduces all multiple image positions. We have implemented a similar
scheme with the intention to check how our parametric mass profile
needs to be modified in order to exactly reproduce the image
positions. The usual parametric modelling is used as a basis after
which we calculate residual mass map needed for a perfect fit. In
calculating the residual mass maps we have not considered the relative
magnifications of the multiple images but only their positions.

In practice the residual mass map is found by calculating the residual
deflection angle $\Delta\vec{\alpha}(\vec{\theta}_{i,j})$ at each
multiple image position $\vec{\theta}_{i,j}$:

\begin{equation}
  \Delta\vec{\alpha}(\vec{\theta}_{i,j}) =\ \vec{\beta}_{i,j} - \langle
  \vec{\beta} \rangle_{i},
\end{equation}
where $\vec{\beta}_{i,j}$ is the source position of image $j$ in
system $i$ as obtained with the parametrised model and $\langle
\vec{\beta} \rangle_{i}$ is the mean source position of the images in
system $i$.

The residual deflection angle $\Delta\vec{\alpha}(\vec{\theta})$ at
any other point is then calculated using a Thin Plate Spline
interpolation \citep[TPS, for details on the topic see
e.g.][]{bookstein:89}. The interpolated surface has the property that
it minimises the bending energy of the function
$\Delta\vec{\alpha}(\vec{\theta})$ ,

\begin{equation}
  E = \int\!\!\int \Bigg[
  \frac{ \partial^2 \Delta\vec{\alpha} }{\partial\theta_{1}^2}
  + 2\ \Bigg( \frac{ \partial^2 \Delta\vec{\alpha}
  }{\partial\theta_{1}\partial\theta_{2}} \Bigg)^2
  + \frac{ \partial^2 \Delta\vec{\alpha} }{\partial\theta_{2}^2}\
  \Bigg]\ \mathrm{d}\theta_{1}\mathrm{d}\theta_{2},
\end{equation}
where $\theta_{1}$ and $\theta_{2}$ are the Cartesian components of
the position vector $\vec{\theta}$.

The amount and location of residual mass $\Delta\kappa(\vec{\theta})$
that is needed to perfectly reproduce the multiple image positions is
then simply calculated from the residual deflection angle. Since the
surface mass density is related to the deflection angle via $\kappa =
\vec{\nabla} \vec{\alpha} / 2$, the use of TPS interpolation
essentially minimises the gradients of $\Delta\kappa$.

\begin{figure}
  \centering
  \includegraphics[width=1.0\columnwidth]{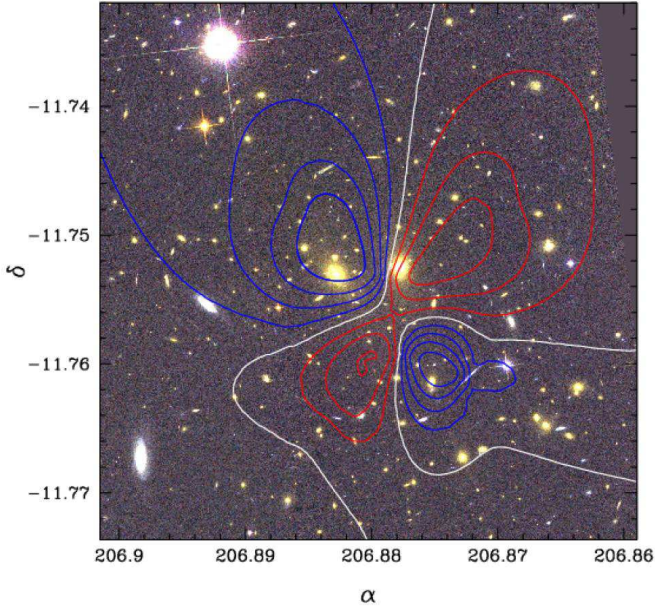}
  \caption{The residual surface mass density contours for a source at
    redshift 2. The red and blue contours show positive and negative
    contributions respectively while the white line shows the
    $\Delta\kappa=0$ level. The contours are separated by 0.005 in
    $\kappa$ indicating that only a very small correction to the mass
    map is enough to perfectly reproduce the multiple image
    positions.}
  \label{fig:kappa_tps}
\end{figure}

The 2-dimensional distribution of the residual mass $\Delta\kappa$ is
shown in Fig. \ref{fig:kappa_tps}. The red and blue contours show
positive and negative contributions, respectively, while the white
line shows the $\Delta\kappa=0$ level. The very small correction
needed is represented by the very small separation of the contours of
only 0.005. Although the correction to the mass is small, it is clear
from Fig. \ref{fig:kappa_tps} that the correction has a bipolar
distribution so that south-east and north-west quadrants have
generally a mass deficit in the parametric modelling while the
south-west and north-east quadrants have a slight mass surplus.

The radial mass profile of the residual mass map is shown on the top
panel in Fig. \ref{fig:mass} as a percentage of the total mass at any
given radius. In the strong lensing region of the cluster the average
residual mass at any radius is consistent with 0\% with a scatter of
only around 1\%.

\subsection{Lensing redshift of multiple image system 1}
\label{sec:results:lens_z}

\begin{figure}
  \centering \includegraphics[width=1.0\columnwidth,bb=38 174 592 718]{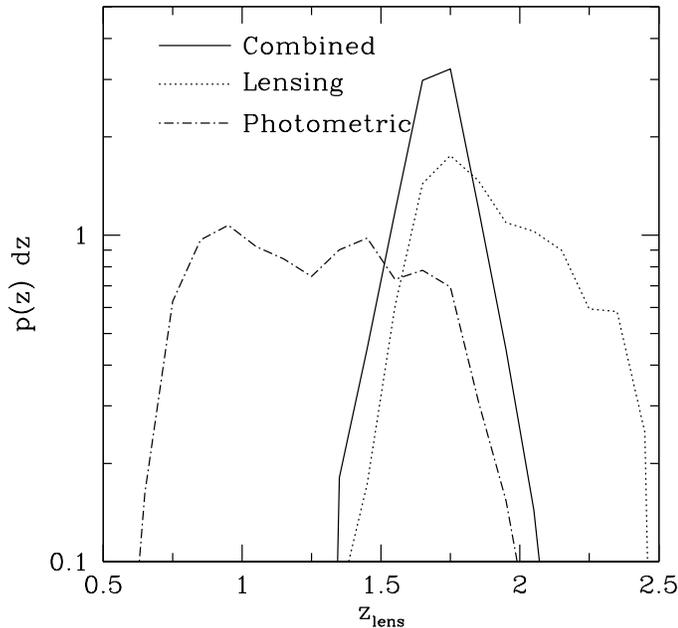}
  \caption{The redshift distribution of image system 1 in the MCMC
    sampling of the parameter space. The dotted line shows the lensing
    redshift probability distribution. The photometric redshift
    probability distrubution is given by the dot$-$dashed line.
    Combining the two probability densities results in the solid line.
    The lensing redshift probability densities have a clear cutoffs at
    $z \sim 1.5$ and $z \sim 2.5$. The lensing redshift for image
    system 1 is $z_{\mathrm{lens}}$=1.90$^{+0.21}_{-0.31}$, with the
    best combined estimate being 1.70$^{+0.13}_{-0.12}$.}
  \label{fig:results:lens_z:img1}
\end{figure}

The lensing models allow us to also constrain the redshifts of
multiple image systems. This can be done by looking at the
distribution of the redshifts of the multiple image systems in the
MCMC sampling of the parameter space.  The redshift probability
density of multiple image system 1 is shown in Fig.
\ref{fig:results:lens_z:img1}. The dotted line shows the redshift
distribution from the lensing and the dot-dashed from the photometric
redshift analysis, while the solid line shows the total distribution.
The lensing redshift distributions do not show any dependence on the
profile used in modelling the smooth mass component of the cluster.
The lensing redshift for image system 1 is
z$_{\mathrm{lens}}$=1.90$^{+0.21}_{-0.31}$. The photometric redshift
was not used as a prior in the lensing modelling, and we only use it
to compare to the redshift obtained with the lensing models.  The two
estimates for the source redshift agree in the redshift range 1.5-2.0
with a combined mean redshift of 1.70$^{+0.13}_{-0.12}$.

\subsection{Constraints from all multiple image systems}
\label{sec:results:all_images}

All the previous analysis has been based on models that use only image
systems 1 and 2 to constrain the mass distribution of the cluster. We
have repeated the analysis using all multiple image systems to check
how the mass distribution changes in the presence of more constraints,
and to predict redshifts for the remaining multiple image systems. The
source redshifts are added as free parameters to the models without
priors from photometric redshifts.

\subsubsection{Mass distribution}
\label{sec:results:all_images:mass}

\begin{figure}
  \centering
  \includegraphics[width=1.0\columnwidth]{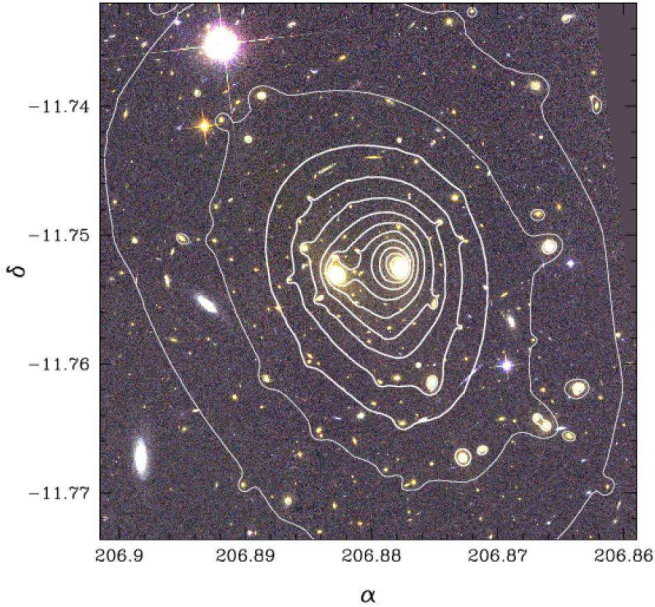}
  \caption{The average scaled surface mass density contours for a
    source at redshift 2. All image systems were used to constrain the
    mass distribution. The thick contours show the $\kappa$ levels
    greater or equal to one, and the thin lines for $\kappa$ less than
    one (the contours are separated by $\Delta\kappa$=0.25, first
    thick contour level is at $\kappa$=1), these are the same as in
    Fig. \ref{fig:kappa_avg} for easier comparison. The differences
    are very small, mostly noticeable by small deviations near cluster
    galaxies.}
  \label{fig:kappa_avg_all}
\end{figure}

The mass distribution obtained with all the multiple images are
virtually indistinguishable from that obtained with only the two first
ones. The surface mass density contours for the mass distribution
derived with constraints from all multiple images systems are shown in
Fig. \ref{fig:kappa_avg_all}. The contour levels are the same as those
used in Fig. \ref{fig:kappa_avg}. The differences are most noticeable
around some of the cluster galaxies. Also the critical lines are only
marginally affected when all image systems are used.

The reason for the similarity of the mass maps is most likely the free
redshifts of the multiple images combined with the geometry of the
images. With the exception of image systems 9 and 11 all of the image
systems are only on one side of the cluster.  The unconstrained
redshifts leave the mass scale free while the geometries cannot
constrain the position of the mass concentration. All of the
information is hence contained in the first two image systems which
constrain the position of the mass concentration through the 5 images
in image system 1 while the scale is fixed by the spectroscopic
redshift of image system 2.

\subsubsection{Lensing redshifts of multiple image systems}
\label{sec:results:all_images:lens_z}

\begin{figure}
  \centering
  \includegraphics[width=1.0\columnwidth,bb=38 154 592 718]{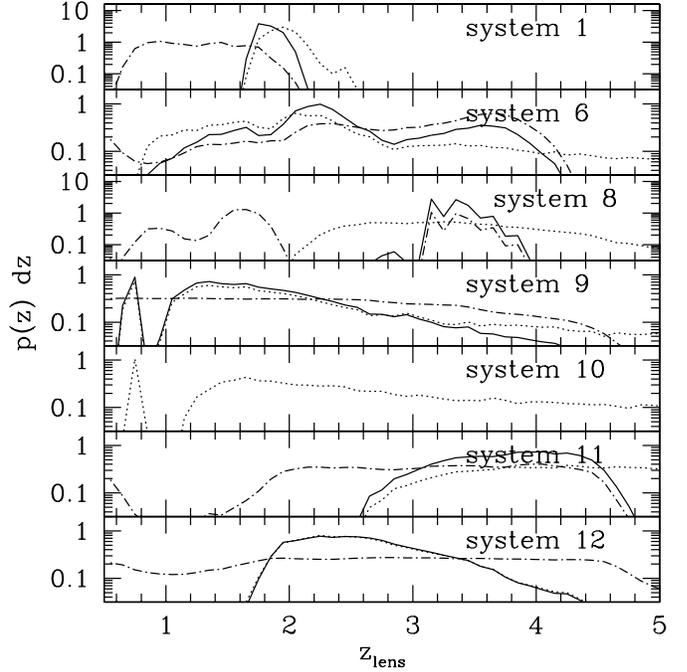}
  \caption{The redshift distributions of the image systems for which
    the lensing models are able to constrain the redshift. The dotted
    line shows the lensing redshift probability distribution, the
    dot-dashed line the photometric redshift probability distribution
    and the solid line the combined redshift probability distribution.
    All multiple images are used in the lensing modelling. For image
    systems 9 and 10 the model requirement is that a critical curve
    should pass between the two images. This produces the low redshift
    bump in the redshift probability when the tangential critical
    curve and not the radial critical curve passes between the two
    images.  As it is clear from the positions of the multiple images
    that they form a radial arc, the lower redshifts can be ignored.}
  \label{fig:results:all_images:lens_z:mcmc}
\end{figure}

The redshift probability densities for image systems 1, 6, 8, 9, 10,
11, and 12 are shown in Fig. \ref{fig:results:all_images:lens_z:mcmc}
as dotted lines. The redshifts of the other multiple image systems are
poorly constrained. The lower redshift bump on the probability
densities for the image systems 9 and 10 is a consequence of their
modelling. We have only required that a critical line should pass
between the two images in these systems. For a source at low redshift
it is the tangential critical line, and not the radial one, that
passes through the two images. The dot-dashed line shows the
probability distribution of the photometric redshifts, the solid line
the probability density combining both the lensing and photometric
redshifts probability densities. The photometric redshift probability
distribution of image system 8 has two distinct peaks, one between 0.6
and 2.0 and another between redshifts 4 and 5. The lensing redshift
probability distribution for this image system is broad but excludes
the lower redshift bump seen in the photometric redshift probability
distribution. The photometric redshift probability densities for the
other image systems are very broad and constrain the redshifts very
little. The low redshift bump in image systems 9 and 10 can be
excluded since in those cases the image systems would define a
tangential critical curve where as it is clear from the ACS images
that the images are separated by a radial critical curve.

\section{Comparison to previous mass estimates}
\label{sec:comparison}

\begin{figure}
  \centering
  \includegraphics[width=1.0\columnwidth,bb=38 174 592 718]{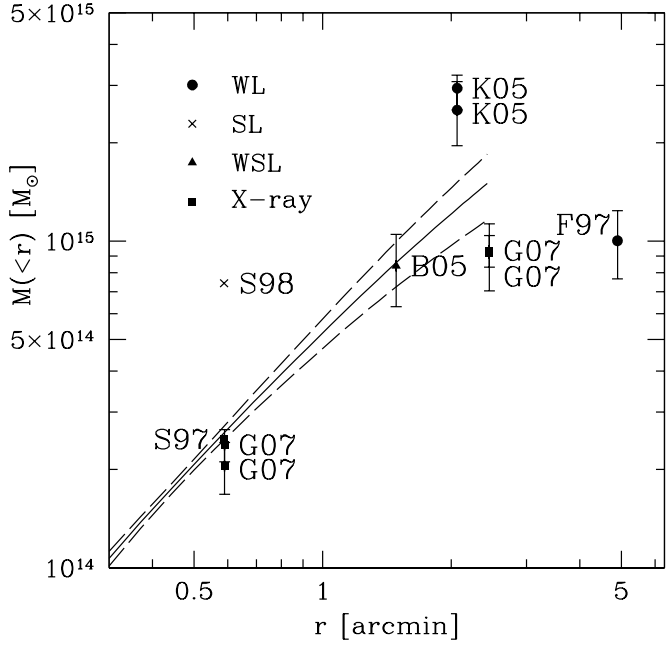}
  \caption{Projected mass profile of RX J1347 from various published
    results. The solid line shows the mass profile obtained in this
    work (the dashed lines show the 1-sigma confidence region in
    mass). The different symbols in the figure denote different
    methods, weak lensing (WL), strong lensing (SL), strong and weak
    lensing (SWL) and finally X-rays shown by filled circles, crosses,
    a triangle, and squares respectively (also labelled in the
    figure).  The authors are marked next to the points, full details
    can be found in the text. We have plotted only X-ray masses where
    projected masses are given in literature.}
  \label{fig:mass:comparison}
\end{figure}

There are several previous mass estimates for \rxj in the literature.
In this section we compare the masses from different methods, namely
from the kinematics of the cluster galaxies, \mbox{X-ray} emission
from the cluster gas, weak and strong lensing separately (WL \& SL,
analysis respectively) as well as a combined strong and weak lensing
(SWL). We have converted results in the literature to the cosmology
used in this work (H$_{0}$ = 70 km/s/Mpc, $\Omega_{\rm{m}}$=0.3 and
$\Omega_{\Lambda}$=0.7), whenever required.  For the direct mass
comparison we have only used literature values where projected masses
are given. The converted masses are shown in Table
\ref{tab:mass:comparison} and plotted in Fig.
\ref{fig:mass:comparison}.

The strong lensing mass estimate by \citet[S98]{sahu:98} is almost a
factor of 4 higher than what we obtain. This is probably caused by the
low redshift assumed for the arc by \citet{sahu:98} and their
assumption of spherical symmetry. The two purely weak lensing mass
estimates have either higher \citep[K05]{kling:05} or lower
\citep[F97]{fischer:97} masses than what we obtain with strong
lensing. This indicates that there is at least no systematic offset
between weak and strong lensing masses obtained.

The agreement with \mbox{X-ray} mass estimates by both
\citet[S97]{schindler:97} and \citet[G07]{gitti:07} is fairly good,
although all of the \mbox{X-ray} mass estimates are lower than the
ones we obtain. A detailed comparison of lensing mass from a combined
strong and weak lensing analysis to new X-ray data taken with the
Chandra X-ray telescope is presented in Brada{\v c} et al. (2008, in
preparation). We find excellent agreement with the combined strong and
weak lensing analysis of \citet[B05]{bradac:05}.  This is not
surprising since the redshift for the arc used both in
\citet{bradac:05} and this work has not changed, thus fixing the mass
of the cluster at the position of the arc.

\begin{table}
  \centering
  \caption{Comparison of projected mass estimates interior to a given
    radius in the literature. The different methods are weak lensing
    (WL), strong lensing (SL), strong and weak lensing (SWL), X-rays and
    kinematics of the cluster galaxies. We have converted all the
    results to the cosmology used in the this paper (H$_{0}$ = 70
    km/s/Mpc, $\Omega_{\rm{m}}$=0.3 and $\Omega_{\Lambda}$=0.7) if
    necessary. The remarks for \citet{gitti:07} refer to the types of
    profiles fitted, SO for a single $\beta$ model and DDg1 for a double $\beta$ model.}
  \label{tab:mass:comparison}
  \begin{tabular}{lrr@{.}ll}

    \hline
    \bf{Reference}  & \bf{$r$} & \multicolumn{2}{c}{\bf{M($<r$)}}          & \bf{Method} \\
                    & $''$     & \multicolumn{2}{c}{10$^{14}$ M$_{\odot}$} & \bf{(Remark)} \\
    \hline
    \hline
    \citet{gitti:07}     &  35 &  2&38 $\pm$ 0.27         & X-ray (SO)  \\
    \citet{gitti:07}	 &  35 &  2&05 $\pm$ 0.37         & X-ray (DDg1)\\
    \citet{schindler:97} &  35 &  2&94                    & X-ray       \\
    \citet{sahu:98}	 &  35 &  8&82                    & SL          \\
    This work            &  35 &  2&56 $\pm$ 0.12         & SL          \\
    \citet{bradac:05}	 &  89 &  8&40 $\pm$ 2.10         & SWL         \\
    This work            &  89 &  8&57 $\pm$ 1.34         & SL          \\
    \citet{kling:05}	 & 124 & 25&20 $\pm$ 5.60         & WL    (SIS) \\
    \citet{kling:05}	 & 124 & 29&40 $_{-4.20}^{+2.80}$   & WL    (NFW) \\
    This work            & 124 & 12&60 $\pm$ 2.58         & SL          \\
    \citet{gitti:07}	 & 147 &  9&35 $_{-1.03}^{+1.05}$   & X-ray (SO)  \\
    \citet{gitti:07}	 & 147 &  9&17 $\pm$ 2.13         & X-ray (DDg1)\\
    \citet{fischer:97}   & 294 & 11&90 $\pm$ 2.80         & WL          \\
    \hline

  \end{tabular}
\end{table}

In cases where we have not found projected mass estimates for the
cluster we can still compare our results for the parameters of the
commonly used halo profiles, namely the family of isothermal profiles
and the Navarro, Frenk and White profile. These are shown in Table
\ref{tab:param:comparison}. For the comparison we have converted the
literature values of the $r_{\rm{c}}$ of an isothermal profile and the
$r_{200}$ of an NFW profile to our cosmology where applicable. In
addition, if only $r_{\rm{s}}$ of an NFW profile was given in the
literature we have multiplied it (and its error) by the concentration
parameter given to obtain an $r_{200}$. This is only a rough estimate
of the error, however, since the two parameters have very strong
degeneracies with higher concentrations having smaller scale radii.

The high core radius required to fit the mass profile in this work is
compensated by an unusually high velocity dispersion. A singular
isothermal sphere would have a much lower velocity dispersion (in the
range 1450-1600 km/s) as can be seen at least qualitatively by
extending the degeneracy of the two parameters in Fig.
\ref{fig:nsie_conf}. The exact value obtained for the velocity
dispersion depends very strongly on the range of radii in which the
mass is fitted. On the other hand the range of velocity dispersions is
in good agreement with the singular isothermal fits in the literature.
This also includes the X-ray based velocity dispersion measurement by
\citet{allen:02}, once the large core radius found in our analysis is
taken into account (for a given velocity dispersion a larger core
radius lowers the mass of an isothermal sphere).

The concentration parameter we obtain for the NFW profile is in
agreement with the relatively low concentrations expected of cluster
-sized haloes. The $r_{200}$ (3.3 $\pm$ 0.2 Mpc) on the other hand is
rather large when compared to the literature. Since we are only
determining the mass in the strong lensing regime we are not able to
effectively constrain the mass at large radii, e.g. at the virial
radius of the cluster, the value we obtain for $r_{200}$ is therefore
an extrapolation of the NFW profile from the very inner regions of the
cluster. In our case the scale radius NFW profile is already $\sim$600
kpc and outside the region that can be probed directly with strong
lensing only. Weak lensing analysis beyond the field of view of the
ACS is necessary to strongly constrain the $r_{200}$ of the cluster.
This work will be done in Wuttke et al. (2008, in preparation) using
both WFI and MegaCam data.

\begin{table}
  \centering
  \caption{Comparison between the estimated parameters of different
    parametrised mass profiles from the literature. When comparing
    parametrised halo profiles there is no confusion between spherical
    and cylindrical densities/masses making the comparison of different
    methods easier. We have converted the $r_{\rm{c}}$ of an isothermal profile
    and the $r_{200}$ of an NFW profile to our cosmology where
    applicable. If only $r_{\rm{s}}$ of an NFW profile was given in
    literature we have multiplied it (and its errors) by the best fit
    concentration to obtain an $r_{200}$.}
  \label{tab:param:comparison}
  \begin{tabular}{lr@{}lr@{}ll}
    \hline
    \bf{Reference}  & \multicolumn{2}{c}{\bf$\sigma$}  & \multicolumn{2}{c}{\bf $r_{\rm{c}}$} & \bf{Method} \\
                    & \multicolumn{2}{c}{   km/s    }  & \multicolumn{2}{c}{  kpc          } &             \\
    \hline
    \hline
    \citet{cohen:02}   &  910&\ $\pm$ 130        &    &          & kinematic \\
    \citet{allen:02}   & 1590&\ $\pm$ 150        &  38&\ $\pm$ 8   & X-ray     \\
    \citet{kling:05}   & 1400&\ $^{+130}_{-140}$   &    &         & WL        \\
    \citet{fischer:97} & 1500&\ $\pm$ 160        &    &          & WL        \\
    This work          & 1949&\ $\pm$ 40         & 117&\ $\pm$ 12 & SL        \\
    \hline
    \\
    \hline
    \bf{Reference}  & \multicolumn{2}{c}{\bf$c$}  & \multicolumn{2}{c}{\bf $r_{200}$} & \bf{Method} \\
                    & \multicolumn{2}{c}{           }  & \multicolumn{2}{c}{       Mpc     } &             \\
    \hline
    \hline
    \citet{allen:02} & 5.87&\ $\pm$ 1.4      & 1.99&\ $^{+1.19}_{-0.60}$ & X-ray \\
    \citet{gitti:07} &  3.2&\ $\pm$ 0.3      & 2.31&\ $\pm$ 0.36       & X-ray \\
    \citet{kling:05} &   15&\ $^{+64}_{-10}$  & 2.64&\ $^{+0.14}_{-0.57}$ & WL    \\
    This work        &  5.4&\ $_{-0.5}^{+0.7}$ & 3.29&\ $\pm$ 0.20       & SL    \\
    \hline
  \end{tabular}
\end{table}

\section{Conclusion}
\label{sec:conclusion}

In this paper we present the first detailed strong lensing model for
the galaxy cluster \rxj. The models are based on images taken with the
Advanced Camera for Surveys on the Hubble Space Telescope. The high
resolution of the image along with the three colours have allowed us
to identify several new strongly lensed background images in the
cluster. The follow-up spectroscopic observations on the FORS2 at the
VLT have provided us with redshifts for the images in one multiple
image system. This is crucial for an accurate absolute mass
determination of the cluster. Among the new multiple image systems is
a 5 image system with a central image that is very important in fixing
the mass profile at the core of the cluster.

The mass in the cluster is modelled with small-scale mass components
in the cluster galaxies described by truncated isothermal ellipsoids,
and a large-scale component in two parametric mass profiles (Navarro,
Frenk and White profile and non-singular isothermal ellipsoid). The
parameters of the large-scale component are constrained by the
location and magnifications of the multiple images. The total mass
profile of the cluster is well described by both a Navarro, Frenk and
White profile with a moderate concentration of $c$ =
5.3$^{+0.4}_{-0.6}$ and $r_{\rm{200}}$ = 3.3$^{+0.2}_{-0.1}$ Mpc, or a
non-singular isothermal sphere with velocity dispersion
$\sigma_{\rm{nsis}}$ = 1949 $\pm$ 40 km/s and a core radius
$r_{\rm{c}}$ = 20.3 $\pm$ 1.8 $''$. The core radius in the isothermal
profile is necessary in order to reproduce the central image and to
fit the total mass of the profile at the all radii. A singular
isothermal sphere fit to the total mass has too much mass at small
radii and too little at larger radii. The total mass of the cluster
inside the Einstein radius of the main arc ($\sim35''$, or $\sim$200
kpc) is \mbox{(2.6 $\pm$ 0.12) $\times$ 10$^{14}$ M$_{\odot}$}.

A comparison with the X-ray mass estimates by \citet{gitti:07} shows
that the lensing mass is consistently higher than the X-ray mass,
although still within the error bars. The X-ray based velocity
dispersion in \citet{allen:02} on the other hand is in good agreement
with the one determined here once the effect of the different core
radii is taken into account.

\section{Acknowledgements}
A. Halkola and T. Erben acknowledge support by the German BMBF through
the Verbundforschung under grant no.\ 50 OR 0601.  M. Brada{\v c}
acknowledges support for this work provided by NASA through grant
number HST-GO-10492.03A from the Space Telescope Science Institute,
which is operated by AURA, Inc., under NASA contract NAS 5-26555. MB
acknowledges support from the NSF grant AST-0206286 and from NASA
through Hubble Fellowship grant \# HST-HF-01206.01 awarded by the
Space Telescope Science Institute.

This research has made use of the NASA/IPAC Extragalactic Database
(NED) which is operated by the Jet Propulsion Laboratory, California
Institute of Technology, under contract with the National Aeronautics
and Space Administration.

Based on observations made with the NASA/ESA Hubble Space Telescope,
obtained from the data archives at the Space Telescope European
Coordinating Facility and the Space Telescope Science Institute, which
is operated by the Association of Universities for Research in
Astronomy, Inc., under NASA contract NAS 5-26555.




\clearpage
\newpage

\appendix

\section{Multiple image systems}
\label{app:images}

\begin{figure*}
  \vspace{20mm}
  \begin{tabular}{ccccccc}
    \multicolumn{1}{m{2.1cm}}{{\Large RX J1347}}
    & \multicolumn{1}{m{2.2cm}}{\includegraphics[height=2.50cm,clip]{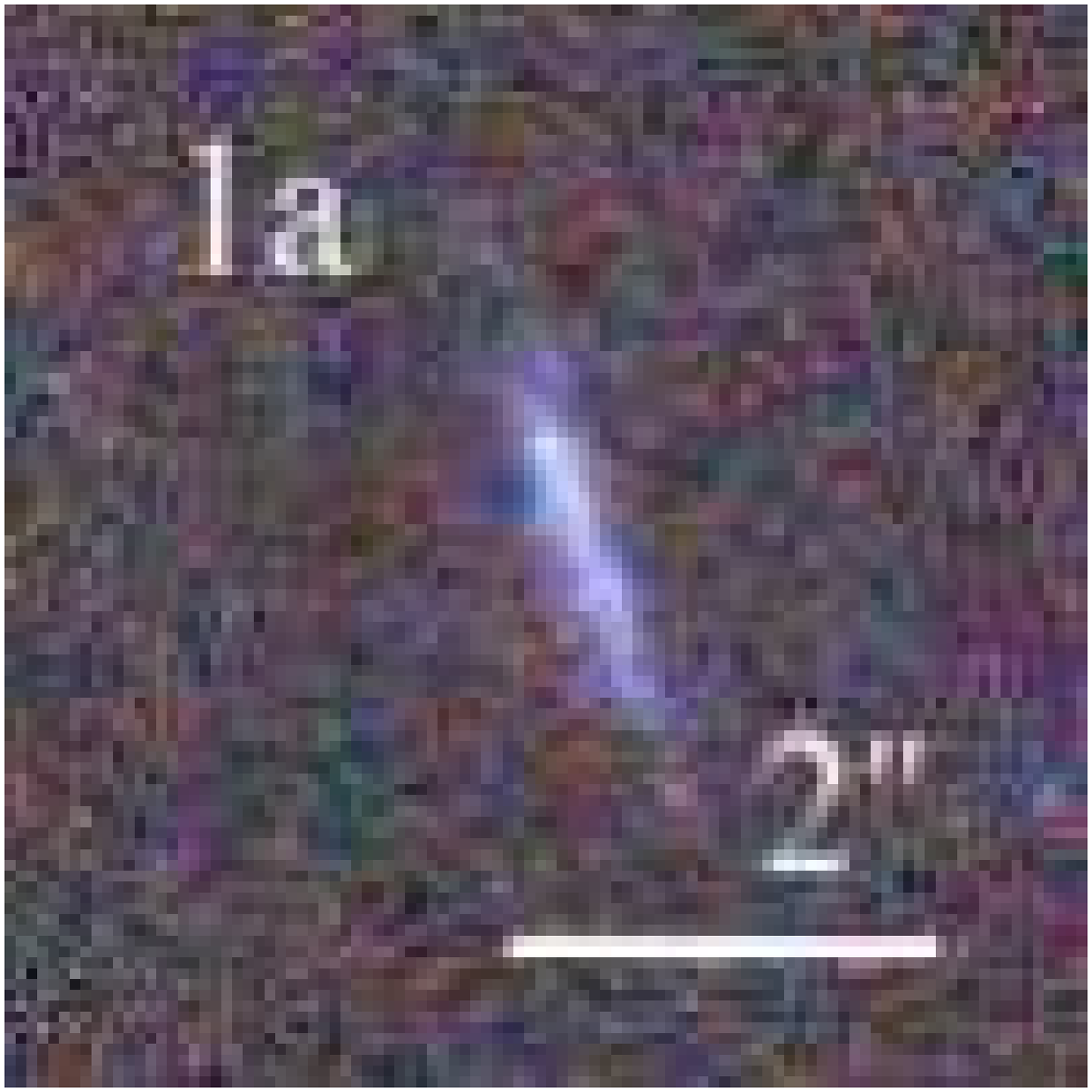}}
    & \multicolumn{1}{m{2.2cm}}{\includegraphics[height=2.50cm,clip]{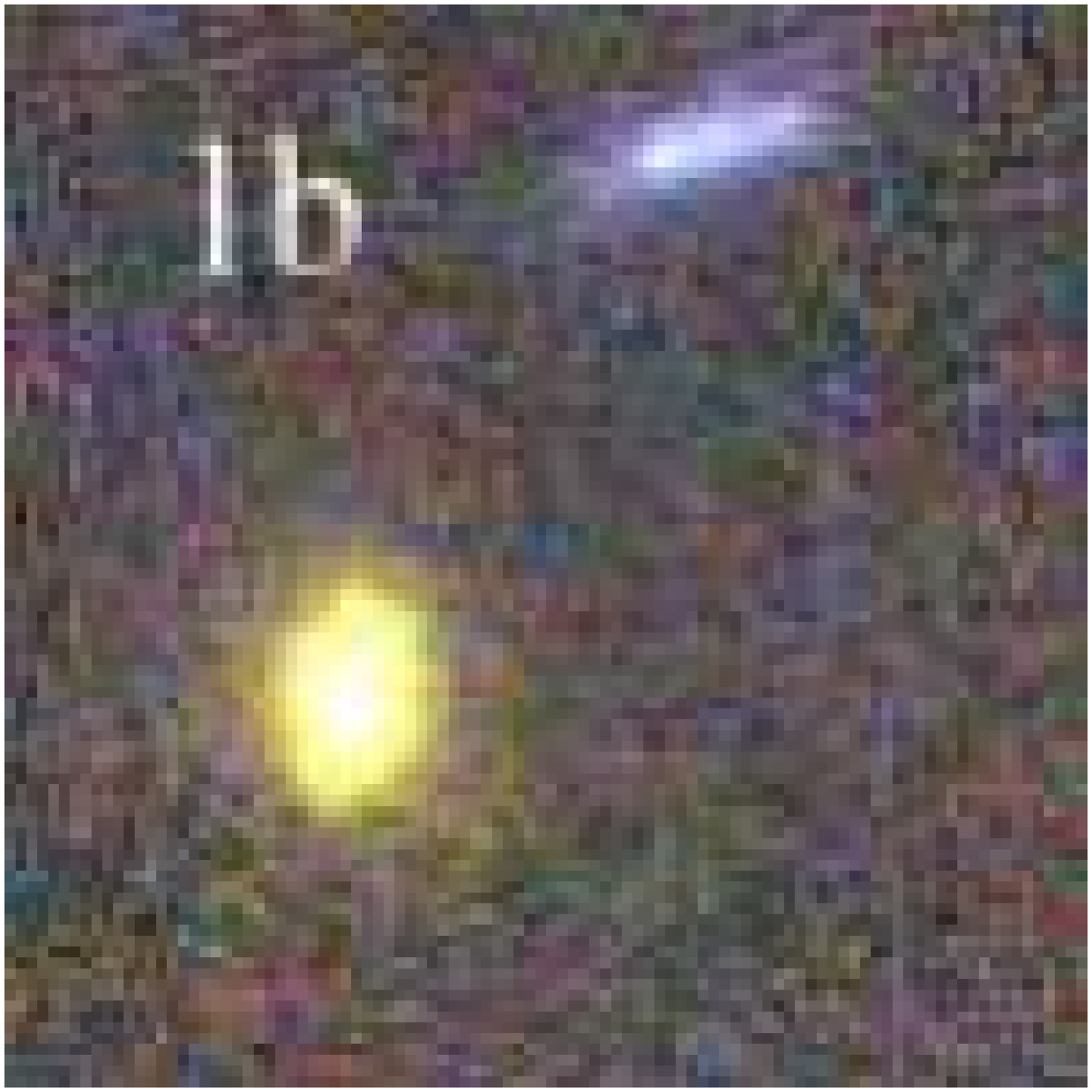}}
    & \multicolumn{1}{m{2.2cm}}{\includegraphics[height=2.50cm,clip]{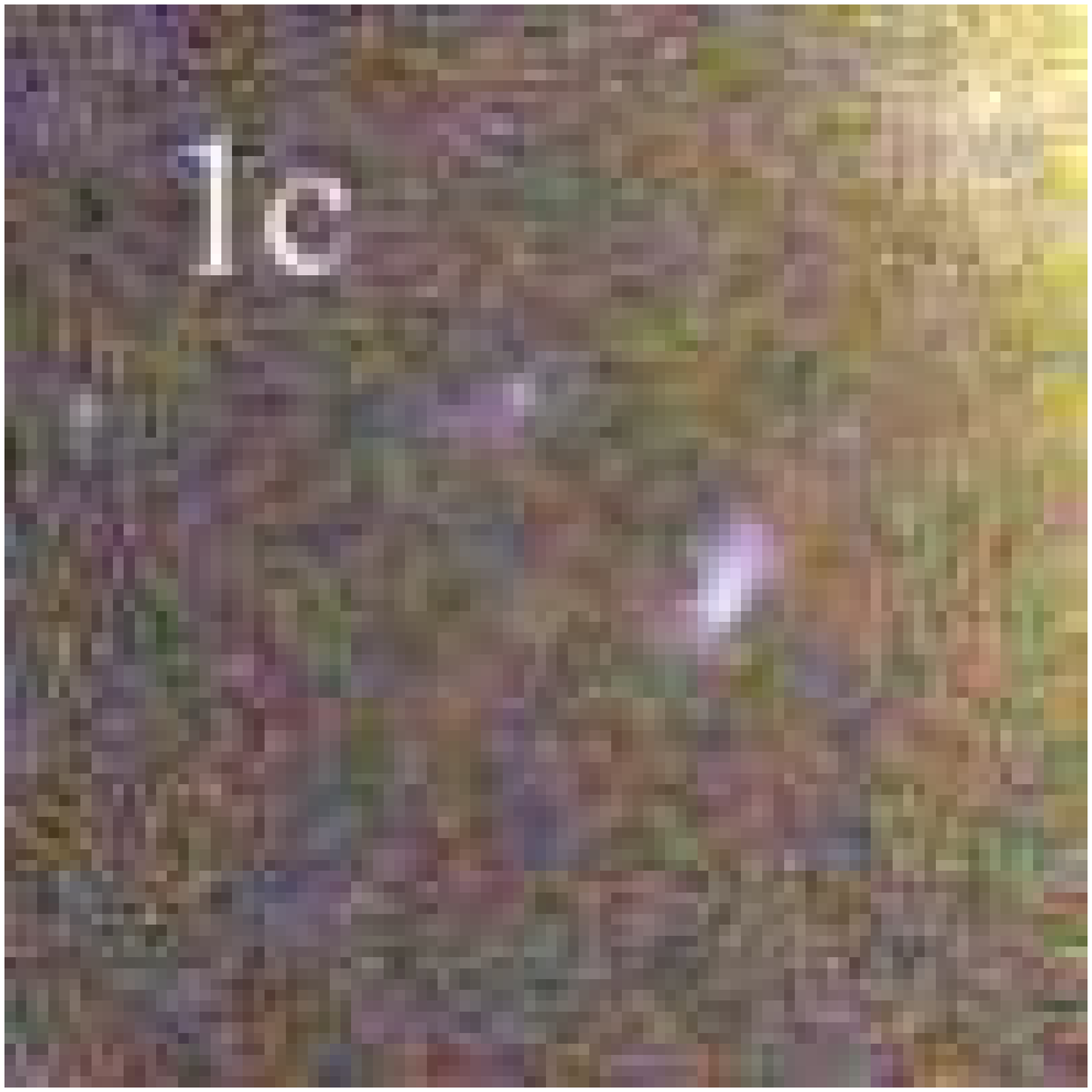}}
    & \multicolumn{1}{m{2.2cm}}{\includegraphics[height=2.50cm,clip]{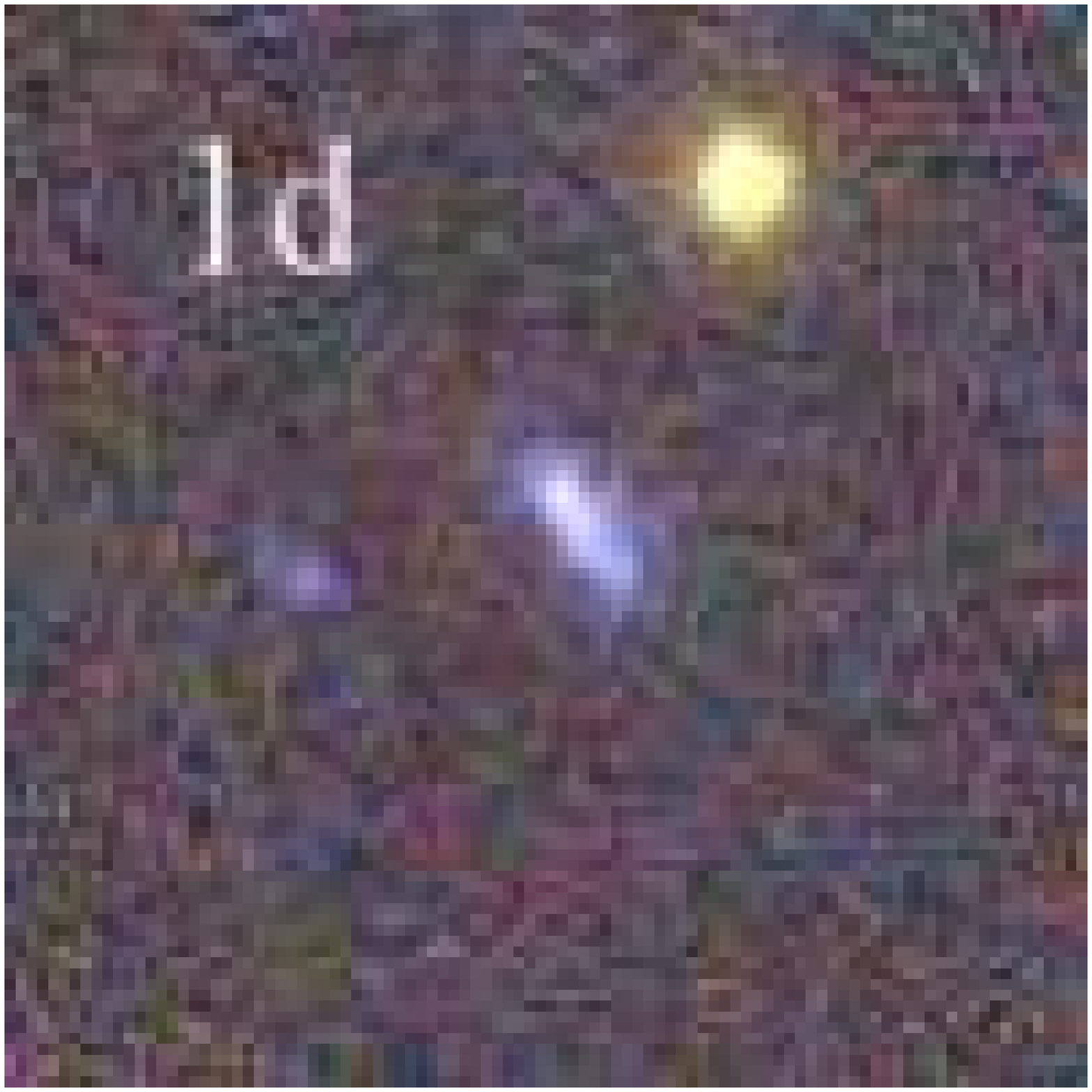}}
    & \multicolumn{1}{m{2.2cm}}{\includegraphics[height=2.50cm,clip]{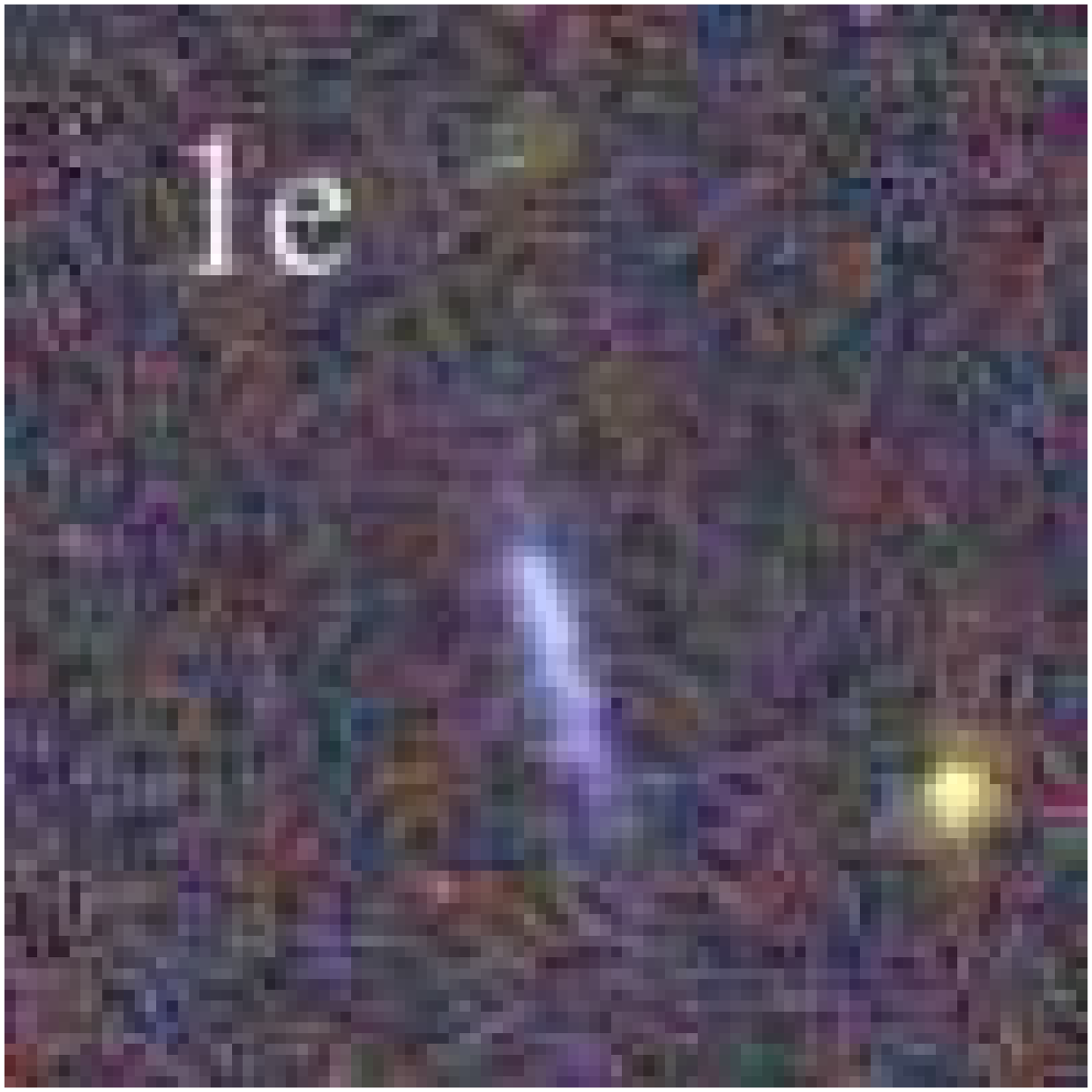}}
    & \multicolumn{1}{m{2.2cm}}{ } \\
      \multicolumn{1}{m{2.1cm}}{{\Large MODEL}}
    & \multicolumn{1}{m{2.2cm}}{\includegraphics[height=2.50cm,clip]{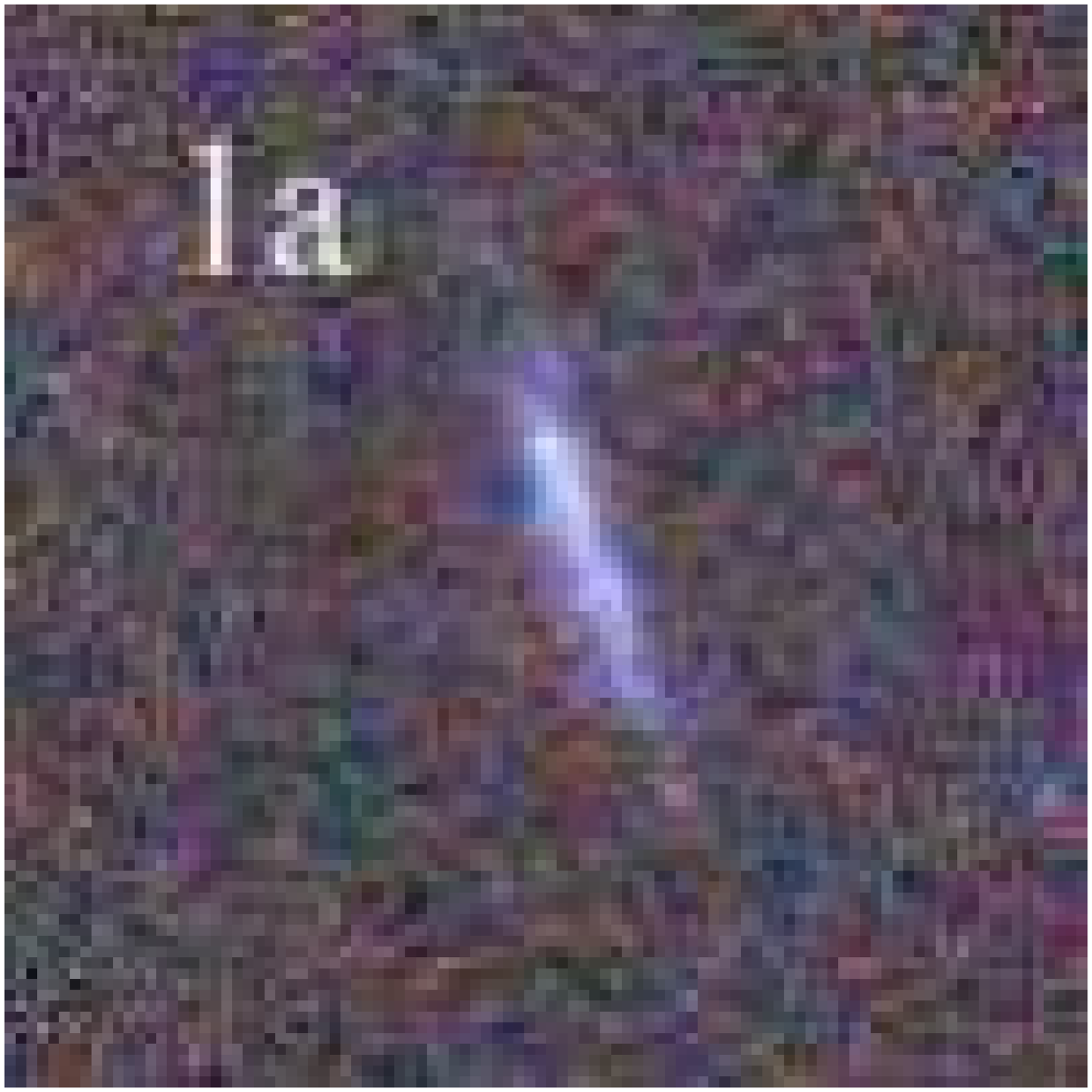}}
    & \multicolumn{1}{m{2.2cm}}{\includegraphics[height=2.50cm,clip]{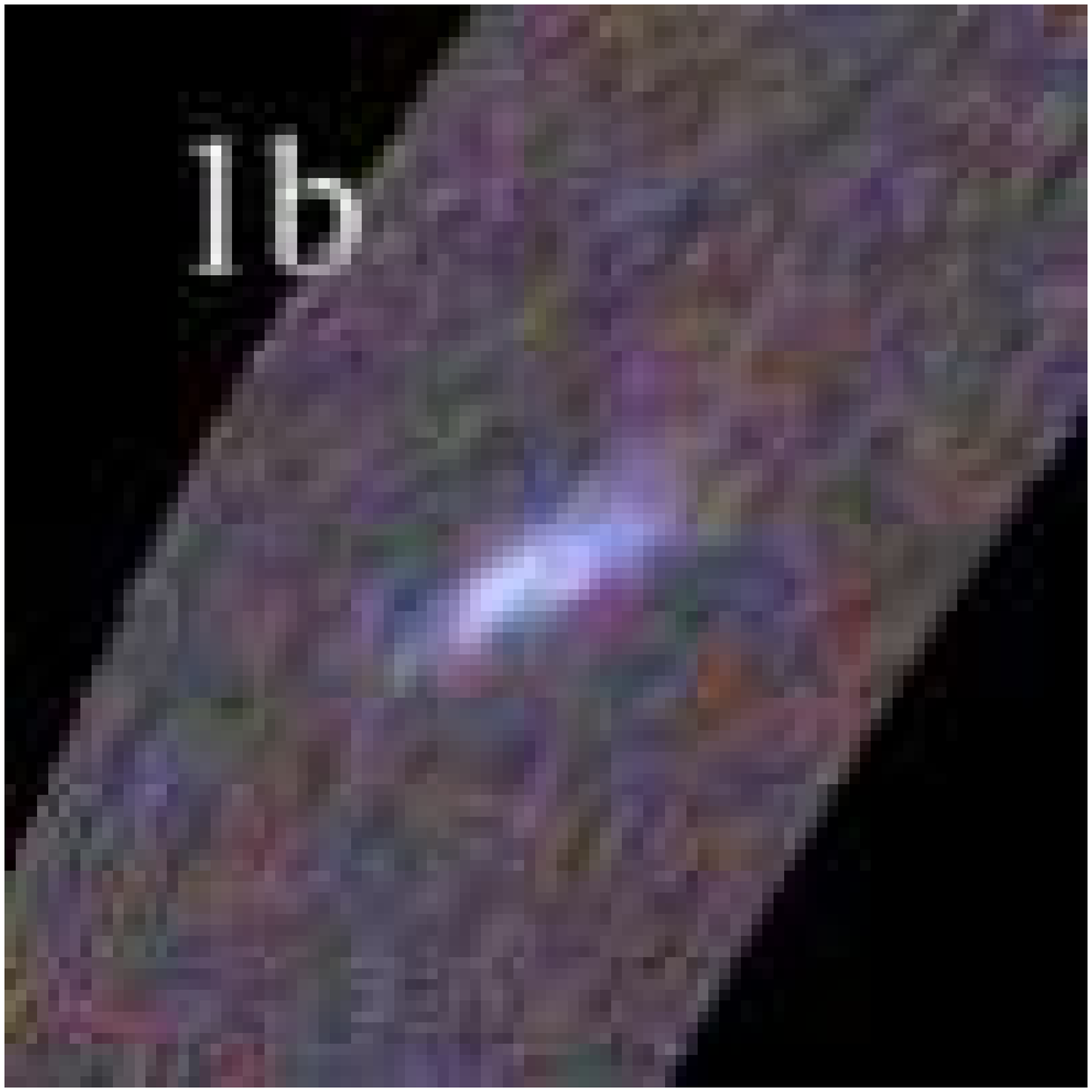}}
    & \multicolumn{1}{m{2.2cm}}{\includegraphics[height=2.50cm,clip]{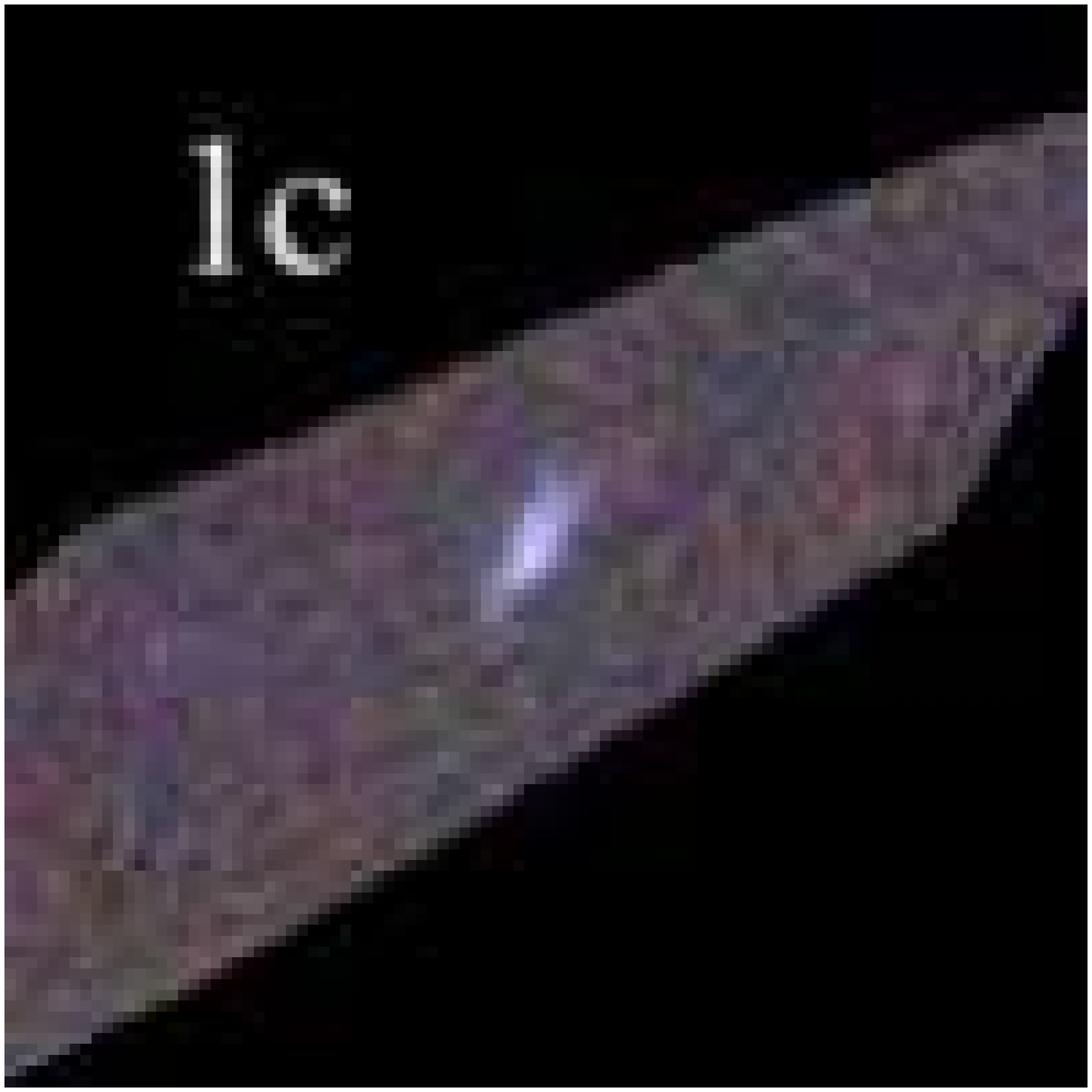}}
    & \multicolumn{1}{m{2.2cm}}{\includegraphics[height=2.50cm,clip]{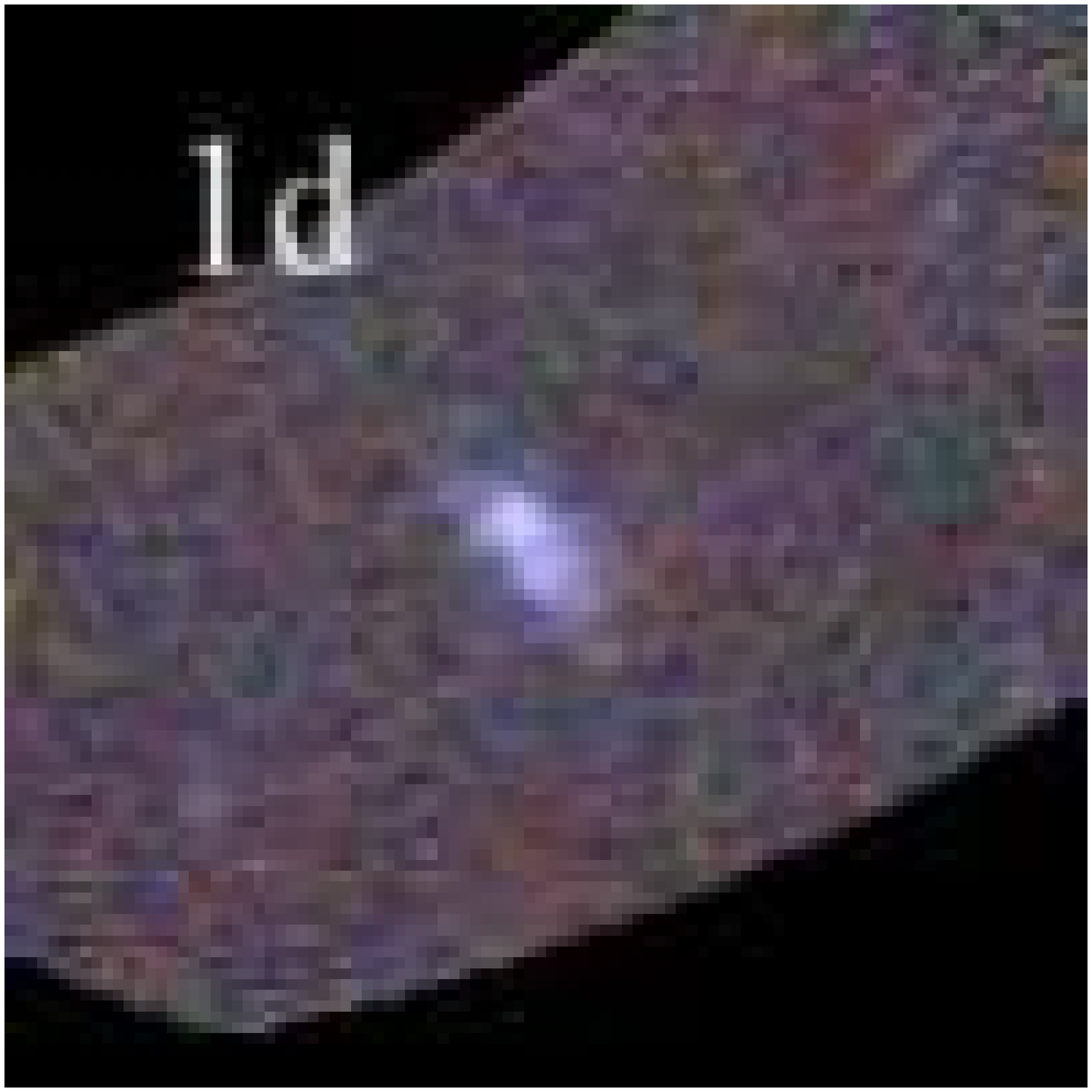}}
    & \multicolumn{1}{m{2.2cm}}{\includegraphics[height=2.50cm,clip]{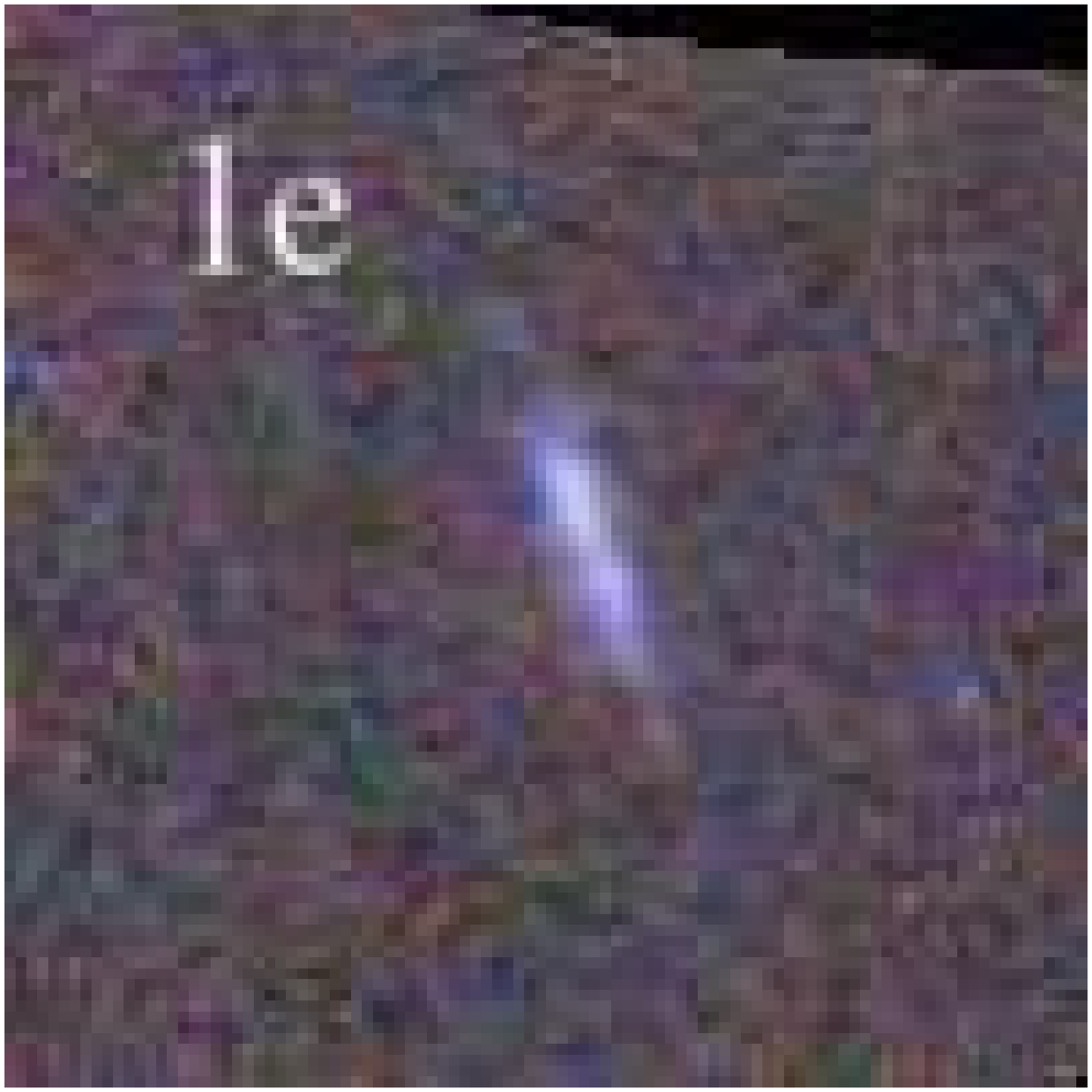}}
    & \multicolumn{1}{m{2.2cm}}{ } \\
  \end{tabular}
  \caption{Image system 1: The offsets between the observed and
    predicted image positions represent the quality of the models. The
    relatively large offset present in multiple image 1b can result
    from the nearby cluster galaxy. }
  \vspace{0mm}
  \label{fig:images:1}
\end{figure*}

\begin{figure*}
  \begin{tabular}{ccccccc}
    \multicolumn{1}{m{2.1cm}}{{\Large RX J1347}}
    & \multicolumn{1}{m{2.2cm}}{\includegraphics[height=2.50cm,clip]{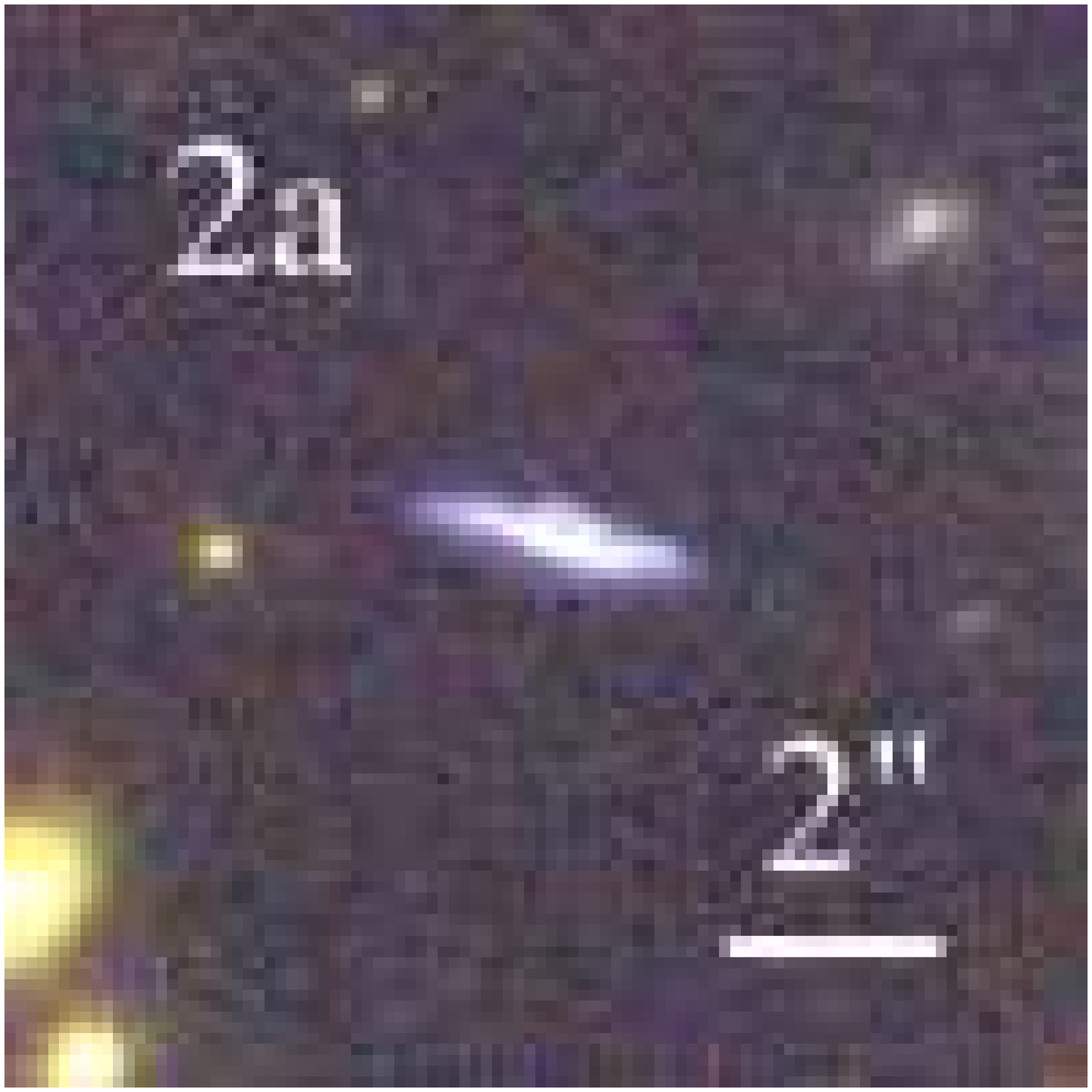}}
    & \multicolumn{1}{m{2.2cm}}{\includegraphics[height=2.50cm,clip]{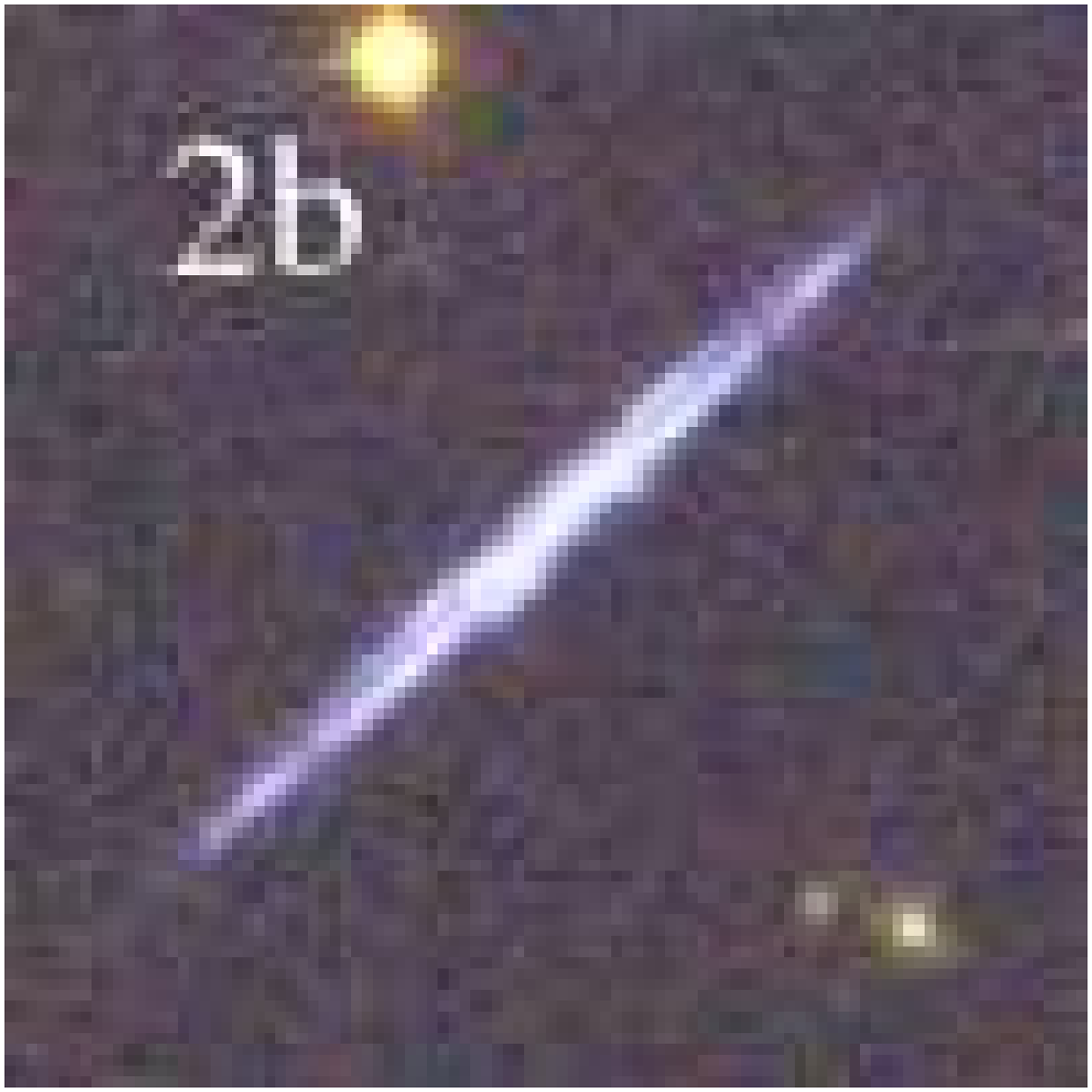}}
    & \multicolumn{1}{m{2.2cm}}{ } \\
      \multicolumn{1}{m{2.1cm}}{{\Large MODEL}}
    & \multicolumn{1}{m{2.2cm}}{\includegraphics[height=2.50cm,clip]{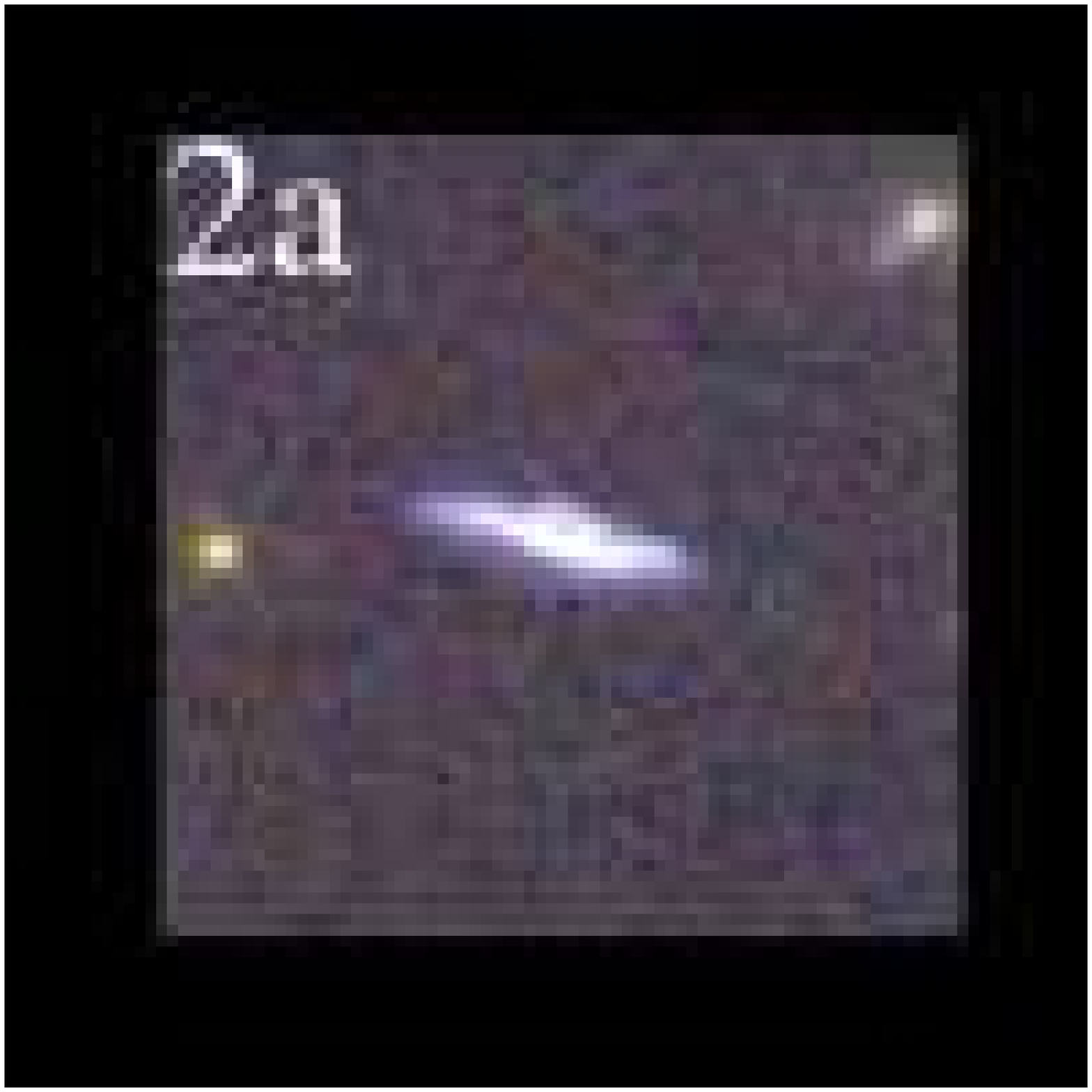}}
    & \multicolumn{1}{m{2.2cm}}{\includegraphics[height=2.50cm,clip]{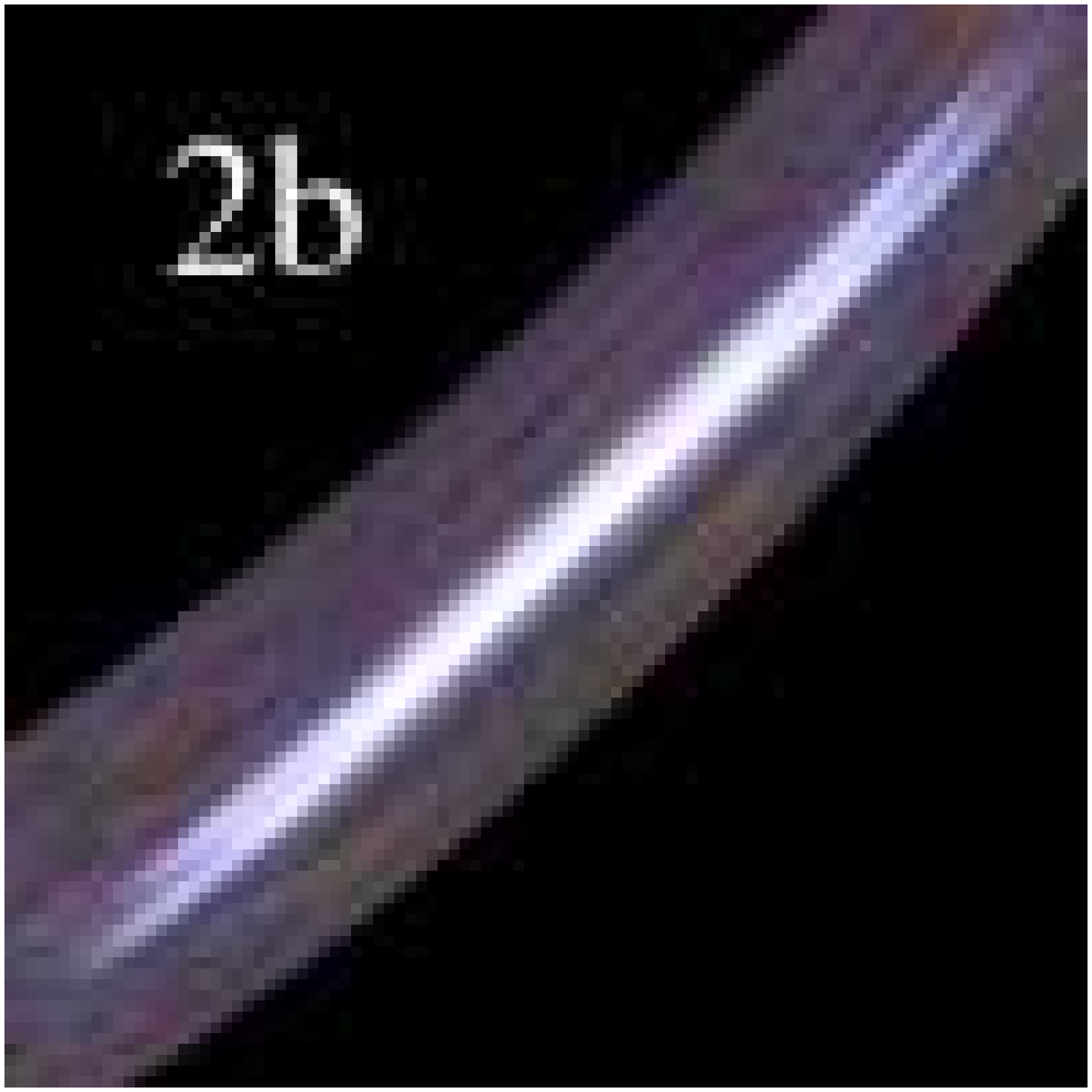}}
    & \multicolumn{1}{m{2.2cm}}{ } \\
  \end{tabular}
  \caption{Image system 2:}\vspace{0mm}
  \label{fig:images:2}
\end{figure*}

\begin{figure*}
  \vspace{5mm}
  \begin{tabular}{ccccccccccc}
    \multicolumn{1}{m{2.1cm}}{{\Large RX J1347}}
    & \multicolumn{1}{m{2.2cm}}{\includegraphics[height=2.50cm,clip]{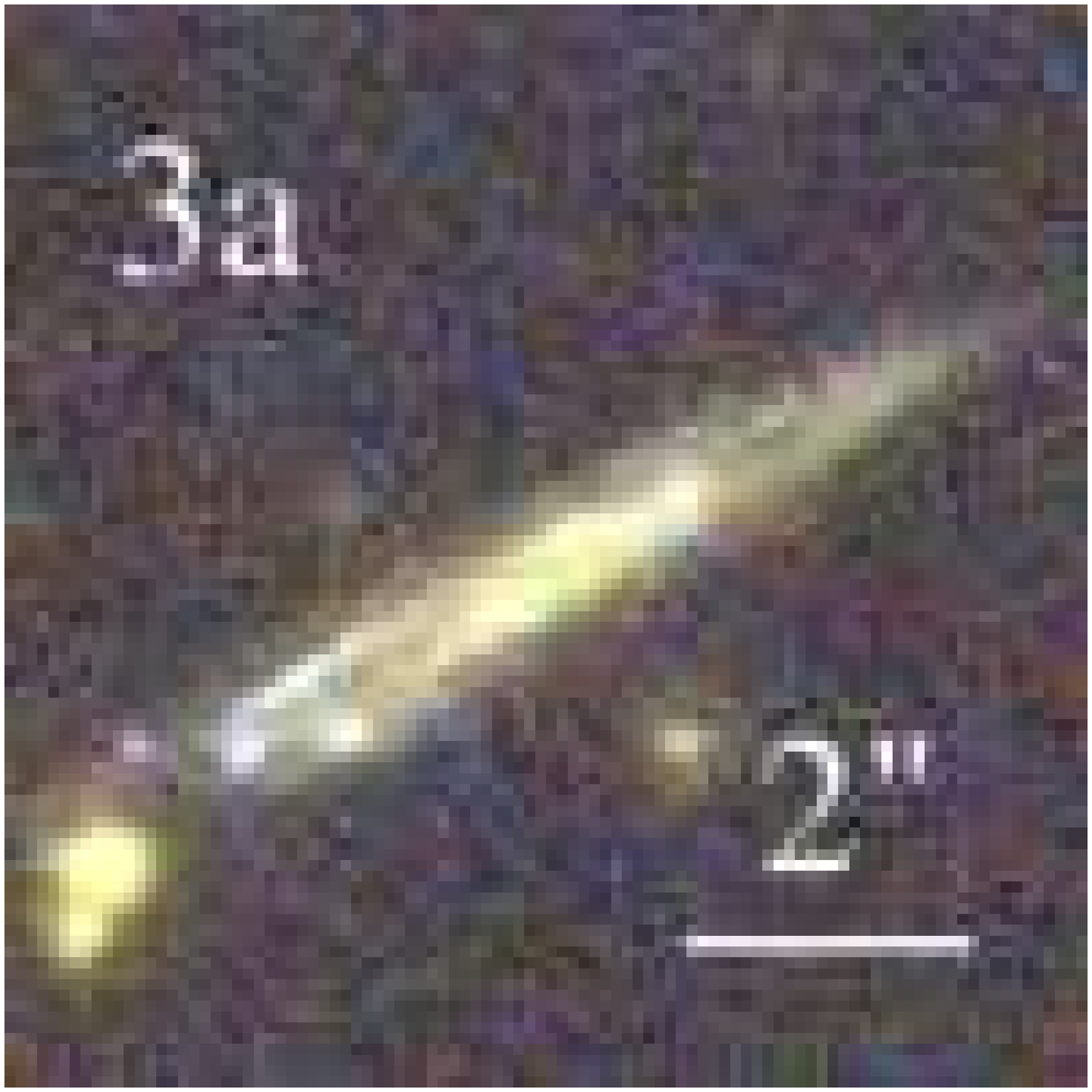}}
    & \multicolumn{1}{m{0,0cm}}{ }
    & \multicolumn{1}{m{2.2cm}}{\includegraphics[height=2.50cm,clip]{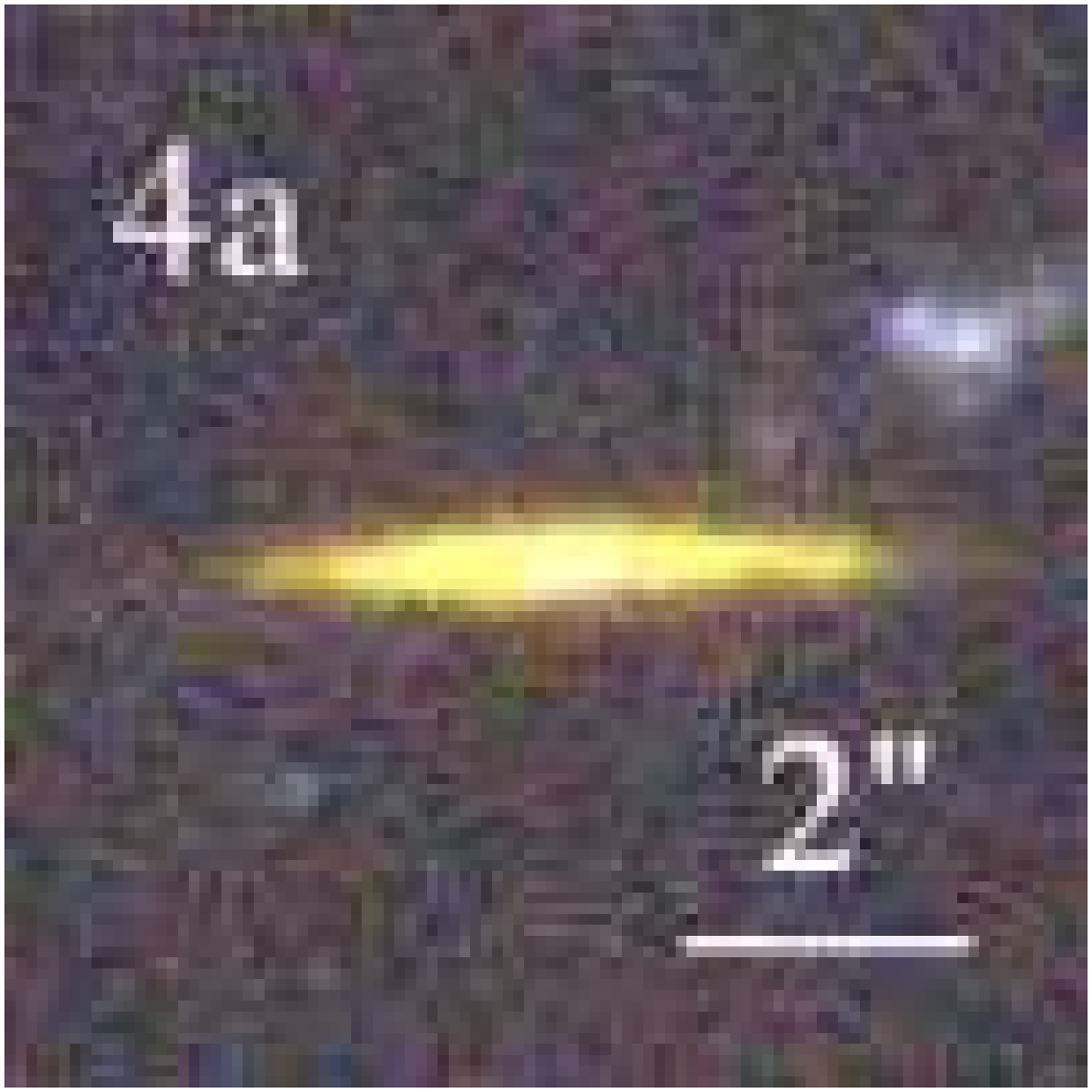}}
    & \multicolumn{1}{m{0,0cm}}{ }
    & \multicolumn{1}{m{2.2cm}}{\includegraphics[height=2.50cm,clip]{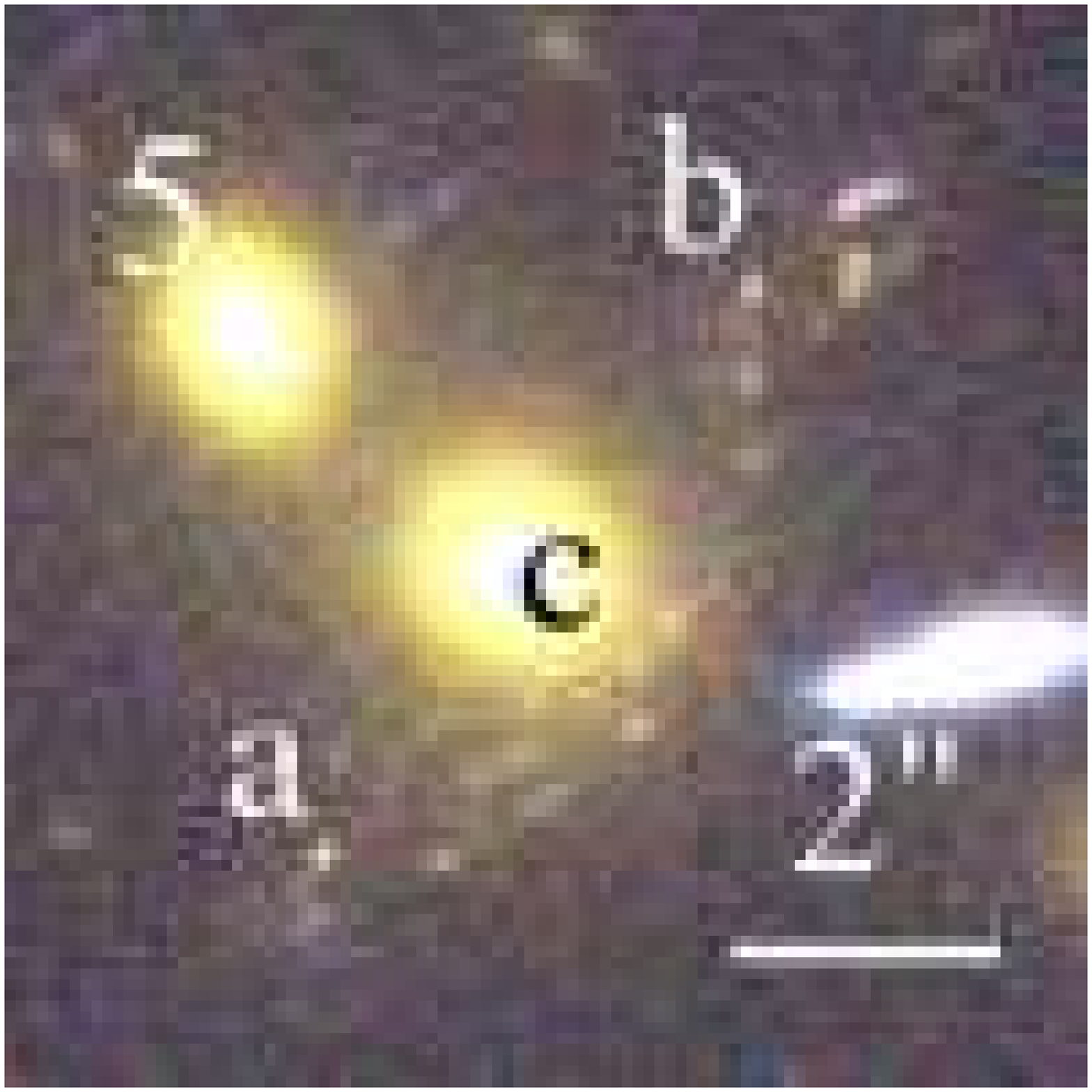}}
    & \multicolumn{1}{m{0,0cm}}{ }
    & \multicolumn{1}{m{2.2cm}}{\includegraphics[height=2.50cm,clip]{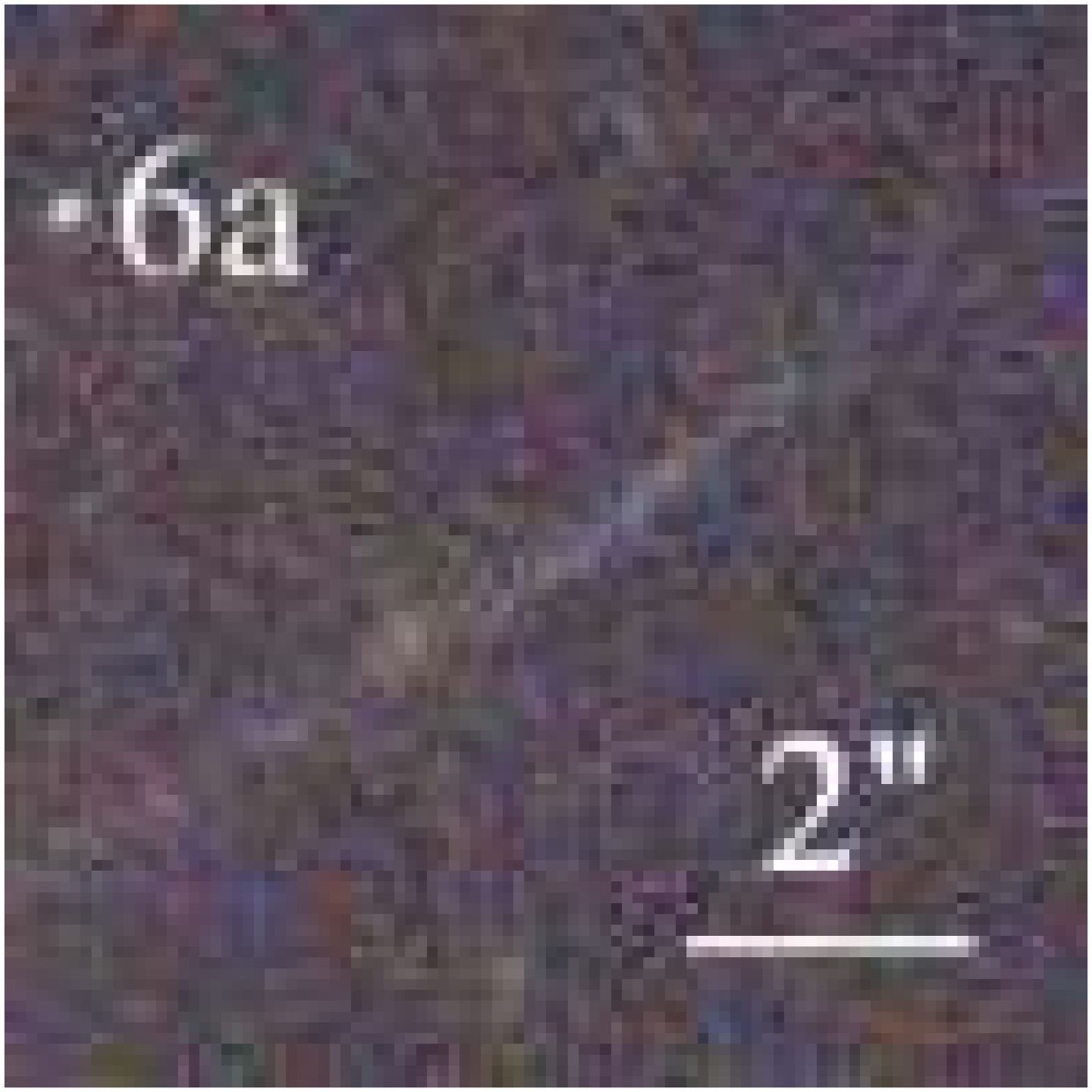}}
    & \multicolumn{1}{m{0,0cm}}{ }
    & \multicolumn{1}{m{2.2cm}}{\includegraphics[height=2.50cm,clip]{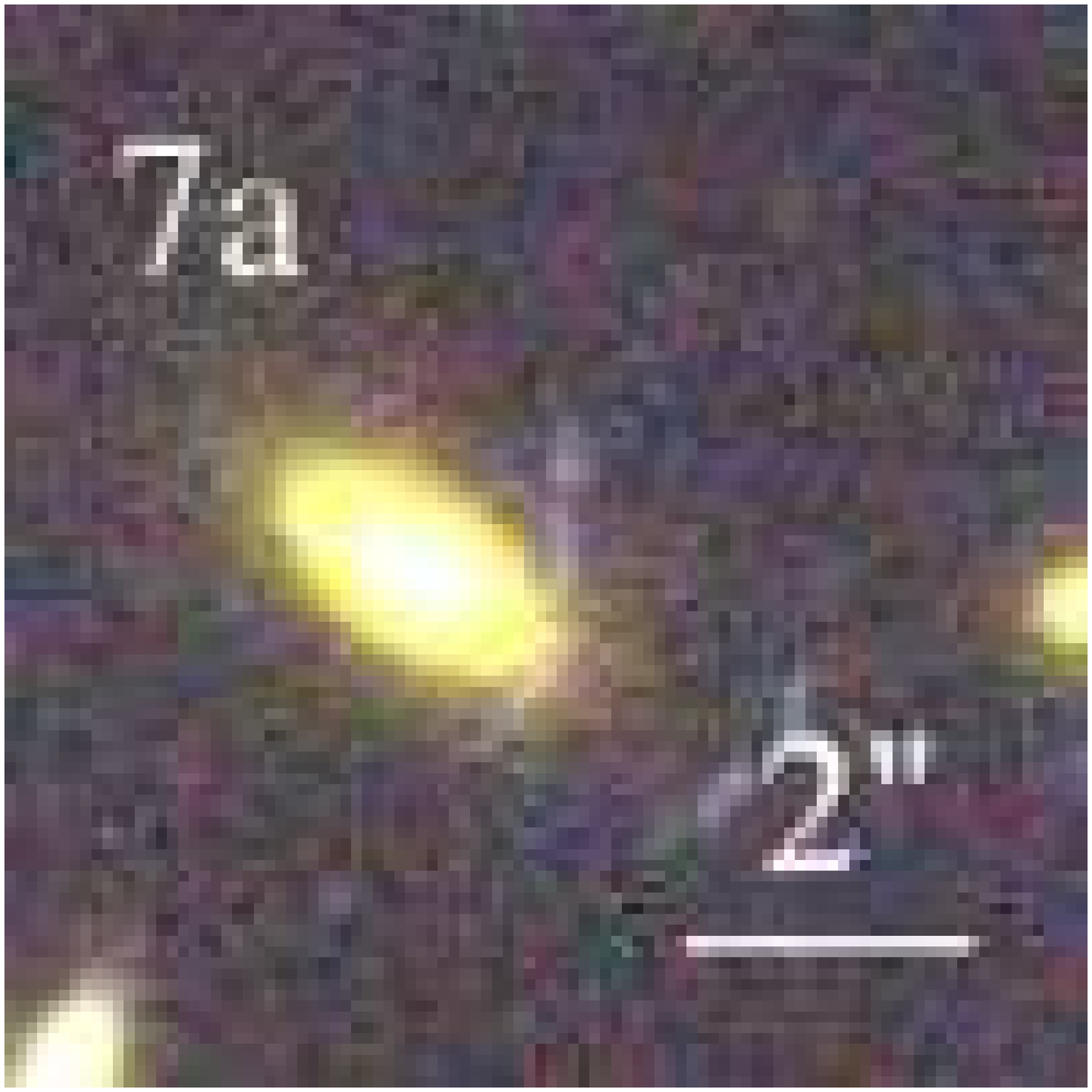}}
    & \multicolumn{1}{m{2.2cm}}{ } \\
  \end{tabular}
  \caption{Image systems 3-7:}\vspace{0mm}
  \label{fig:images:3-7}
\end{figure*}

\begin{figure*}
  \vspace{5mm}
  \begin{tabular}{ccccccccc}
    \multicolumn{1}{m{2.1cm}}{{\Large RX J1347}}
    & \multicolumn{1}{m{2.2cm}}{\includegraphics[height=2.50cm,clip]{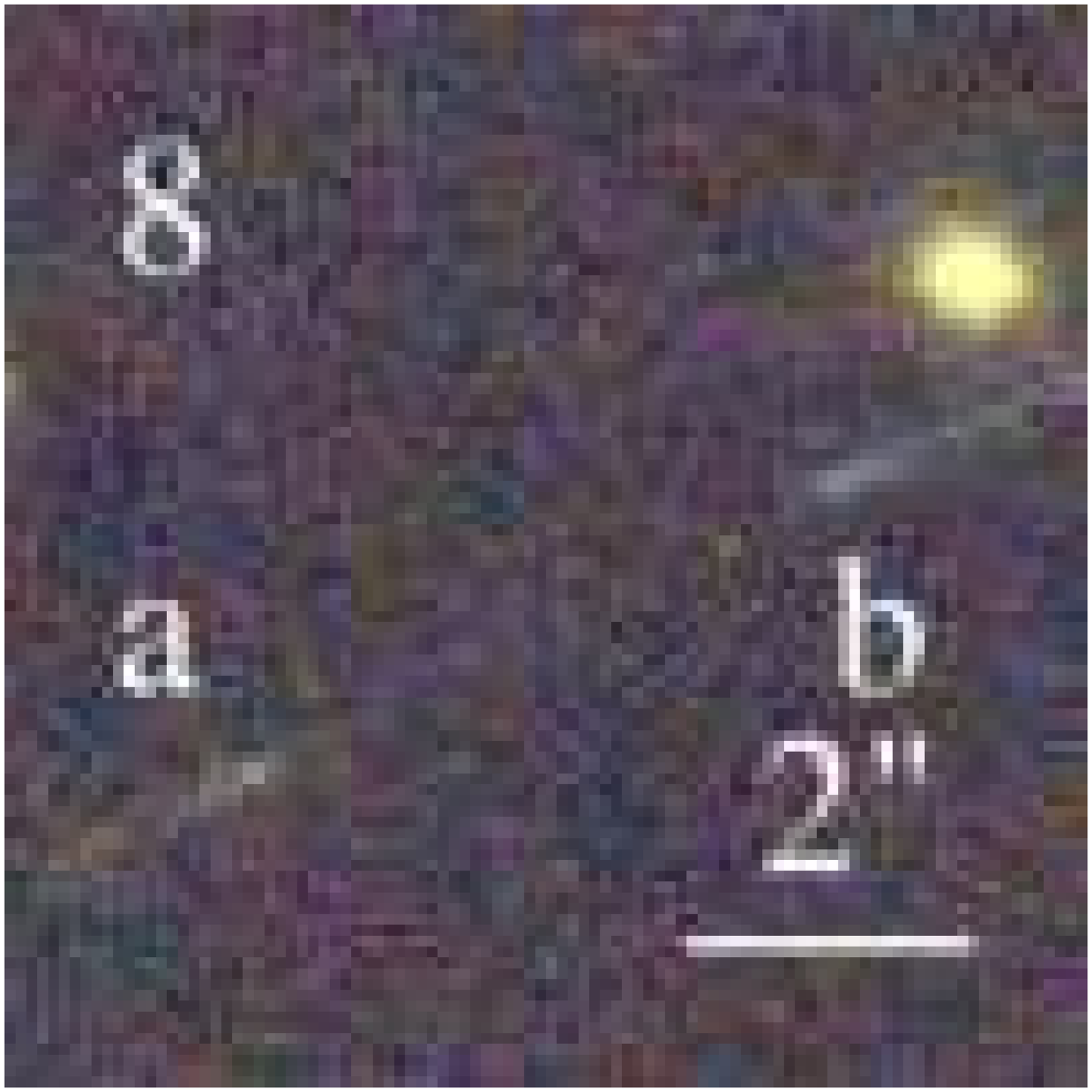}}
    & \multicolumn{1}{m{2.2cm}}{\includegraphics[height=2.50cm,clip]{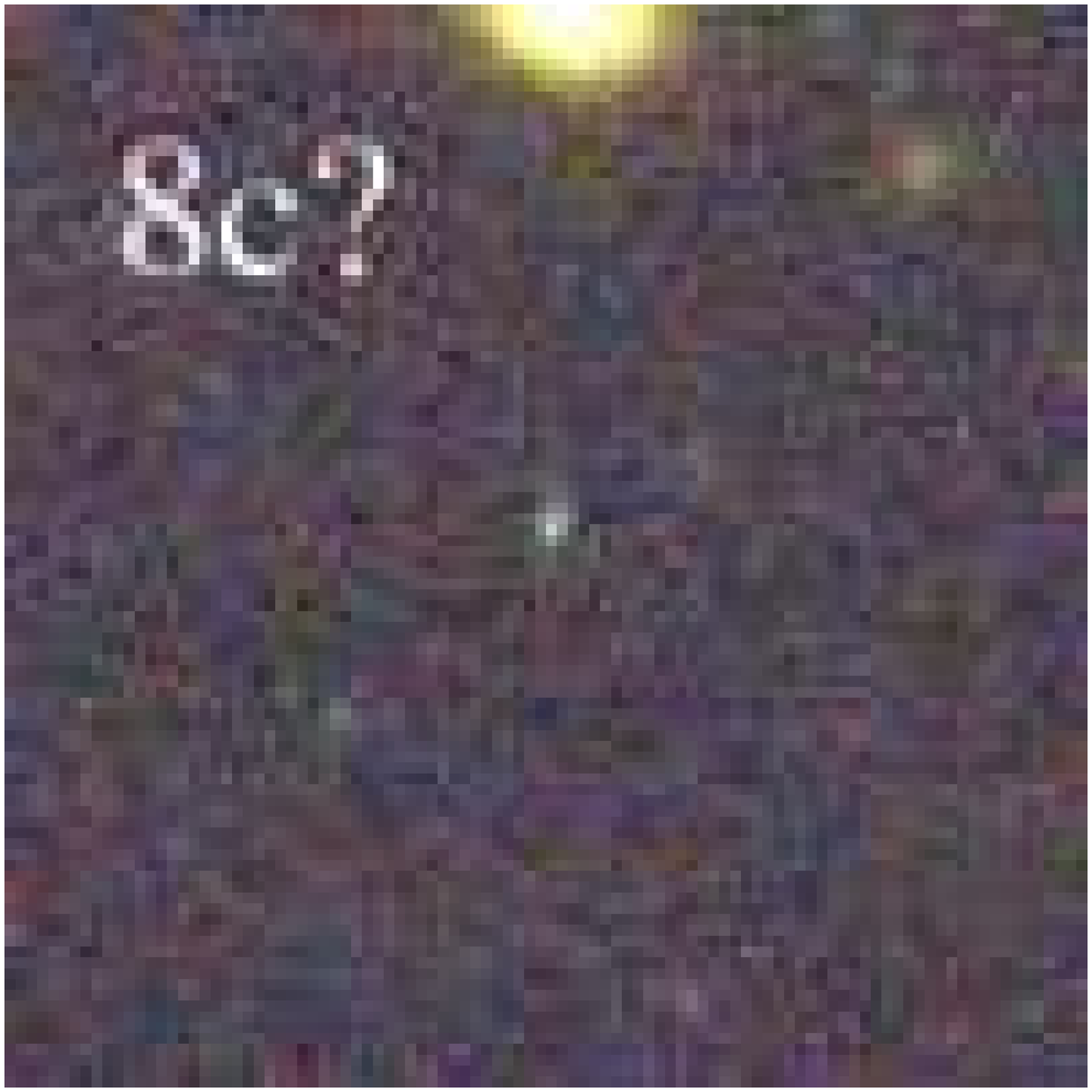}}
    & \multicolumn{1}{m{0,0cm}}{ }
    & \multicolumn{1}{m{2.2cm}}{\includegraphics[height=2.50cm,clip]{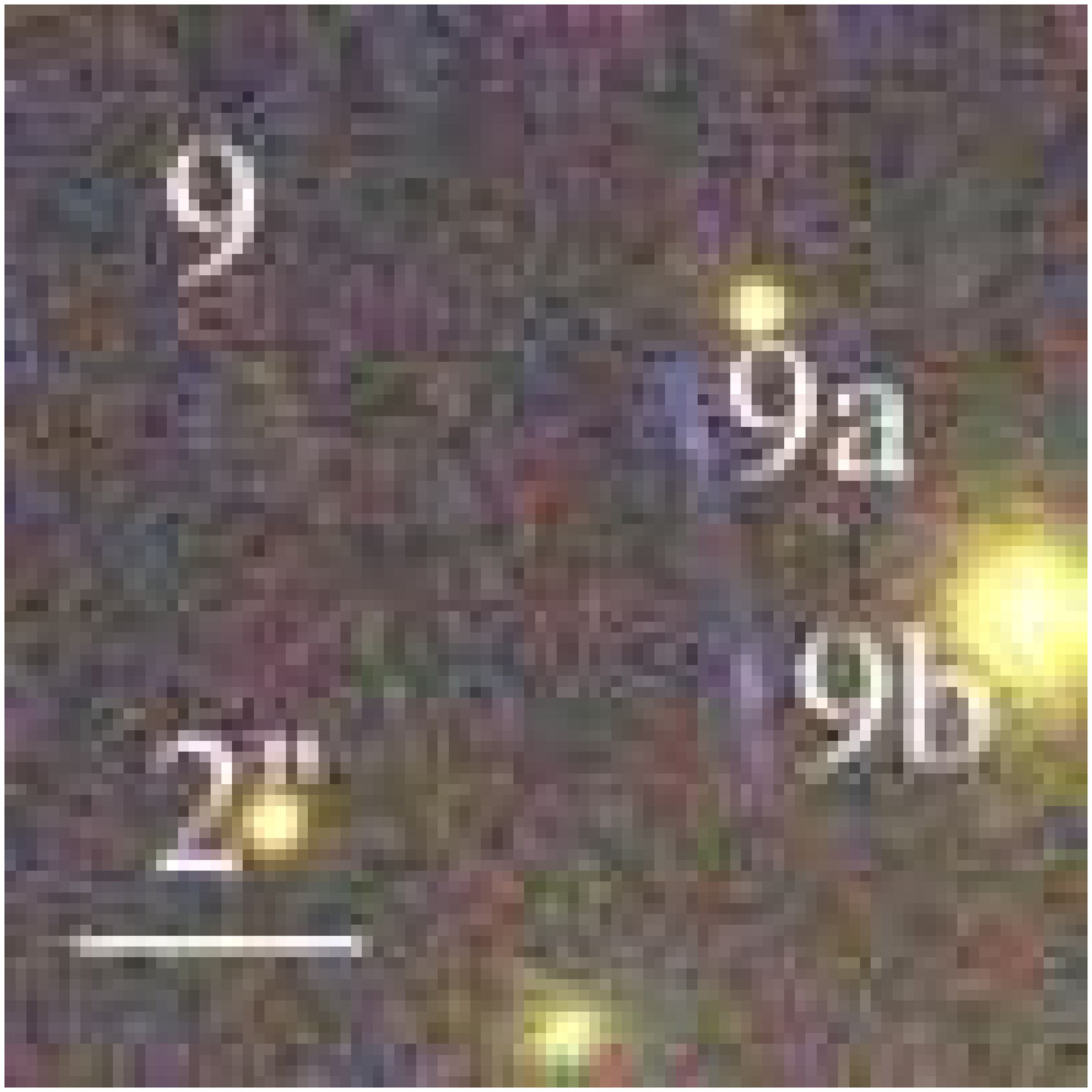}}
    & \multicolumn{1}{m{2.2cm}}{\includegraphics[height=2.50cm,clip]{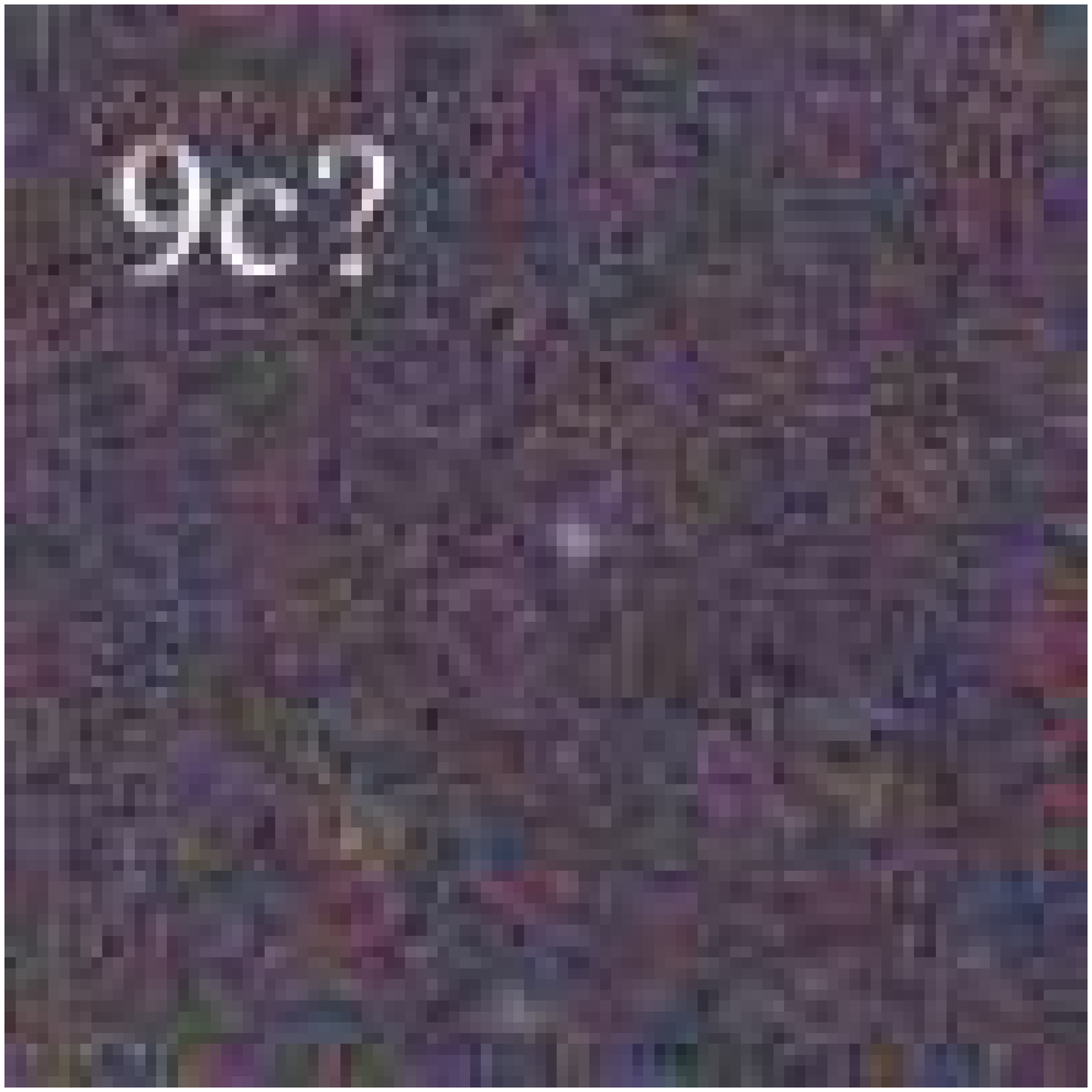}}
    & \multicolumn{1}{m{0,0cm}}{ }
    & \multicolumn{1}{m{2.2cm}}{\includegraphics[height=2.50cm,clip]{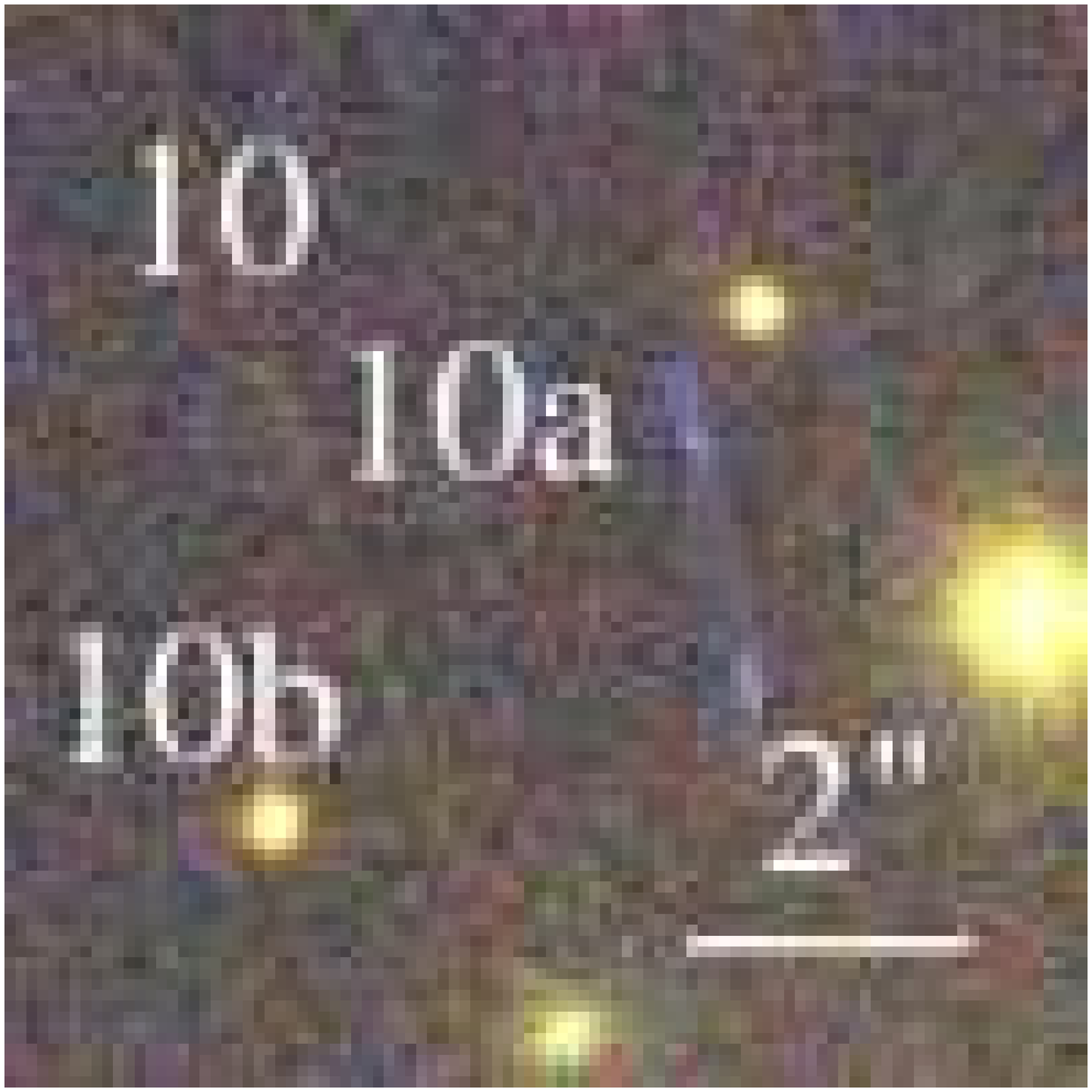}}
    & \multicolumn{1}{m{2.2cm}}{ } \\
  \end{tabular}
  \caption{Image systems 8, 9 and 10:}\vspace{0mm}
  \label{fig:images:8-10}
\end{figure*}

\begin{figure*}
  \vspace{5mm}
  \begin{tabular}{cccccc}
    \multicolumn{1}{m{2.1cm}}{{\Large RX J1347}}
    & \multicolumn{1}{m{2.2cm}}{\includegraphics[height=2.50cm,clip]{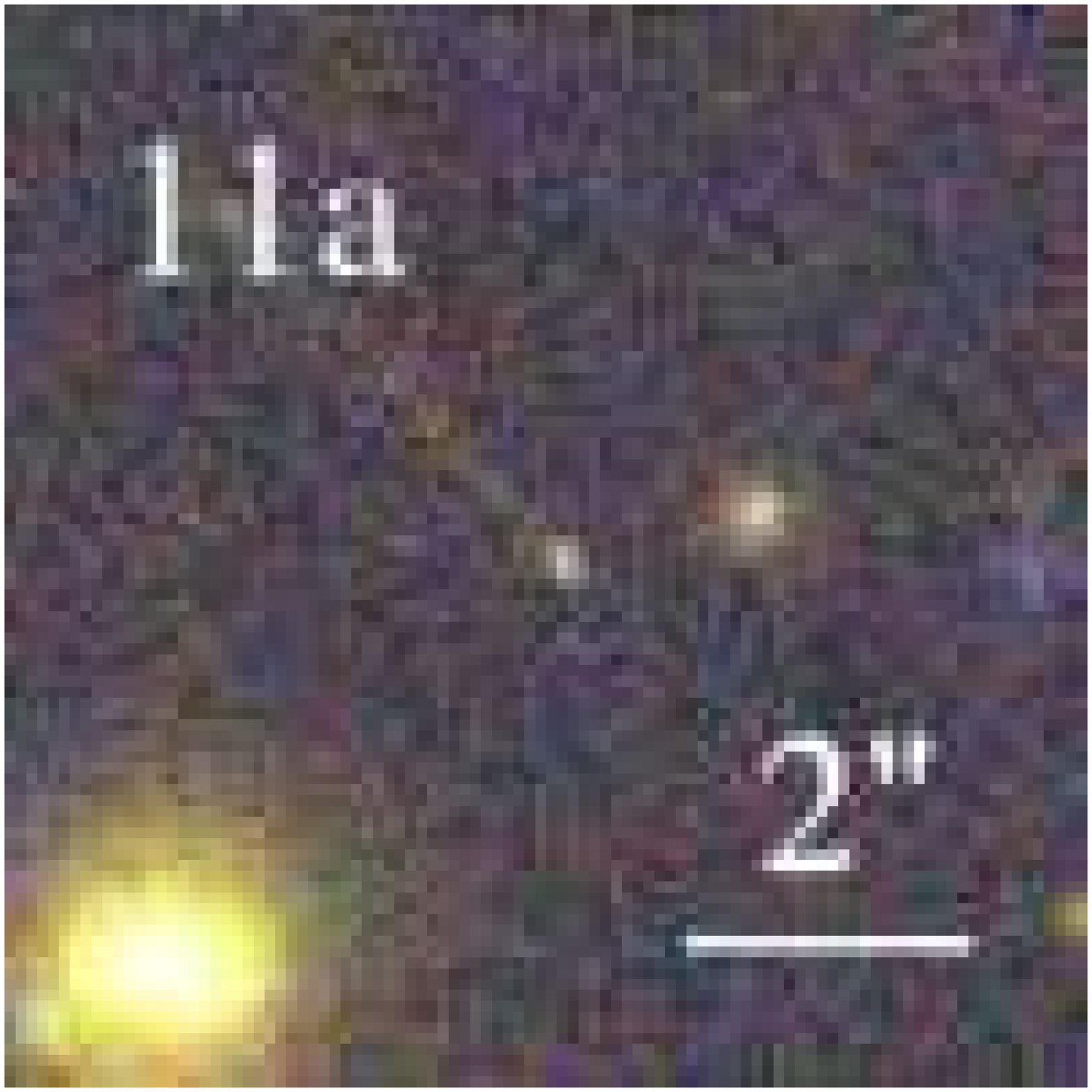}}
    & \multicolumn{1}{m{2.2cm}}{\includegraphics[height=2.50cm,clip]{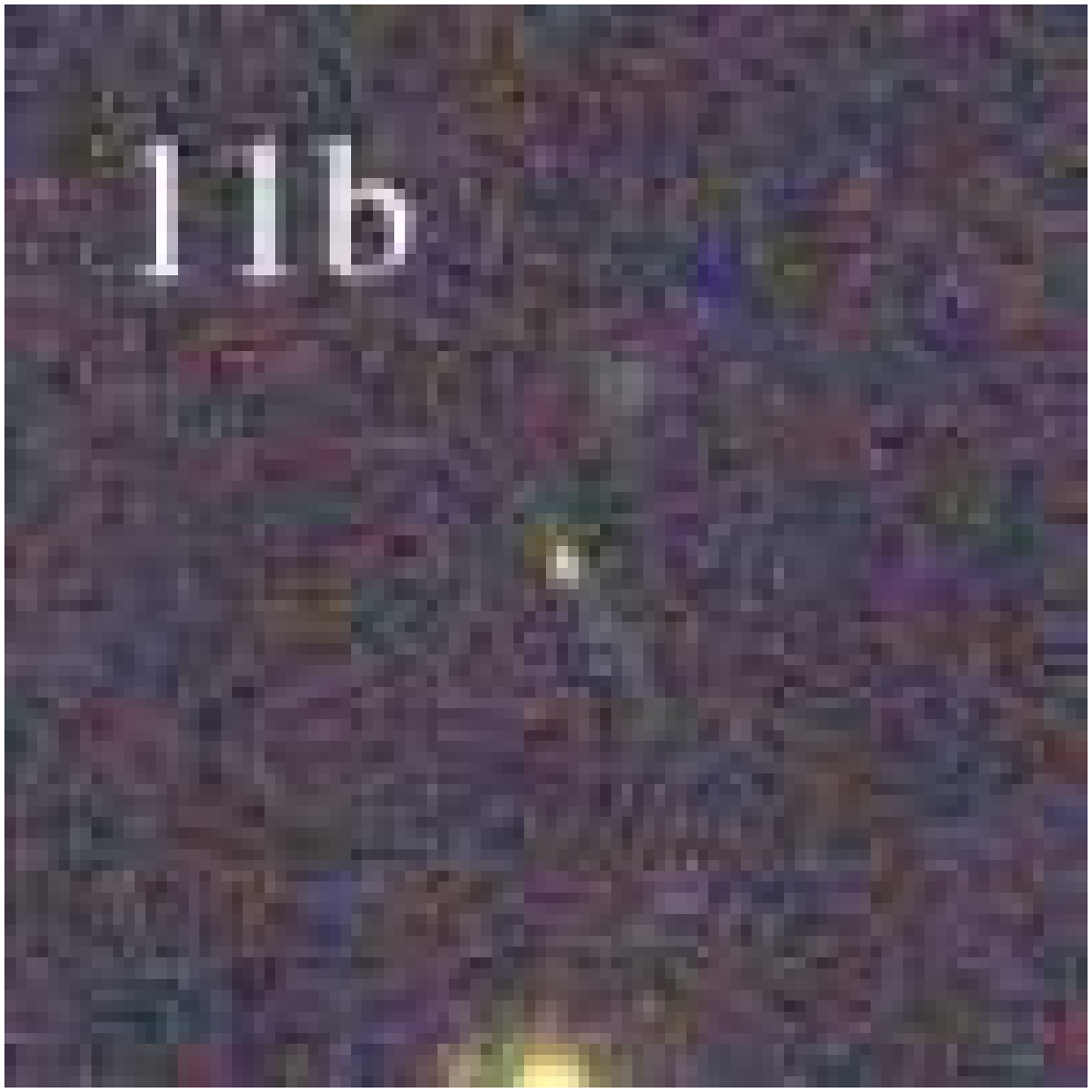}}
    & \multicolumn{1}{m{2.2cm}}{\includegraphics[height=2.50cm,clip]{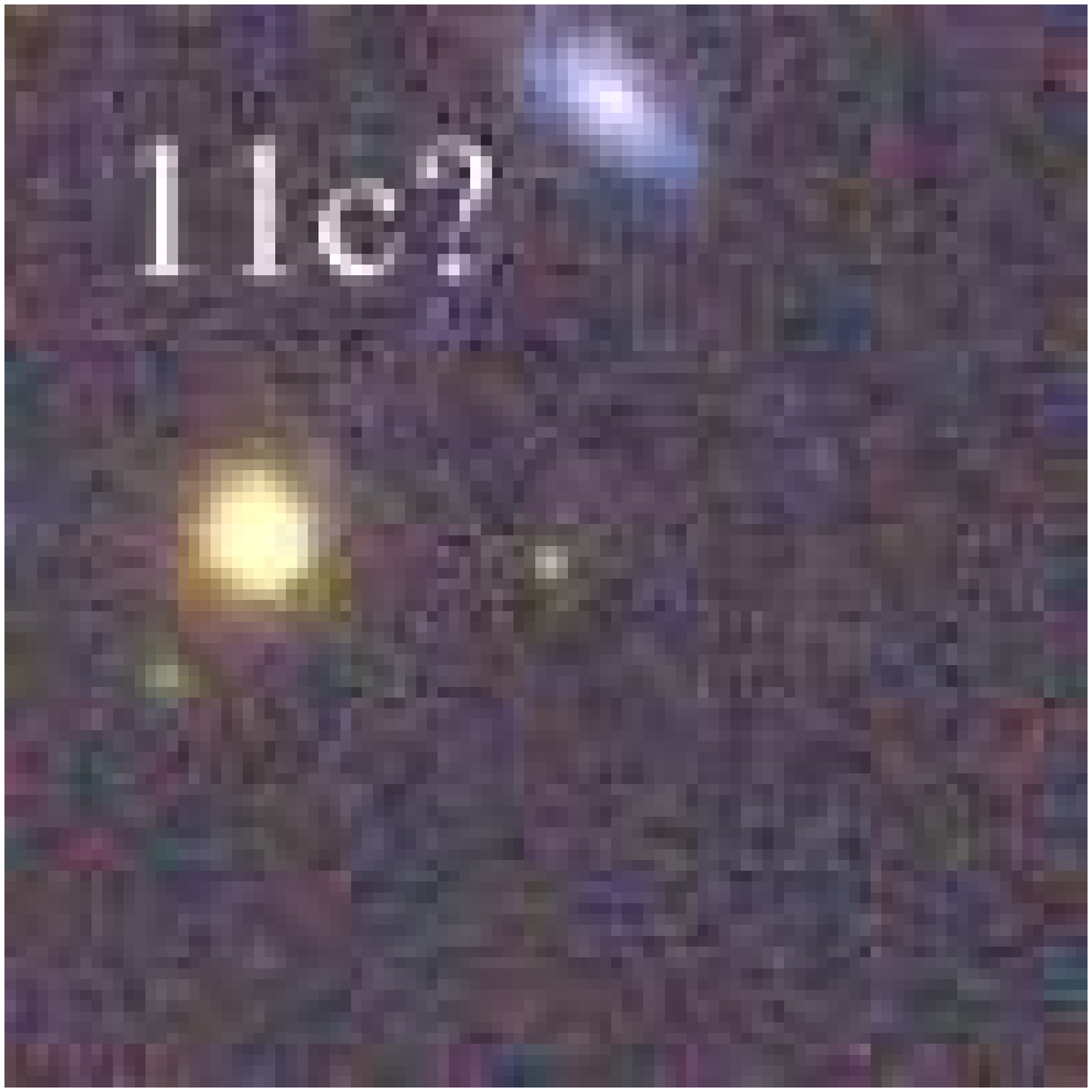}}
    & \multicolumn{1}{m{2.2cm}}{\includegraphics[height=2.50cm,clip]{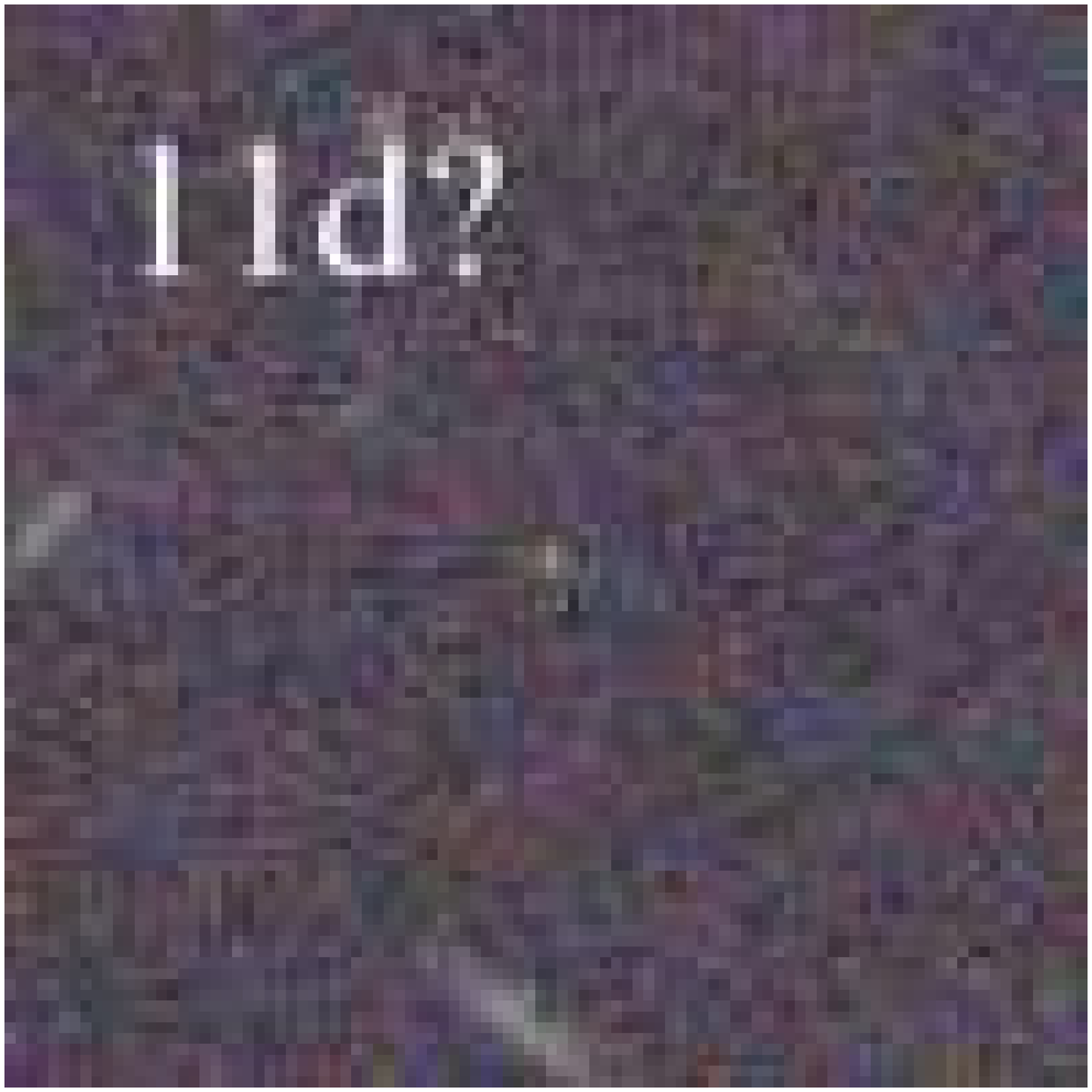}}
    & \multicolumn{1}{m{2.2cm}}{ } \\
  \end{tabular}
  \caption{Image system 11:}\vspace{0mm}
  \label{fig:images:11}
\end{figure*}

\begin{figure*}
  \vspace{5mm}
  \begin{tabular}{ccccccc}
    \multicolumn{1}{m{2.1cm}}{{\Large RX J1347}}
    & \multicolumn{1}{m{2.2cm}}{\includegraphics[height=2.50cm,clip]{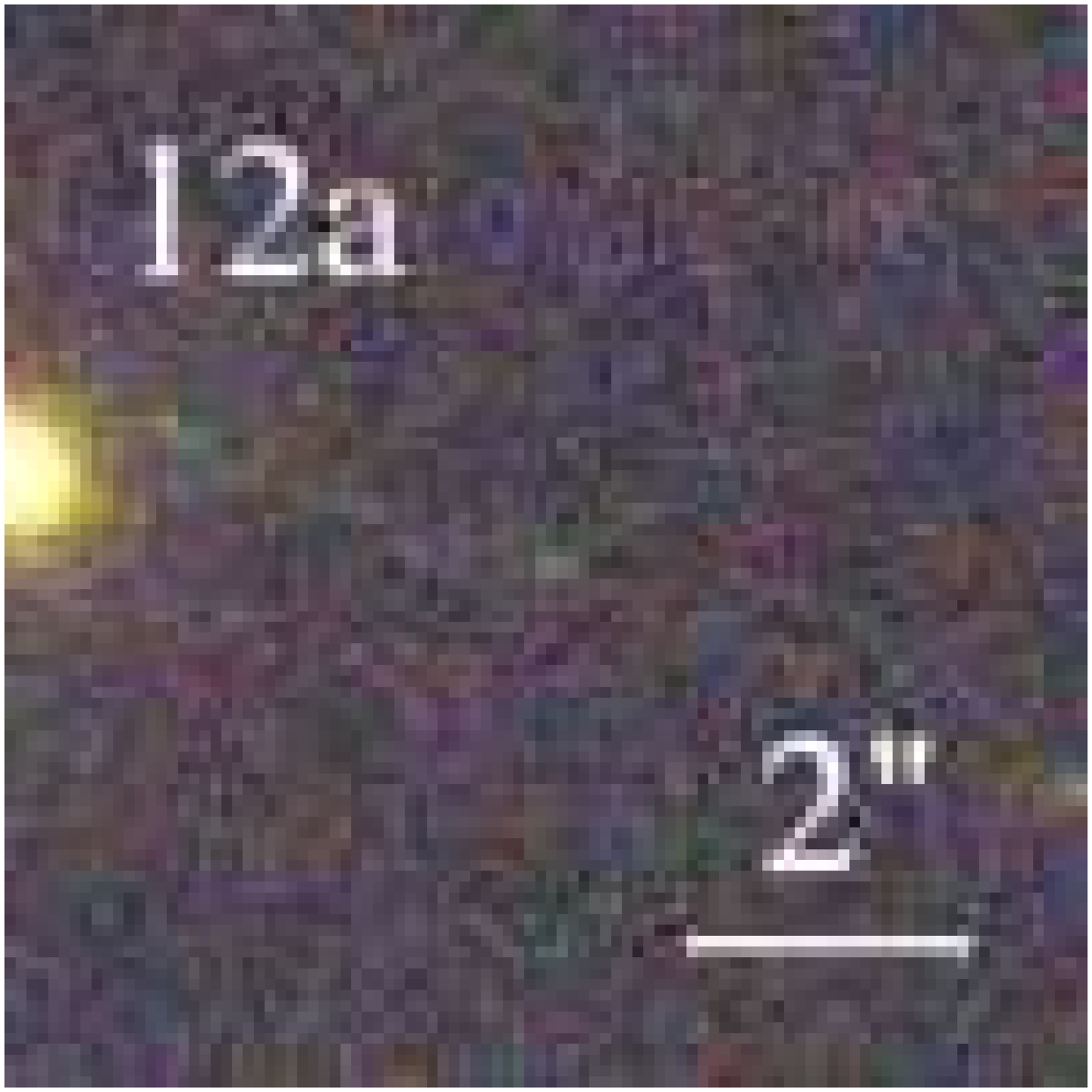}}
    & \multicolumn{1}{m{2.2cm}}{\includegraphics[height=2.50cm,clip]{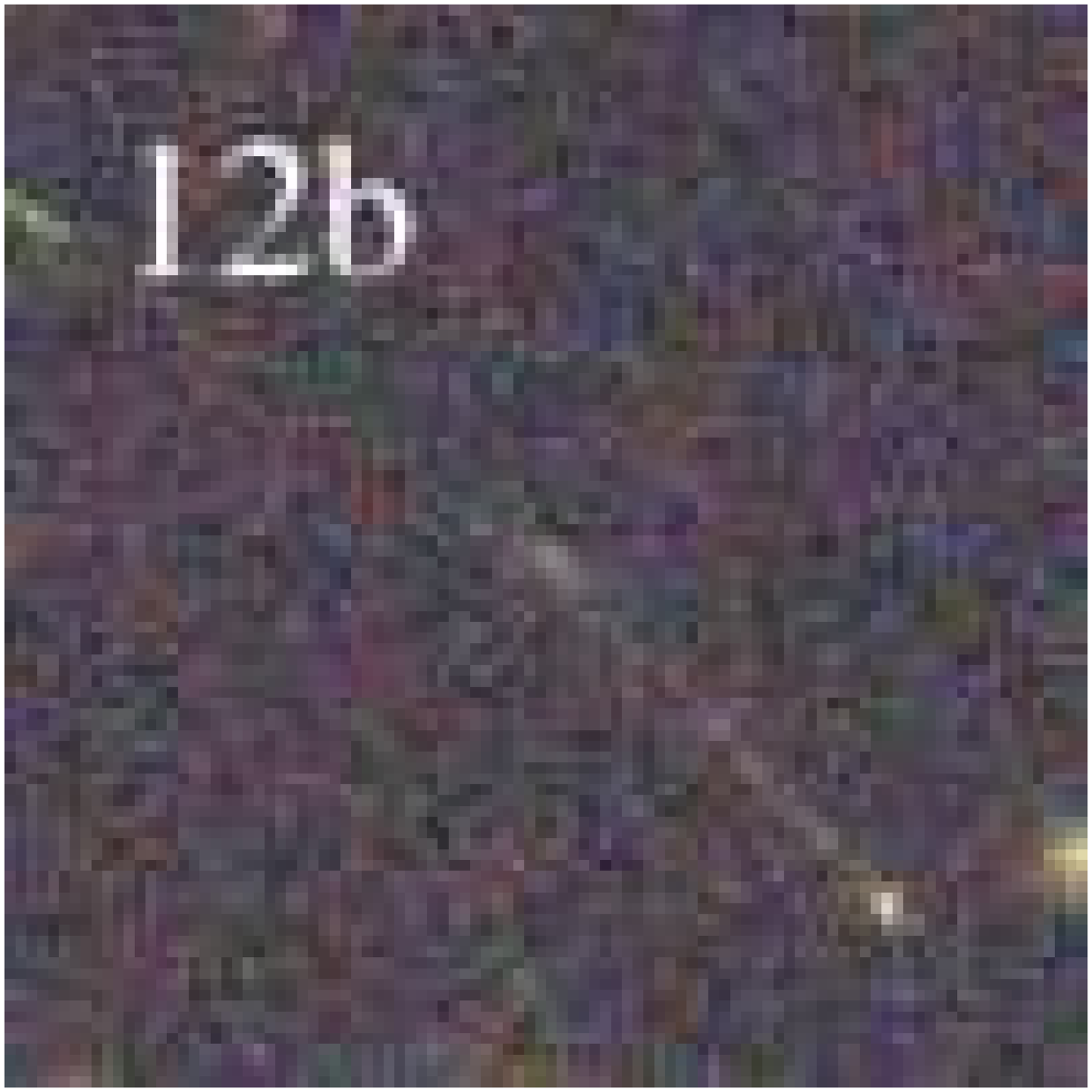}}
    & \multicolumn{1}{m{2.2cm}}{\includegraphics[height=2.50cm,clip]{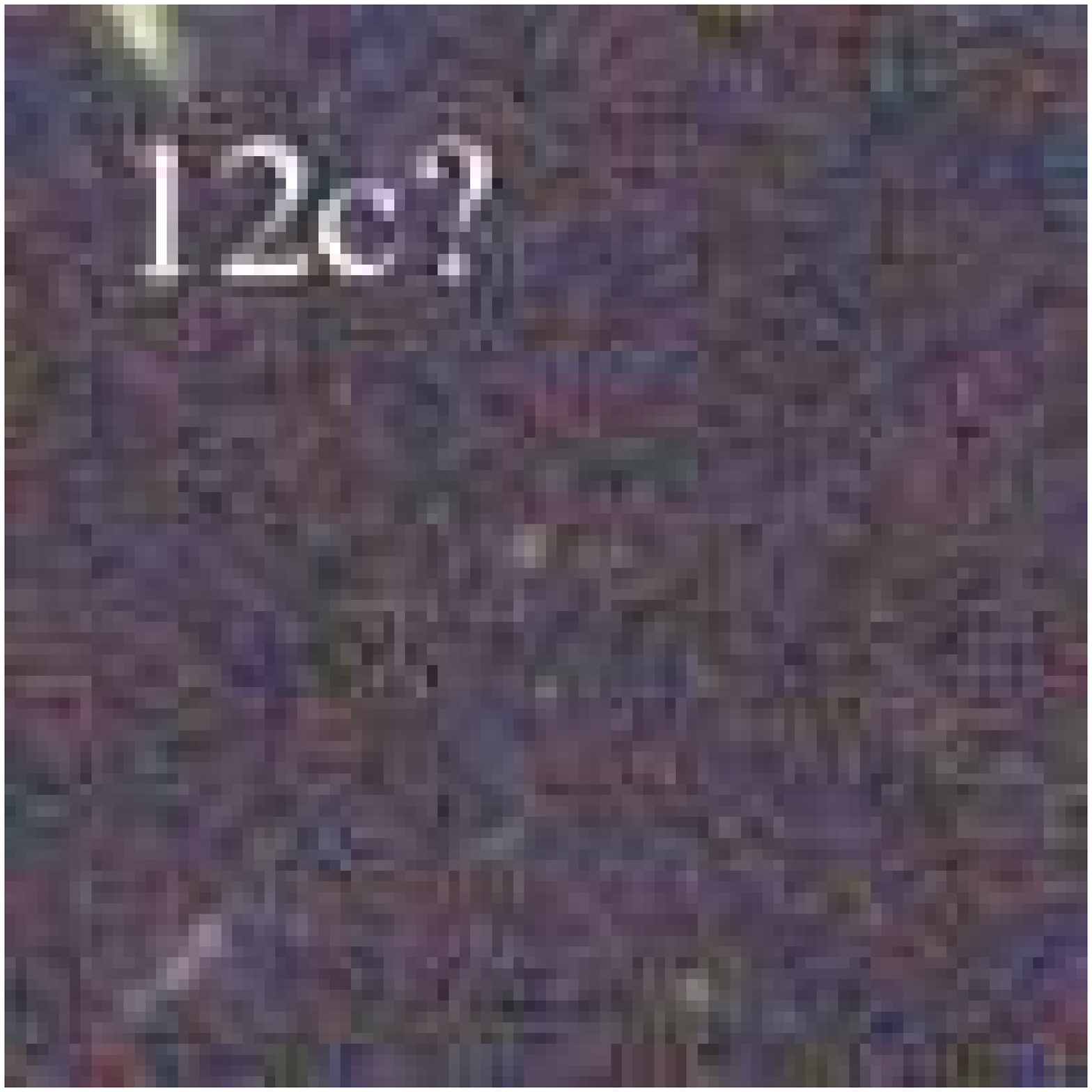}}
    & \multicolumn{1}{m{0,0cm}}{ }
    & \multicolumn{1}{m{2.2cm}}{\includegraphics[height=2.50cm,clip]{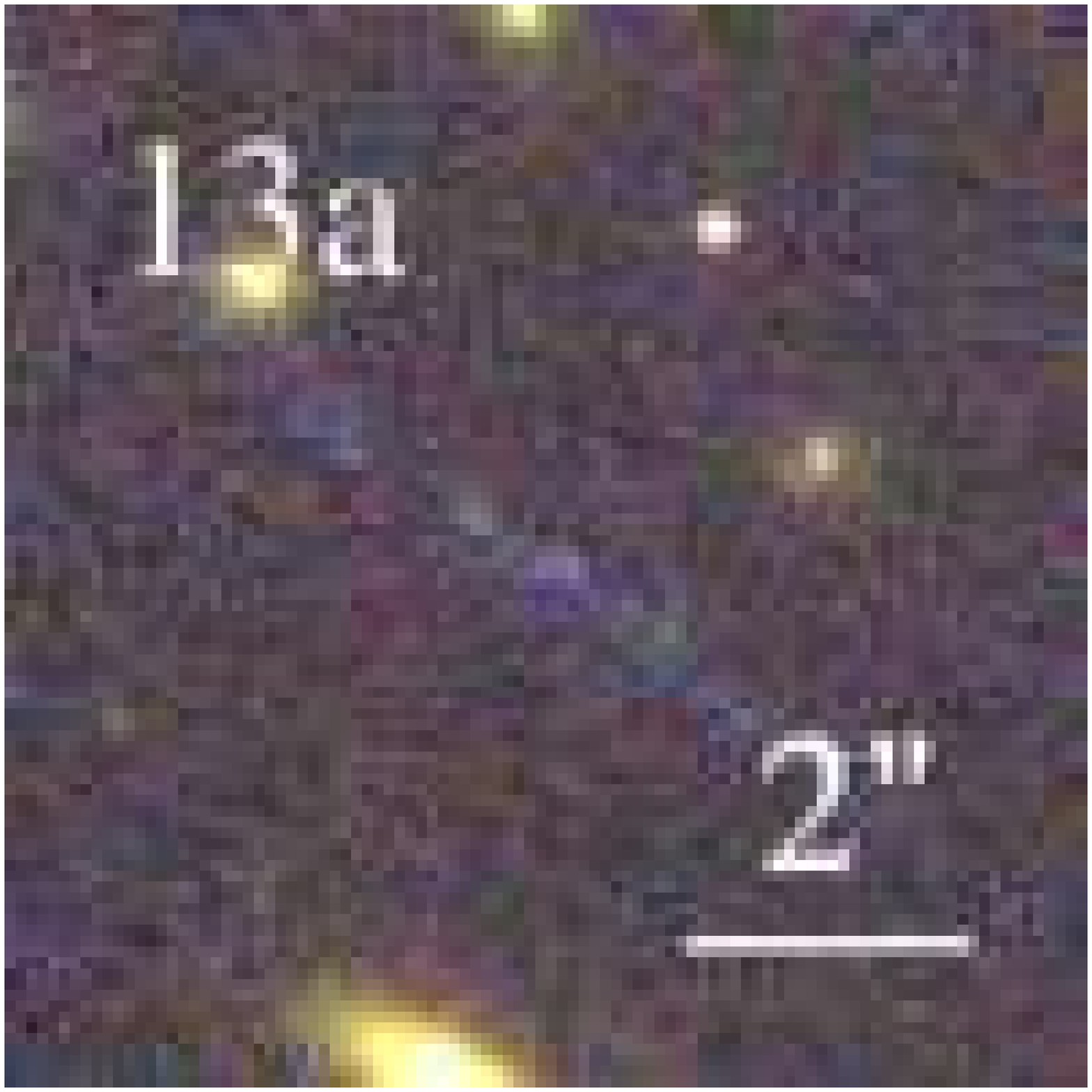}}
    & \multicolumn{1}{m{2.2cm}}{ } \\
  \end{tabular}
  \caption{Image systems 12 and 13:}\vspace{0mm}
  \label{fig:images:12-13}
\end{figure*}

The multiple images with some of their properties are listed in Table
\ref{tab:images}. The images in each image system are shown in Figs.
\ref{fig:images:1}-\ref{fig:images:12-13}. In addition we show for
multiple image systems 1 and 2 the images and predicted images in
Figs. \ref{fig:images:1} and \ref{fig:images:2}. In the figures the
top row shows the original 3-colour image created from the HST ACS
images in F475W (blue), F814W (green) and F850LP (red) pass bands. The
scales are shown on the images. The model prediction on the lower row
are created by taking the first image of an image system, delensing it
into the source plane using the lensing model and then relensing it
back to the image plane at the positions of the multiple images.  The
first multiple image in each system is therefore reproduced exactly
both in shape and position, while the remaining multiple images have
an offset and a distortion relative to the observed multiple image due
to the model. Note that this assumes that the source position of the
multiple image system is at the source position of the first image,
not the average source position which is used in calculating the
$\chi^2$. The sizes and positions are matched well indicating that
both the magnifications (size) and positions are well reproduced by
the model.  Image 1b has somewhat larger discrepancy both in
magnification and position (the positional offset is 3$''$). This can
be at least partly due to the presence of a cluster galaxy near the
multiple image position. A comparison of the orientations of the model
images with the observed ones (see Fig. \ref{fig:images:1}) can be
used to check the quality of the modelling since the orientations have
not been used in determining the mass distribution. The orentations
are in excellent agreement providing strong support for the mass
modelling presented in this paper. The mean separation between an
image in an image system and that of the models is $\sim1''$ in the
MCMC analysis.  Note that this is relatively large since the MCMC
chain does not explicitly try to minimise the $\chi^2$.  The best
models in the MCMC analysis have mean separations below $\sim0.5''$.

\begin{table*}
  \centering
  \caption{Positions, colours and redshifts estimates for the multiple
    images in this study. The mass profile and lensing redshifts are
    constrained by image systems 1 and 2 only. The more tentative
    images in a system are marked with a question mark.}
  \label{tab:images}
  \begin{tabular}{ccccccc}

    \hline
    \bf{Multiple}  &
    $\alpha$ &
    $\delta$ &
    \bf{M$_{F475W}$ - M$_{F814W}$}  &
    \bf{M$_{F814W}$ - M$_{F850LP}$} &
    z$_{\rm{spec}}$ & z$_{phot}$ \\ 

    \bf{image} &
    J2000 &
    J2000 & & & \\ 

    \hline
    \hline

     1a  & 13:47:33.009 & $-$11:45:27.28 &    0.23 $\pm$ 0.03 & $-$1.04 $\pm$ 0.05 &           &        -        \\ 
     1b  & 13:47:32.432 & $-$11:44:54.23 &    0.12 $\pm$ 0.03 & $-$1.11 $\pm$ 0.05 &           &        -        \\ 
     1c  & 13:47:30.906 & $-$11:45:12.14 &    0.11 $\pm$ 0.06 & $-$0.99 $\pm$ 0.09 &           &        -        \\ 
     1d  & 13:47:30.635 & $-$11:45:33.84 &    0.15 $\pm$ 0.03 & $-$1.10 $\pm$ 0.05 &           & $2.19 \pm 0.05$ \\ 
     1e  & 13:47:28.703 & $-$11:44:50.54 &    0.22 $\pm$ 0.03 & $-$1.06 $\pm$ 0.05 &           & $2.20 \pm 0.15$ \\ 
    \hline

     2a  & 13:47:31.834 & $-$11:45:51.80 &    0.24 $\pm$ 0.01 & $-$1.00 $\pm$ 0.01 & 1.75$^1$  & $0.01 \pm 0.05$ \\ 
     2b  & 13:47:29.283 & $-$11:45:39.59 &    0.17 $\pm$ 0.01 & $-$1.00 $\pm$ 0.01 & 1.75$^1$  & $0.01 \pm 0.02$ \\ 
    \hline

     3a  & 13:47:32.027 & $-$11:44:41.98 &    1.39 $\pm$ 0.01 & $-$0.95 $\pm$ 0.01 & 0.806$^2$ & $1.26 \pm 0.02$ \\ 
    \hline

     4a  & 13:47:31.148 & $-$11:44:38.87 &    2.59 $\pm$ 0.03 & $-$0.76 $\pm$ 0.01 & 0.785$^2$ & $1.32 \pm 0.02$ \\ 
    \hline

     5a  & 13:47:27.988 & $-$11:45:56.48 &    1.68 $\pm$ 0.16 & $-$1.07 $\pm$ 0.10 &           &        -        \\ 
     5b  & 13:47:27.720 & $-$11:45:51.28 &    1.77 $\pm$ 0.16 & $-$1.02 $\pm$ 0.09 &           &        -        \\ 
     5c  & 13:47:27.793 & $-$11:45:54.65 &            -       & $-$        -       &           &        -        \\ 
    \hline

     6a  & 13:47:29.256 & $-$11:45:54.36 &    0.55 $\pm$ 0.12 & $-$1.63 $\pm$ 0.25 &           & $0.57 \pm 0.14$ \\ 
    \hline

     7a  & 13:47:34.797 & $-$11:45:01.50 &    2.45 $\pm$ 0.01 & $-$0.83 $\pm$ 0.00 &           & $0.44 \pm 0.06$ \\ 
    \hline

     8a  & 13:47:32.199 & $-$11:44:30.48 &    0.98 $\pm$ 0.13 & $-$1.50 $\pm$ 0.18 &           & $1.88 \pm 0.19$ \\ 
     8b  & 13:47:31.901 & $-$11:44:28.33 &    0.83 $\pm$ 0.10 & $-$1.73 $\pm$ 0.18 &           &        -        \\ 
     8c? & 13:47:34.185 & $-$11:45:05.25 &    1.00 $\pm$ 0.11 & $-$1.48 $\pm$ 0.14 &           & $3.68 \pm 0.11$ \\ 
    \hline

     9a  & 13:47:30.826 & $-$11:44:57.04 & $-$0.43 $\pm$ 0.10 & $-$1.72 $\pm$ 0.31 &          &        -        \\ 
     9b  & 13:47:30.798 & $-$11:44:58.81 & $-$0.16 $\pm$ 0.10 & $-$1.15 $\pm$ 0.19 &          &        -        \\ 
     9c? & 13:47:32.598 & $-$11:45:40.95 &    0.49 $\pm$ 0.14 & $-$1.55 $\pm$ 0.25 &          & $0.67 \pm 1.73$ \\ 
    \hline

    10a  & 13:47:31.031 & $-$11:44:56.44 &            -       & $-$        -       &           &        -        \\ 
    10b  & 13:47:30.974 & $-$11:44:58.38 &            -       & $-$        -       &           &        -        \\ 
    \hline

    11a  & 13:47:29.399 & $-$11:44:47.26 &    2.01 $\pm$ 0.19 & $-$1.04 $\pm$ 0.10 &           & $2.94 \pm 0.23$ \\ 
    11b  & 13:47:29.078 & $-$11:44:54.09 &    1.91 $\pm$ 0.18 & $-$1.29 $\pm$ 0.11 &           & $3.61 \pm 0.20$ \\ 
    11c? & 13:47:29.354 & $-$11:45:26.02 &    1.67 $\pm$ 0.20 & $-$1.13 $\pm$ 0.13 &           & $2.80 \pm 0.73$ \\ 
    11d? & 13:47:33.909 & $-$11:45:37.53 &    2.36 $\pm$ 0.59 & $-$1.12 $\pm$ 0.23 &           &        -        \\ 
    \hline

    12a  & 13:47:30.078 & $-$11:44:40.69 &    0.94 $\pm$ 0.23 & $-$1.47 $\pm$ 0.33 &           & $2.79 \pm 0.75$ \\ 
    12b  & 13:47:29.556 & $-$11:44:44.89 &    1.79 $\pm$ 0.27 & $-$1.55 $\pm$ 0.22 &           &        -        \\ 
    12c? & 13:47:33.989 & $-$11:45:34.45 &    1.74 $\pm$ 0.37 & $-$0.95 $\pm$ 0.21 &           & $1.75 \pm 1.09$ \\ 
    \hline

    13a  & 13:47:32.308 & $-$11:45:30.59 & $-$0.42 $\pm$ 0.13 & $-$1.13 $\pm$ 0.27 &           & $1.77 \pm 0.14$ \\ 
    \hline
    \hline
    
    \\
    \multicolumn{7}{l}{$^1$ Redshift from Lombardi et al. (2008, in preparation).}\\
    \multicolumn{7}{l}{$^2$ Redshift from \citet{ravindranath:02}.}\\

  \end{tabular}
\end{table*}


\clearpage

\label{lastpage}

\end{document}